\documentclass[aps,pra,10pt,twocolumn,groupedaddress,notitlepage,showpacs,floatfix]{revtex4-1}
\pdfoutput=1
\usepackage{graphicx,graphics,epsfig,subfigure,times,bm,bbm,amssymb,amsmath,amsfonts,amsthm,mathrsfs,MnSymbol}
\usepackage{gensymb}
\usepackage[matrix,frame,arrow]{xypic}
\usepackage[pdfstartview=FitH]{hyperref}
\usepackage{subfigure}
\hypersetup{
    colorlinks=true,       
    linkcolor=red,          
   citecolor=magenta,        
    filecolor=magenta,      
    urlcolor=cyan,           
    runcolor=cyan
}
\usepackage[pdftex]{color}
\usepackage{braket}
\usepackage{enumerate}
\usepackage[normalem]{ulem}
\usepackage[usenames,dvipsnames]{xcolor}
\usepackage{multirow}
\usepackage{mathtools}

\definecolor{orange}{rgb}{1,0.5,0}

\newcommand{\summm}[3]{\sum_{\substack{ #1 \\ #2 \\ #3 }}}
\newcommand{\summ}[2]{\sum_{\substack{ #1 \\ #2}}}
\newcommand{\po}[2]{\hat{\sigma}_{#1}^{#2}}

\newcommand{\inner}[2]{\langle #1 | #2 \rangle}

\newcommand{\ignore}[1]{}

\newcommand{\be}{\begin{equation}}
\newcommand{\ee}{\end{equation}}

\newcommand{\bes} {\begin{subequations}}
\newcommand{\ees} {\end{subequations}}
\newcommand{\bea} {\begin{eqnarray}}
\newcommand{\eea} {\end{eqnarray}}

\definecolor{gold}{rgb}{0.85,.66,0}

\newcommand{\Z}{\ensuremath{\mathbb{Z}}} 
\newcommand{\e}{\ensuremath{{e}}} 
\newcommand{\abs}[1]{\ensuremath{\left|#1\right|}} 

\newcommand{\beq}{\begin{equation}}

\newcommand{\eeq}{\end{equation}}
\newcommand{\avg}[1]{\left< #1 \right>}

\newcommand{\mc}[1]{\mathcal{#1}}

\def\id{\mathbf{1}}
\def\gn{g}
\def\gt{\tilde{g}}
\def\gb{\bar{g}}

\def\bb{\bar{b}}
\def\fz{\mathcal{Z}}

\def\zt{\tilde{z}}
\def\zn{z}
\def\up{\ket{\Uparrow}}

\def\tr{\mathrm{Tr}}

\def\del{\partial}

\def\g{\gamma}

\def\e{\epsilon}

\def\k{\kappa}

\def\s{\sigma}

\def\>{\rangle}
\def\<{\langle}
\def\Tr{\mathrm{Tr}}

\def\s0{I}

\newcommand{\ig}[1]{}

\begin{document}

\title{Local convertibility of the ground state of the perturbed Toric code}
\author{Siddhartha Santra}

\affiliation{Department of Physics and Center for Quantum Information Science \&
Technology, University of Southern California, Los Angeles, California
90089, USA}
\author{Alioscia Hamma}

\affiliation{Center for Quantum Information, Institute for Interdisciplinary Information Sciences, Tsinghua University, Beijing 100084, P.R. China}

\author{Lukasz Cincio}
\affiliation{Perimeter Institute for Theoretical Physics, 31 Caroline St. N, N2L 2Y5, Waterloo ON, Canada}

\author{Yigit Subasi}
\affiliation{Joint Quantum Institute and Maryland Center for Fundamental Physics, University of Maryland, College Park, Maryland 20742}

\author{Paolo Zanardi}
\affiliation{Department of Physics and Center for Quantum Information Science and
Technology, University of Southern California, Los Angeles, California
90089, USA}

\author{Luigi Amico}
\affiliation{CNR-MATIS-IMM \& Dipartimento di Fisica e Astronomia Universit\'{a} di Catania, Via S. Sofia 64, 95127 Catania, Italy}
\affiliation{Centre for Quantum Technologies, National University of Singapore, 3 Science Drive 2, Singapore 117543}

\begin{abstract}
We present analytical and numerical studies of the behaviour of the $\alpha$-Renyi entropies in the Toric code in presence of several types of perturbations aimed at studying the simulability of these perturbations to the parent Hamiltonian using local operations and classical communications (LOCC) - a property called local-convertibility. In particular, the derivatives, with respect to the perturbation parameter, present different signs for different values of $\alpha$ within the topological phase. From the information-theoretic point of view, this means that such ground states cannot be continuously deformed within the topological phase by means of catalyst assisted local operations and classical communications (LOCC). Such LOCC differential convertibility is on the other hand always possible in the trivial disordered phase. The non-LOCC convertibility is remarkable because it can be computed on a system whose size is independent of correlation length. This method can therefore constitute an experimentally feasible witness of topological order.

\end{abstract}
\maketitle

\section{Introduction}

In recent years a central thrust of research in quantum many-body theory and quantum information science has been the identification and characterization of novel phases of matter which cannot be adequately described by the Landau symmetry breaking mechanism \cite{goldenfeld}. These phases are generically exhibited by ground states of strongly interacting systems in two spatial dimensions. Quantum spin liquids \cite{yan:2011}, topological insulators \cite{hasan:2010}, and anyonic systems \cite{wen},  are examples that are of immediate interest to the condensed matter community and important for quantum information processing tasks as well \cite{flammia:2009,wen,briegel:2001,freedman:2002,nayak:2008}. Because the low energy states of these gapped systems do not break any symmetry of the Hamiltonian there exists no local observable whose expectation values may be taken as an order parameter denoting the phase \cite{goldenfeld}; however despite sharing the same symmetries there may exist phases that exhibit different physical properties \cite{stormer:1999}. The non-symmetry-breaking \emph{quantum} order \cite{wen} in such systems thus needs careful definition and characterization. To this end, methods of varying reliability and feasibility have been proposed \cite{wen, hamma:2005a, hamma:2005b,kitaev:2006,levin:2006,gu:2009,wen:1995}.

Here we focus on the class of spin liquids featuring topologically ordered  phases of matter. 
According to the most common definition, Gapped topological phases of matter have a ground state degeneracy, protected by the topology of the lattice on which the spin Hamiltonian is defined, that cannot be resolved by local observables  \cite{wen,kitaev:2003,wen:1990} and a gap above the ground state. These states are very non-trivial from the point of view of entanglement. One defines a state as trivially entangled if it is possible to deform it to a completely factorized state in an adiabatic way by means of a local Hamiltonian. In the language of quantum circuits, this is the same as limiting oneself to  unitary circuits (with finite range) of constant depth (not scaling with lattice size). A topological state then, cannot be completely disentangled by  local unitary quantum circuits. For this reason, one says that topological states possess \emph{long range} entanglement \cite{chen:2010}. The order in such states can then be detected through the values (zero for topologically trivial states) of carefully constructed quantities, such as the topological entanglement entropy \cite{wen, hamma:2005a, hamma:2005b,kitaev:2006,levin:2006,gu:2009, grover}, which characterize the correlation between different subregions of the many-body system, or via Wilson loop operators \cite{fradkin,confinement}. We note that such figures of merit  for Topological Order (TO) are reliable, provided that length scales of the system much larger than the correlation length $\xi$ are inspected. This makes the detection of the topological order experimentally challenging, because it involves a state tomography of a macroscopic portion of the system.


In this paper we elaborate on the idea that the detection of topological quantum phases is possible through the study of its local convertibility properties  \cite{hamma:2013,cui:apr2013}. One starts by imagining the manifold of ground states for a many body quantum system formed by the continuous set of ground states $\ket{\psi(\lambda)}$, for all possible values of a control parameter $\lambda$ of the Hamiltonian. The controllability of the Hamiltonian is assumed to arise from the addition of a tunable perturbation. One then asks whether it would be possible to simulate the effect of this perturbation on the ground state by using LOCC operations, restricted to two parts in which the system has been partitioned, to convert a ground state at one point to another nearby ground state in the manifold. If the LOCC class of operations is sufficient to effect such a conversion then we call the ground state locally convertible w.r.t. the perturbation {\it and} the bipartition and non locally-convertible otherwise. This notion of local-convertibility can be translated in terms of the behaviour of the entire set of R\'{e}nyi entropies of the reduced state on either of the subsystems \cite{turgut:2007,klimesh:2007}. Equivalently, because the R\'{e}nyi entropies are analytic functions of the eigenvalues of the reduced density matrix, the set of which is called the entanglement spectrum, local-convertibility can also be studied via the nature of the entanglement spectral flow as one tunes the perturbation strength \cite{cirac,turgut:2007,klimesh:2007,abanin:2012,marshall,aubrun:2008,bandyo:2002,duan:2005,yuan:2006,bhatia}. A ground state is locally-convertible if and only if {\it all}  R\'{e}nyi entropies (parametrized by the continuous real parameter $\alpha>0$) show the same monotonicity with varying perturbation strength \cite{turgut:2007,klimesh:2007}.

Note that the LOCC class of operations is a restricted subset of general coherent quantum operations on the whole system \cite{nielchuang,jonathanD:1999,sanders:2009}. An example of the latter would be the adiabatic tuning of the Hamiltonian which would of course be capable of implementing the conversion. On the other hand the LOCC we refer to, involves coherent operations local to the two parts into which the whole system has been bipartitioned, and thus can include portions of the system that, indeed, can be very non-local on the scale fixed by the interactions in the Hamiltonian. In particular, therefore, the notion of local-covertibility  we will examine is very distinct to the one implied in the  ideas involved  in  the Local Unitary Transformations protocols (LUTs) \cite{chen:2010}. There two gapped states are said to be in the same phase if and only if they are related by a local unitary evolution defined as a unitary operation resulting from the evolution of a local (range of the terms does not scale with the system size) Hamiltonian for finite time.

Our findings suggest that topologically ordered ground states are non-locally convertible with respect to {\it generic} perturbations and bipartitions (see sections \ref{sec:horizfield},\ref{summary},\ref{discussion} for a precise meaning of the term `generic'). Once the perturbation strength gets strong enough to take the system out of the topologically ordered phase the ground states become locally-convertible. Exploiting the above mentioned connection with the properties of the R\'{e}nyi entropies, we show that for \emph{generic} bipartitions and systems  with non-constant correlation length, while certain R\'{e}nyi entropies (with Renyi's parameter $\alpha\geq\alpha_c$)  decrease as the Hamiltonian is tuned towards the quantum critical point within a TO phase, others ($0\leq\alpha\leq\alpha_c$) show an increase - the `splitting phenomenon'. In the topologically trivial phases, like paramagnetic and symmetry breaking phases \cite{cui:2011,franchini:2013}, however, all entropies increase monotonically as the critical point is approached.


The intuition behind our result is that the property of non local-convertibility is associated with topological order because the global nature \cite{globalnature} of correlations characterizing the latter poses constraints on the locality of operations that may be used to convert one topologically ordered ground state to another at a different parameter value of the Hamiltonian. In a way, our work bridges between the ideas that TO is indeed a property of the wave function \cite{levin:2006} with  the classical analysis of the topological phases based on dynamical properties (quasi-particle statistics, edge excitations etc) \cite{arovas:84,wen:95}. Our approach may be seen to complement the analysis based on the topological entanglement entropy which relies on constraints on the boundary degrees of freedom for sufficiently large subsystems. There the large size of subsystems is required to cancel the contribution from local correlations - bulk contributions are rejected by design. Here we show that for the class of quantum double models \cite{wen}, of which the Toric code is an example, the response of the Renyi entropies to a Hamiltonian perturbation depends on how many and how much the degrees of freedom within the bulk of the subsystems contribute to the entanglement spectrum.


We comment that, despite the fact that the set of R\'{e}nyi entropies by itself does not provide any extra universal information, compared to the Topological entanglement entropy at any fixed value of the Hamiltonian parameter \cite{flammia:2009}, the `splitting' of the Renyi entropies we discussed above provides a  faithful indicator of Topological order,  even for   Renyi entropies of very small (sub)systems. In other words, our approach  has an added value,  in that it  involves the analysis on subsystems whose sizes need not scale with the correlation length of the physical system. This implies an obvious reduction of the complexity involved in the operation to trace the topological order in the system, opening the way to much simpler  experimental protocols.

The structure of the paper is as follows: In section \ref{sec:matprem} we explain our basic strategy, lay down the notation and quickly review the basic theory of majorization of probability vectors along with criteria for LOCC convertibility of ground states. In section \ref{models} we present the different models, a couple of which are amenable to exact analytical treatment while the most general case is dealt with numerically using 2D DMRG. In section \ref{summary} we summarize our results and conclude with comments and discussion in section \ref{discussion} about the scope of this line of inquiry. We place in appendix all calculations that we reference in the main text to ease the readibility.

\section{General strategy and mathematical preliminaries}
\label{sec:matprem}
\subsection{General strategy}
As a concrete example of a spin Hamiltonian with TO in the ground state, we choose Kitaev's Toric code \cite{kitaev:2003} with a perturbation $V(\lambda)$, that may be tuned through to the topologically trivial phase. Here all perturbed Hamiltonians $H_{TC}+V(\lambda)$ have a unique quantum critical point. We choose the perturbation $V$ so that it can drive a quantum phase transition to either a disordered paramagnetic phase, or a ferromagnet. Phase transitions of this kind have been studied in \cite{trebst:2007,hamma:2008,jahromi:2013,dusuel:2011, dusuel2}.  Because we want statements about local convertibility within a phase to be generic, our aim is to obtain the reduced density matrix (specifically its eigenvalues or trace of arbitrary powers) in full generality. We then analyze the behaviour of the R\'{e}nyi entropies w.r.t. $\lambda$. These entropies are functions of the eigenvalues of the reduced density matrix and the monotonicity of the \emph{entire} set of entropies depends on their relative majorization, which is a partial order on the set of probability vectors (the vector of eigenvalues) \cite{nielsen:1999}. Finally we check if \emph{all} R\'{e}nyi entropies show monotonic behaviour within a phase or does a subset of them show opposing behaviour from the rest. In order to achieve this, we need to solve for the ground state $|\psi(\lambda)\rangle$ and then obtain the reduced density matrix as a function of the parameters $\lambda$.

We employ both analytical and numerical methods to find the ground state and compute the R\'enyi entropies of the model. Analytically, we resort to two models. One, the Castelnovo-Chamon model, possesses an exact form for the ground state. We are able to compute exactly all the R\'enyi entropies by using group theoretic methods \cite{castelnovo:2008}. We also study the toric code in an external magnetic field, where the field is only acting on a subset of spins. This model maps into free fermions \cite{wen, halasz:2012a,yu:2008}, and is thus exactly solvable. In \cite{halasz:2012a, halasz:2012b}, an expression was derived for the $2-$R\'enyi entropy for a particular subsystem in terms of correlation functions. Here, we achieve a general expression for the $2-$R\'enyi entropy of a generic subsystem of this model. These results are actually more general and can be applied to any lattice gauge theory. Finally, we study the toric code in presence of Ising couplings in both the $x$ and $z$ direction. This model is non exactly solvable. We attack the problem numerically using a version of infinite DMRG  in two dimensions \cite{white:1992,culloch:2008,crosswhite:2008}, based on a Matrix Product State (MPS) representation of the ground state manifold for  a cylinder of infinite length and finite width. This method has proven very useful to study topological phases \cite{cincio:2013}.

\subsection{R\'{e}nyi Entropies}
Consider a multipartite pure quantum state $\ket{\psi}\in \otimes_i^N\mathcal{H}_i$. The entanglement spectrum $\bar{\nu}=\{\nu_1,\nu_2,...,\nu_{d_A}\}$ of the state, is defined as the set of eigenvalues of the reduced density matrix $\rho_A=\Tr_{\bar{A}}(\ket{\psi}\bra{\psi})$, where $A$ is a subset of local Hilbert space indices, $A\subset [N]$, with the associated Hilbert space given by $\mathcal{H}_{A}=\otimes_{j\in A}\mathcal{H}_j$. We call $A$ the subsystem. The complement of the subsystem $A$ then is $\bar{A}=[N\backslash A]$ with its associated Hilbert space $\mathcal{H}_{\bar{A}}=\otimes_{j\in\bar{A}}\mathcal{H}_j$. 

The entanglement spectrum of a state is the crucial ingredient in the definition of R\'{e}nyi entropies for the reduced density matrix $\rho_A$ defined as:
\begin{align}
S_{\alpha}(\rho_A):=\frac{1}{1-\alpha}\log\Tr(\rho^\alpha_A)=\frac{1}{1-\alpha}\log(\sum_{j}\nu_j^\alpha)~~~\forall \alpha\geq0
\label{renyient}
\end{align}

Knowledge about the entire set of R\'{e}nyi entropies $S_{\alpha}(\rho_A)~~\forall \alpha\in[0,\infty)$ is equivalent to complete knowledge about the spectrum of the state itself. At specific values of the continuous parameter $\alpha$, the R\'{e}nyi entropies provide operationally important information about the state: $S_{\alpha=0}=\log R$ - $R$ being the Schmidt rank is a measure of bipartite entanglement for the state that serves as a criteria for efficient classical representation of the state \cite{vidal:2003} while $\lim_{\alpha\to1}S_{\alpha}=S_{VN}$ is the entanglement entropy of the pure state $\ket{\psi}$, that is a measure of its distillable entanglement, entanglement cost and that of formation, relative entropy of entanglement and squashed entanglement \cite{nielchuang}. Also a linear combination of  2-R\'{e}nyi entropies $S_2$ calculated for suitably chosen bipartitions, can be used as a probe of topological order \cite{halasz:2012a,halasz:2012b}. For product states $\ket{\psi}=\ket{\psi}_A\otimes\ket{\psi}_{\bar{A}}$, the entanglement spectrum collapses to unity for one eigenvalue and zero for all others: $\rho_A^2=\rho_A$, which means that all R\'{e}nyi entropies are zero as well.


\subsection{Manifold of topologically ordered ground states}
We define the ground state manifold, $\mathcal{M}$, of a Hamiltonian $H(\lambda)$ as the continuous set of ground states $|\psi(\lambda)\rangle$ (in a particular topological sector) for all possible values of the control parameters $\lambda$. So $\mathcal M =\{|\psi(\lambda)\rangle~ \text{s.t.}~ |\psi(\lambda)\rangle ~\text{is the ground state of~} H(\lambda)~\forall  \lambda =(\lambda^1,...,\lambda^n)\in\mathbb R^n\}$. As the Hilbert space is endowed with a definite tensor product structure $\mathcal H = \mathcal H_A \otimes \mathcal H_B$, which defines a bipartition of the system, we can consider the set of reduced density matrices $\rho_A(\lambda)$ to the subsystem $A$ as a function of $\lambda$, and study the behaviour of the set of R\'{e}nyi entropies $S_\alpha(\lambda)$ with $\lambda$ and $\alpha$:
\begin{align}
S_{\alpha}(\lambda)&:=S_{\alpha}(\rho_A(\lambda))\nonumber\\
&=S_{\alpha}(\Tr_{\bar{A}}(\ket{\psi(\lambda)}\bra{\psi(\lambda)}))~~\forall\alpha\geq0,
\end{align}
In the next section,  we show, on the back of specific examples, that the monotonicity of the entire set $S_{\alpha}(\lambda)~\forall\alpha$ is a characteristic of the phase unless the perturbation and/or the choice of bipartition is fine tuned. The collective behaviour can be captured succinctly by the sign of the derivative $\text{Sign}[{\partial_\lambda S_\alpha(\lambda)}]~\forall \alpha$, which  remains constant in the topologically disordered phase - negative as the perturbation is tuned away from the critical point; whereas in the ordered phase $\partial_\lambda S_\alpha(\lambda)<0~ \text{for}~\alpha<\alpha_c$, while it is positive for $\alpha\geq\alpha_c$, as we move away from the quantum critical point.


\subsection{Differential local convertibility on the ground state manifold}

The class of Local Operations and Classical Communications \cite{chitambar:2012} - LOCC operations - are general quantum operations augmented with classical communication. The operations allowed are local in the sense of being restricted separately to the two parts of some bipartition of the system while potentially unlimited two-way classical communication (CC) is allowed between observers of the two regions so that operations conditioned on outcomes of the other region may be implemented. This class of operations is motivated by current technological capabilities as generating quantum coherences becomes exponentially more difficult with increasing system size as well as the difficulty in quantum data communication.

Differential local convertibillity (dLOCC) is a property of a submanifold $\mc{M}_i\subset\mc{M}$ of the ground state manifold $\mc{M}$ that determines whether LOCC operations may be used to transform from $\ket{\psi(\lambda)}\in\mathcal{M}_i$ to another $\ket{\psi(\lambda+\delta\lambda)}\in \mathcal{M}_i$.  Mathematically we say that,
\begin{align}
& \mc{M}_i~\text{is dLOCC}~\text{iff},\nonumber\\
&~~~~~~~~~~~~~~~~~\text{Sign}[{\partial_\lambda S_\alpha(\lambda)}]=\text{constant}~\forall \alpha\geq0~\forall~\ket{\psi(\lambda)}\in\mc{M}_i
\label{dlocc}
\end{align}

A negative sign in the R.H.S of the condition above implies dLOCC property of $\mc{M}_i$ in the direction of increasing $\lambda$. 
  In this work we focus on submanifolds $\mc{M}_i$ that are regions of the ground state manifold pertaining to the different phases, labelled by $i$, for the different Hamiltonian models we consider. Thus we frequently refer to a phase being dLOCC as well.

 The quantity: $\text{Sign}[{\partial_\lambda S_\alpha(\lambda)}]~\forall \alpha$, has operational significance w.r.t. traversing $\mathcal{M}_i$ using LOCC. The results of  \cite{nielsen:1999,klimesh:2007,jonathanD:1999,marshall,aubrun:2008,bandyo:2002,duan:2005,yuan:2006}, imply that one can use LOCC operations to transform a ground state $\ket{\psi(\lambda)}\in\mathcal{M}_i$ to another $\ket{\psi(\lambda+\delta\lambda)}\in \mathcal{M}_i$, which may require access to a shared entangled state $\ket{\phi}$ (entanglement catalyst) between $A,\bar{A}$ (bipartition), with probability 1, at proximal values of $\lambda,\lambda+\delta\lambda$, within a phase, iff the vector of Schmidt coefficients of the product state $\ket{\psi(\lambda+\delta\lambda)}\ket{\phi}$ at the target parameter value $\lambda+\delta\lambda$, majorizes the vector of Schmidt coefficients of the state $\ket{\psi(\lambda)}\ket{\phi}$ at the initial point. 

Majorization is a partial order on the set of positive vectors  $\bar{\nu}_{\lambda}$,~$\bar{\nu}_{\lambda+\delta\lambda}$ which, for our purposes here, are the vectors of Schmidt coefficients of the states $\ket{\psi(\lambda)}\ket{\phi}$ and $\ket{\psi(\lambda+\delta\lambda)}\ket{\phi}$ respectively w.r.t. the $A,\bar{A}$ bipartition. It compares the \emph{disorder} in one vector w.r.t. another. Arranging the entries of the vectors $\bar{\nu}_{\lambda+\delta\lambda},\bar{\nu}_{\lambda}$ in a non-increasing manner: $(\nu_{\lambda+\delta\lambda})_1\geq(\nu_{\lambda+\delta\lambda})_2\geq(\nu_{\lambda+\delta\lambda})_3....\geq(\nu_{\lambda+\delta\lambda})_{d}$ and $(\nu_{\lambda})_1\geq(\nu_{\lambda})_2\geq...\geq(\nu_{\lambda})_{d}$, we say $\bar{\nu}_{\lambda+\delta\lambda}$ majorises $\bar{\nu}_{\lambda}$, i.e. $\bar{\nu}_{\lambda}\prec\bar{\nu}_{\lambda+\delta\lambda}$ iff:
\begin{align}
\sum_{j=1}^k(\nu_{\lambda})_j&\leq\sum_{j=1}^k(\nu_{\lambda+\delta\lambda})_j~\forall k=1,2,...,d_A
\label{majorize}
\end{align}
Which may be called the \emph{catalytic majorization} relation since the vectors represent the Schmidt coefficients of states that are a tensor product with the catalyst state $\ket{\phi}$.

It should be clear that not all pairs of states $\ket{\psi(\lambda+\delta\lambda)}$ and $\ket{\psi(\lambda)}$ will require a catalyst for dLOCC conversion. For such states their respective vectors of Schmidt Coefficients $\gamma_{\lambda+\delta\lambda}$,$\gamma_{\lambda}$ follow a majorization relation $\gamma_{\lambda}\prec\gamma_{\lambda+\delta\lambda}$ without the need for the  ancilliary entanglement catalyst $\ket{\phi}$. The necessary and sufficient condition for dLOCC conversion, with or without the need for a catalyst is succintly captured by the condition \cite{klimesh:2007}:
\begin{align}
 S_{\alpha}(\bar{\gamma}_\lambda)&\geq S_{\alpha}(\bar{\gamma}_{\lambda+\delta\lambda})~~\forall \alpha
\label{catmaj}
\end{align}

which implies Eq.~(\ref{dlocc}). In words, one can use LOCC transformations, possibly assisted by entanglement catalysis, to transform from $\ket{\psi(\lambda)}$ to $\ket{\psi(\lambda+\delta\lambda)}$ provided \emph{all} R\'enyi entropies show monotonically decreasing behavior in going from the initial parameter value to the final one.

 Thus catalytic majorization and monotonic behaviour (in $\alpha$) of the whole set of R\'{e}nyi entropies are mutual implications. For $\alpha=1$ for e.g. Ineq.~(\ref{catmaj}) implies that a necessary condition for LOCC operations to be used to transform to the new state $\ket{\psi(\lambda+\delta\lambda)}$ is for it to have a lower value of the entanglement entropy w.r.t. the underlying bipartition \cite{nielchuang}.

\subsection{The models}
\label{models}

Here, we present the models we will be dealing with in the rest of the paper. We consider three different perturbations $V(\lambda)$, to Kitaev's Toric code (TC) model $H_{TC} $\cite{kitaev:2003}.

\begin{figure}[t]
 \centering
\includegraphics[width=.75\columnwidth]{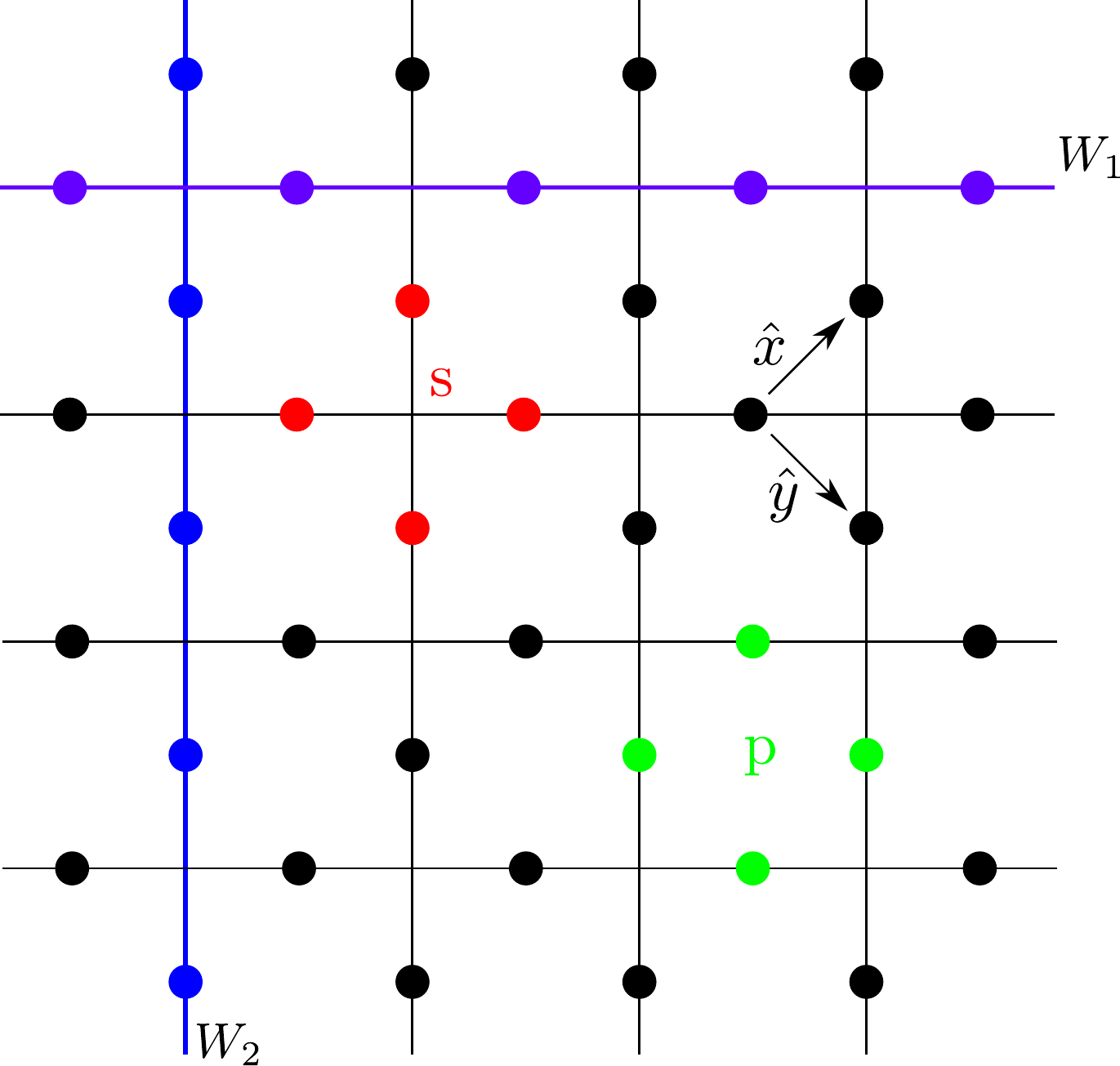}
 \caption{(color online) The spin-1/2s (all filled circles) in the Toric Code model live on the edges of a square lattice with periodic boundary conditions. The star operator at vertex labelled $s$ involves the product of $\hat{\sigma}^x$ operators on the four spins (red circles) of the edges joined at the vertex. The plaquette operator for the unit cell labelled $p$ involves the product of $\hat{\sigma}^z$ operators on the four spins (green circles) on edges that form the cell. $W_1,W_2$ are spin flips along the two non-contractible directions of a torus (blue circles).}
  \label{TClattice}
\end{figure}
 The TC Hamiltonian $H_{TC}$ is defined on a 2-D system of spin-1/2 particles living on the edges of a square lattice with periodic boundary conditions in both directions, Fig.~(\ref{TClattice}). 
 The Hilbert space size of the system defined on a square lattice of size $L\times L$ is $N=2^{2L^2}$. 
 There are two different kinds of mutually commuting operators that appear in the Hamiltonian: stars $A_s=\prod_{i\in s}\hat{\sigma}^x_i$ defined at the vertices of the lattice that are the products of Pauli matrices $\hat{\sigma}^x_i$ acting on the 4 edges shared by a vertex and plaquettes $B_p=\prod_{j\in p}\hat{\sigma}^z_j$ that are products of $\hat{\sigma}^z_j$ on the 4 edges of a unit cell. The operators $A_s,B_p$ have eigenvalues $\pm1$. All our Hamiltonians then have the form
\begin{align}
H&=H_{TC}+V(\lambda):=-\sum_s A_s-\sum_p B_p+V(\lambda)
\label{tchamil}
\end{align}
Note that, because $\prod_s A_s=\prod_p B_p=\openone$, there are only $L^2-1$ independent operators of each kind. They constitute a complete set of commuting operators with $H_{TC}$, and therefore all excitations of the unperturbed Hamiltonian $H_{TC}$ may be labelled by the $\pm1$ eigenvalues of the $2\times(L^2-1)$ operators.
 This means that there are $2^{2L^2-2}$ excited states corresponding to each of the $2^{2L^2}/2^{2L^2-2}=4$ degenerate ground states which is consistent with the fact that the ground state degeneracy for a topologically ordered Hamiltonian of spin-1/2s defined on a torus is $4^g$ with $g=1$ being the genus of the surface. For our purposes though, one can work in a gauge fixed  sector with all $B_p=+1$, that corresponds to an effective low energy theory with $\Z_2$-gauge symmetry since $[A_s,\prod_i\hat{\sigma}^z_i]=0~\forall s$, and the only excitations are those of stars, so that in this sector the Hilbert space dimension is $2^{L^2-1}$ again with 4 degenerate ground states. In this gauge fixed sector all eigenstates of $H_{TC}$ are superpositions of loop operators $g=\prod_{i\in s}\hat{\sigma}^x_i$ that are products of spin-flips on spins that are crossed by contractible closed loops in the dual lattice. The loop operators are elements of the group $G$ that is generated by the stars. The four degenerate ground states, $\ket{\psi},W_1\ket{\psi},W_1\ket{\psi},W_1W_2\ket{\psi}$, each define a particular topological sector within the gauge fixed sector and are related to each other by spin flips on non-contractible loops $W_1,W_2$, along the two non-contractible directions of the Torus.

 In our work we focus on the simplest ground state $\ket{\psi}$ i.e. a fixed topological sector within the gauge. Restricting our attention to this sector, which we call $TS_1$, essentially captures all the phenomenology we want to highlight as well as simplifies the calculations. Thus our analytical results pertain to this sector where in subsections \ref{sec:chamon}, \ref{sec:horizfield} we consider gauge invariant perturbations to $H_{TC}$ that take drive the system across a quantum critical point between a topologically ordered and disordered phase. For a discussion of the critical point see \cite{jvidal:2009,jvidal:2009b,wu:2012}. The more general perturbation \ref{sec:toricising} is studied
numerically. The tool used here is a two dimensional density matrix
renormalization group extended to infinite cylinders
\cite{longDMRGtopo}. The ability to study a Hamiltonian on an infinite
cylinder allows us to obtain the entire set of quasi-degenerated
ground states. From that set we chose a ground state in a given
topological sector and make sure that the same choice was made for
every value of $\lambda_x$ and $\lambda_z$ in Eq. (\ref{V3}). This can
be done by looking at the expectation value of certain loop operators
around the cylinder. For small perturbations studied here, they are
close to $\pm 1$, which allows one to identify the topological sector.
 All DMRG results presented here are converged in bond dimension, which is a refinement parameter in this calculation.
 
Here we list the perturbations studied in the current paper:

 a.~~~\emph{The Castelnovo-Chamon model}\\
This perturbation has an exponential form,
\be
 V_1 (\lambda) = \sum_{s}e^{-\lambda\sum_{i\in s}\hat{\sigma}^z_i},
\ee

that commutes with all the plaquette operators $[B_p,V_2(\lambda)]=0~\forall p$ i.e. it is a gauge invariant perturbation. This system shows a phase transition from a topologically ordered phase to a paramagnetic phase at the critical value of $\lambda\approx 0.44$.

b.~~~\emph{Toric code Hamiltonian with magnetic field along spins on rows.}  
The perturbation here is a $\hat{\sigma}_z$ magnetic field applied only to the spins along the rows of the square lattice (we call this direction the horizontal direction),
\be
 V_2 (\lambda) = -\lambda\sum_{h\in~ \text{horiz}}\hat{\sigma}^z_h. 
 \ee 
Since $[B_p,V_3(\lambda)]=0~\forall p$ this is a gauge invariant perturbation as well that drives the TC model from a topologically ordered phase across the critical point at $\lambda=1$ to a paramagnetic one.

 c.~~~\emph{The Toric-Ising Model} \\
Here the perturbation,
 \be
 V_3 (\lambda_x, \lambda_z) = -\sum_{i,\mu=\hat{x},\hat{y}} (\lambda_x \hat{\sigma}^x_i \hat{\sigma}^x_{i+\mu} + \lambda_z \hat{\sigma}^z_i \hat{\sigma}^z_{i+\mu}),
\label{V3}
 \ee
describes the interplay between topological and antiferromagnetic orders. For generic $\lambda_x$ and $\lambda_z$, the perturbation breaks the $Z_2$ gauge symmetry. The latter is preserved for either $\lambda_x=0$ or $\lambda_z=0$. When $\lambda_x(\lambda_z)=0$,  the topological and antiferromagnetic orders  are separated by a continuous quantum phase transition occuring at the critical value of $\lambda_z(\lambda_x)=\lambda_c\sim 1/6$ \cite{karimipour:2013}.

\section{Results}
In this section we present analytical and numerical results that exhibit the relationship between differential local convertibility and correlation length for Hamiltonians $H=H_{TC}+V(\lambda)$, where $V(\lambda)=V_1,V_2,V_3$ described in the previous section. 

\subsection{The Castelnovo-Chamon model, $V=V_1$}
\label{sec:chamon}

We start by observing here that the perturbation $V_1$ is such that the spin-spin correlation function $\braket{\hat{\sigma}^x_i\hat{\sigma}^x_j}_{\lambda}$ in a ground state within the topological sector $TS_1$ of the Hamiltonian $H=H_{TC}+ \sum_{s}e^{-\lambda\sum_{i\in s}\hat{\sigma}^z_i}$ is zero for all values of $\lambda$. In the sector $TS_1$, we pick a ground state $\ket{\xi}$ given by \cite{castelnovo:2008}:
\begin{align}
\ket{\xi}&=\frac{1}{\sqrt{Z}}\sum_{g\in G}e^{(\lambda/2)\sum_{i\in \Lambda}\sigma^z_i(g)}\ket{g}
\label{gs}
\end{align}

where $g\ket{0}$, is the state obtained by acting with $g=\prod_{i}A_{s_i},g\in G$, that is the product of star operators, on the totally polarized all spins-up (in the z-basis) reference state $\ket{0}$ and the term $\sigma^z_i(g)=\bra{g}\hat{\sigma}^z_i\ket{g}$ in the exponent takes the value of $-1$ if the spin at edge $i$ has been flipped and $+1$ otherwise. $Z=Z(\lambda)=\sum_{g\in G}e^{\lambda\sum_{i}\sigma^z_i(g)}$ is a normalization constant. Note that with $\Lambda$ denoting the set of all spins, $\sum_{i \in \Lambda}\sigma^z_i(g)=N-L(g)$, i.e. the sum counts the total number of spins in a state less the number that have been flipped by the operator $g\in G$ which are closed loops or products of closed loops in the dual lattice.

In order to analyze the DLOCC properties of this model we need the reduced density matrix for a subset of spins $A$, on the whole lattice $\Lambda=A\cup B$, when the whole system is in state (\ref{gs}):
\begin{align}
\rho_{A}(\lambda)&=\frac{1}{Z}\sum_{\substack{g\in G\\g'\in G_A}}e^{\frac{\lambda}{2}(N-L(g))}e^{\frac{\lambda}{2}(N-L(gg'))}x^g_A\ket{0}_A{}_A\bra{0}x^g_Ag'_A
\label{redmat}
\end{align}
where the group $G_A=\{g\in G|g=g_A\otimes \openone_B\}$  is the subgroup of $G$ generated by stars operators acting non-trivially only on the spins in $A$ and $x^g_A$ is the restriction of the operators $g\in G$ to just the subsystem $A$ (for details see \cite{hamma:2005a,hamma:2005b}). We will also  need the subgroup $G_B=\{g\in G|g=\openone_A\otimes g_B\}$ which includes all products of star operators that act  non-trivially only on the spins in $B$. Then the $\alpha$-R\'{e}nyi entropy is given by:
\begin{align}
S_{\alpha}(\rho_A)&=\frac{1}{(1-\alpha)}\log \frac{1}{Z^{\alpha}}\sum_{g\in G}e^{-\lambda E_g}(\sum_{h\in G_A,g\in G_B}e^{-\lambda E_{hgk}})^{\alpha-1}\nonumber\\
&=\frac{1}{(1-\alpha)}\log \frac{1}{Z^{\alpha}(\lambda)}\sum_{g\in G}e^{-\lambda E_g}w^{\alpha-1}(\lambda,g)
\label{ccrenyi}
\end{align}

where $E_g=L(g)-N$ and $w(\lambda,g):=\sum_{h\in G_A,h\in G_B}e^{-\lambda E_{hgk}}$ with all the $\lambda$ dependence made explicit.

After a straighforward but tedious calculation one can obtain the derivative of Eq.~(\ref{ccrenyi}) w.r.t. the parameter $\lambda$ and it is given by the expression:
\begin{align}
\partial_{\lambda}S_{\alpha}(\lambda)&=\avg{\avg{E_g}_{w(\lambda,g)}}_{\tilde{Z}(\lambda,\alpha)}+\frac{\alpha}{(1-\alpha)}\avg{E_g}_{Z(\lambda)}\nonumber\\
&~~~~~~~~~~~~~~~~~~~~~~~~~~~~~~~~~~~~~~~~~~~~~~~~-\frac{1}{(1-\alpha)}\avg{E_g}_{\tilde{Z}(\lambda,\alpha)}
\label{avgsum1}
\end{align}

Here $\tilde{Z}(\lambda,\alpha):=\sum_{g\in G}e^{-\lambda E_g}w^{\alpha-1}(\lambda,g)$ and we use averages w.r.t. the functions $f(g)=w(\lambda,g),\tilde{Z}(\lambda,\alpha),Z(\lambda)$ defined as usual: $\avg{E(g)}_{f(g)}=\sum_g(f(g)E(g))/\sum_g(f(g))$.
One can now evaluate the R.H.S. of Eq.~(\ref{avgsum1}) in the limit $\lambda\to0$ which corresponds to small perturbations of the TC model and find that $\partial_{\lambda}S_{\alpha}(\lambda)\leq0~\forall\alpha$. This implies that \emph{all} R\'{e}nyi entropies decrease as we move away from the point in the phase diagram with a flat entanglement spectrum. Under the assumption that the slopes of R\'{e}nyi entropies for fixed $\alpha$ do not change within a phase we find that this model has DLOCC within the topologically ordered phase. Similarly if one considers the $\lambda\to\infty$ limit one finds that all the slopes are negative as well implying that the particular form of the perturbation $V_1$ leads to DLOCC in both, the TO and the paramagnetic, phases of the model.

\subsection{Toric code  with magnetic field along spins on rows, $V=V_2$}
\label{sec:horizfield}
\begin{figure}
 \centering
\includegraphics[width=.75\columnwidth]{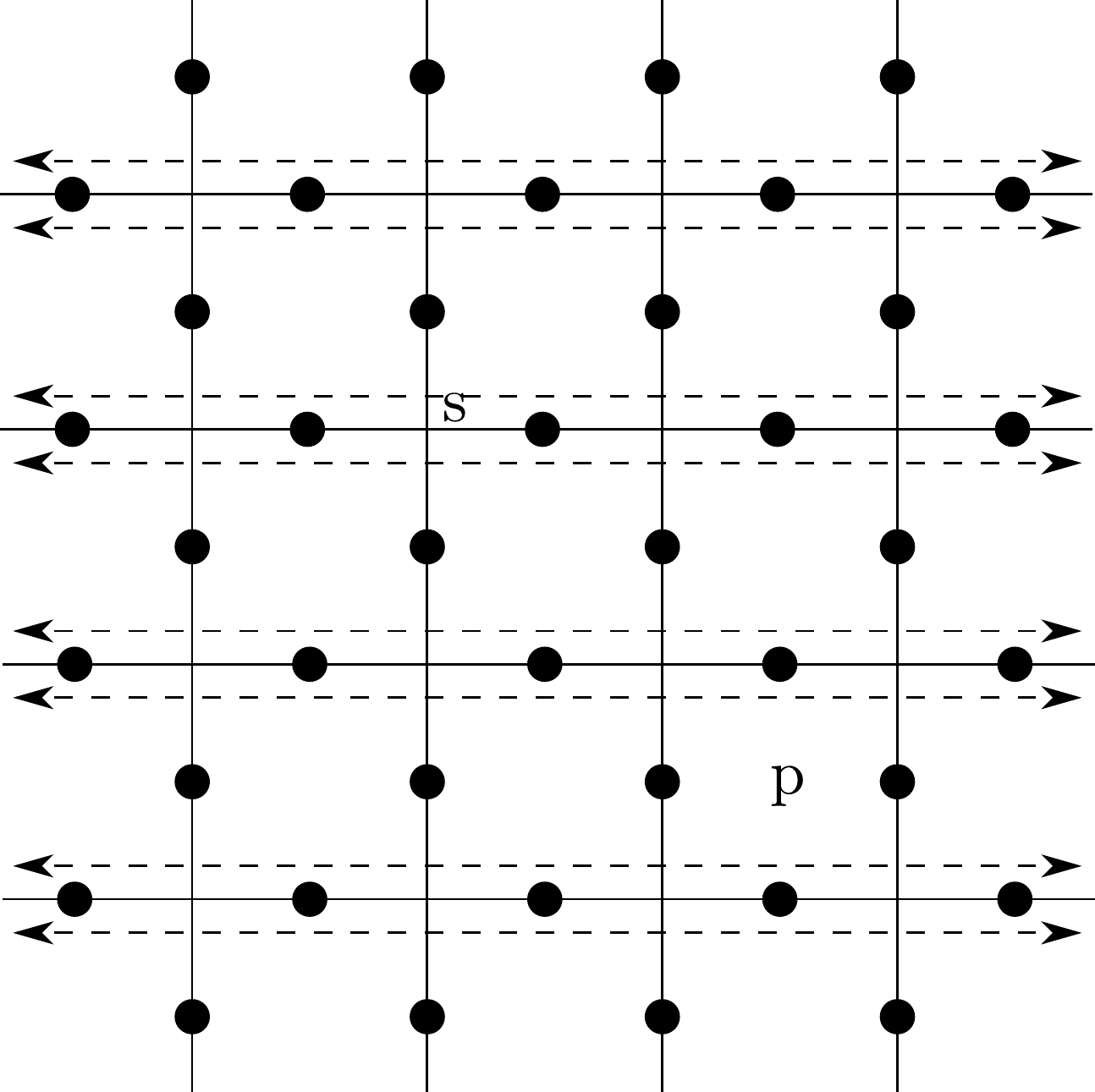}
 \caption{(color online) An artist's rendition of the lattice of spins (dark filled circles) for the Toric code model with a magnetic field on spins along only the horizontal direction (shown by spins within dotted arrows)}
  \label{TChoriz}
\end{figure}

The gauge invariant perturbation $V_2(\lambda)$ lets us analyse a model with a non-constant correlation length $\xi(\lambda)$.
The Gauge fixed ($B_p=1~\forall p$) Hamiltonian (\ref{tchamil}) upto a constant offset  is thus:
\begin{align}
H&=-\sum_{s}A_s-\lambda\sum_{h\in~ \text{horiz}}\hat{\sigma}^z_h
\label{horizhamil}
\end{align}
where by $h\in\text{horiz}$, we mean that the external field is applied only to spins on edges along the rows that we take to be the horizontal direction, Fig.~(\ref{TChoriz}).

To solve Eq.~(\ref{horizhamil}) we map it to an exactly solvable model that preserves the local algebra of the terms. We first observe that the star operators have eigenvalues $\pm1$. Then we note that each $\hat{\sigma}^z_h$ operator on a horizontal link has two neighboring star operators acting on the vertices connected by the edge. Because the action of $\hat{\sigma_h}^z$ is to flip the sign of both the star operators that share the spin `h', $\{A_s,\hat{\sigma}^z_h\}=0$ for these neighboring stars and we can move to an alternate picture where the star operators at a vertex are replaced by pseudo-spin operators, $\hat{\tau}^z_s$, at the same vertex with eigenvalues $\pm 1$. The action of $\hat{\sigma}^z_h$ then corresponds to the action of $\hat{\tau}^x_i\hat{\tau}^x_{i+1}$ when the vertices $s,s+1$ share the edge labelled `h' i.e. it flips both neighboring pseudo-spins. We will call $A_s,\hat{\sigma}^z_h$ operators in the `$\sigma$-picture' in contrast to the `$\tau$-picture' for operators in terms of the pseudo-spin operators $\hat{\tau}$ .  The map is thus given by:

\begin{align}
A_s&\to \hat{\tau}^z_s\nonumber\\
\hat{\sigma}^z_h&\to\hat{\tau}^x_{i}\hat{\tau}^x_{i+1}
\label{opmap}
\end{align}

\noindent which maps the Hamiltonian (\ref{horizhamil}) to:
\begin{align}
\tilde{H}&=-\sum_{s\in \text{~all vertices}}\hat{\tau}^z_s-\lambda\sum_{\text{~all rows}}\sum_{s\in\text{row}}\hat{\tau}^x_{s}\hat{\tau}^x_{s+1}\nonumber\\
&=-\sum_{ \text{~all rows}}\sum_{s\in\text{row}}\hat{\tau}^z_s-\lambda\sum_{\text{~all rows}}\sum_{s\in\text{row}}\hat{\tau}^x_{s}\hat{\tau}^x_{s+1}\nonumber\\
&=\sum_{ \text{~all rows}}(-\sum_{s\in\text{row}}\hat{\tau}^z_s-\lambda\sum_{s\in\text{row}}\hat{\tau}^x_{s}\hat{\tau}^x_{s+1})\nonumber\\
&=\oplus_{\text{~all rows}}H_{\text{row}},~~H_{\text{row}}=-\sum_{s=1}^{L}\hat{\tau}^z_s-\lambda\sum_{s\in\text{row}}\hat{\tau}^x_{s}\hat{\tau}^x_{s+1}
\label{tfim}
\end{align}

Eq.(\ref{tfim}) implies that the new Hamiltonian is a direct sum of 1-D quantum Ising Hamiltonians on the $L$ rows. The ground state of $\tilde{H}$ is thus given by the tensor product of the ground states of each individual row i.e. $\ket{\psi}=\otimes_{j\in \text{~all rows}}\ket{\psi_j}$. Each row Hamiltonian $H_{\text{row}}$ in the expression above is solved by mapping the Pauli spins via the Jordan-Wigner transformation to Fermions and then a Bogoliubov transformation diagonalizes the Hamiltonian to a free Fermionic form \cite{pfeuty:1970}. In the present paper,  we consider the symmetric ground state enjoying the global spin flip symmetry of the Hamiltonian and thus  $ <\hat{\tau}^x_i>=0$ in the ground state.

This model exhibits two phases as well: a topologically ordered one for weak magnetic field and a disordered one beyond the critical value $\lambda=1$ \cite{halasz:2012a,yu:2008}. The results of this model, which follow in the next subsections, demonstrate that for fine-tuned perturbations one might indeed obtain differential local convertibility for specially chosen bipartitions. We remark that although we considered the symmetric ground state of the system this does not result in a loss of generality and at the same time eases the analytical presentation.

For special choices of subsystems (we call these `thin' subsystems for reasons that become clear in the following) we can determine the exact eigenvalues of the reduced density matrix for all values of the perturbing field $\lambda$ and hence all the R\'{e}nyi entropies $S_\alpha$ which show monotonic perturbative behaviour for all $\alpha$. On the other hand, for systems with a `bulk' some R\'{e}nyi entropies have a different behaviour with increasing $\lambda$ than others.


\subsubsection{\it{\textbf{`Thin' subsystems}}}
\label{thinsubsys}
\begin{figure}
 \centering
\includegraphics[width=0.5389\columnwidth]{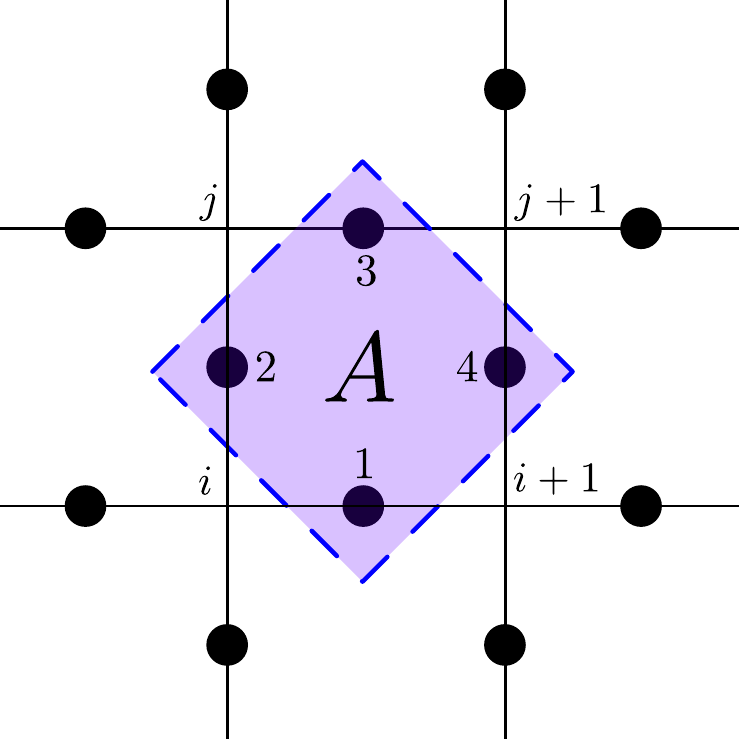}
 \caption{(color online) Subsystem $A$, shown in the shaded region, of one plaquette with the spins $1,2,3,4$, on the edges. The eigenvalues of the reduced density matrix $\rho_A$, involves calculating expectation values of operators on the 4 pseudo-spins $i,i+1,j,j+1$, at the shown vertices (see Appendix~(\ref{app:thinsys})).}
  \label{plaquette}
\end{figure}

A drastic simplification in the exact calculation of the R\'enyi entropies for the ground state of gauge theories (of which the toric code is the simplest example, the $Z_2$ gauge theory) can be obtained by choosing some particular partitions\cite{halasz:2012a,halasz:2012b}.
A `thin' subsytem $A$, in the lattice for the Toric code model is one where there are no star operators that can act on spins which exclusively belong to $A$. For example, the bipartition of spins on the lattice where subsytem $A$ is comprised only of rows (columns) with the columns (rows) forming the complement $B$. Mathematically this means that the group $G_A$ only contains the identity, $\openone$, which in turn implies  that the reduced density matrix, $\rho_A$, is diagonal in the $z$-basis of the $\sigma$-spins \cite{hamma:2005b}. All loops on the real lattice are other examples, the shortest such loop being a plaquette, Fig.~(\ref{plaquette}). Intuitively, `thin' subsystems are those wherein all the degrees of freedom are maximally entangled, even in the unperturbed toric code model, while respecting the gauge constraints. Thus increasing correlation length cannot lead to newer non-zero values appearing in the entanglement spectrum.

 Such is the subsystem $A$ that we now investigate. The reduced density matrix for a plaquette with 4 spins is a matrix of size $2^4\times 2^4$. However because of the gauge constraint, $B_p=1$, only three spins are independent which means that the maximal rank of the reduced density matrix is $2^3=8$. The diagonal entries of this matrix, $(\rho_A)_{\bar{s}\bar{s}}$, correspond to expectation values of the projector onto the different spin configurations,  $\bar{s}=(s_1,s_2,s_3)\in\{-1,1\}^{3}$, in the ground state of the Hamiltonian (\ref{horizhamil}) of the three independent spins i.e.:
\begin{align}
(\rho_A)_{\bar{s},\bar{s}}&=\frac{1}{2^3}\bra{\psi}(1+s_1\hat{\sigma}^z_1)(1+s_2\hat{\sigma}^z_2)(1+s_3\hat{\sigma}^z_3)\ket{\psi}\nonumber\\
&=\frac{1}{2^3}(1+s_1\braket{\hat{\sigma}^z_1}+s_2\braket{\hat{\sigma}^z_2}+s_3\braket{\hat{\sigma}^z_3}\nonumber\\
&+s_1s_2\braket{\hat{\sigma}^z_1\hat{\sigma}^z_2}+s_2s_3\braket{\hat{\sigma}^z_2\hat{\sigma}^z_3}+s_3s_1\braket{\hat{\sigma}^z_3\hat{\sigma}^z_1}\nonumber\\
&~~~~~~~~~~~~~~~~~~~~~~~~~~~~~~~~~~~~~~~~~~~~~~~~~~+s_1s_2s_3\braket{\hat{\sigma}^z_1\hat{\sigma}^z_2\hat{\sigma}^z_3})\nonumber\\
&=\frac{1}{2^3}(1+s_1\braket{\hat{\tau}^x_i\hat{\tau}^x_{i+1}}+s_2\braket{\hat{\tau}^x_i}^2+s_3\braket{\hat{\tau}^x_j\hat{\tau}^x_{j+1}}\nonumber\\
&+s_1s_2\braket{\hat{\tau}^x_i}^2+s_2s_3\braket{\hat{\tau}^x_i}^2+s_3s_1\braket{\hat{\tau}^x_i\hat{\tau}^x_{i+1}}^2\nonumber\\
&~~~~~~~~~~~~~~~~~~~~~~~~~~~~~~~~~~~~~~~~~~~~~~~+s_1s_2s_3\braket{\hat{\tau}^x_i}^2)
\label{diagelem}
\end{align}
where in the last line above we have used the mapping (\ref{opmap}) to express the diagonal entries in terms of the $\tau$-spins (see Appendix~\ref{app:thinsys}). Notice that the only non-trivial expectation values of the $\tau$-spins are those of two point functions since  $\braket{\hat{\tau}^x_i}=0$ in the symmetric ground state.
The thermodynamic limit expressions \cite{pfeuty:1970,barouch:1971} in the entire domain of $\lambda$ is: 
\begin{align}
\braket{\hat{\tau}^x_{i}\hat{\tau}^x_{i+1}}&=\frac{1}{\pi}\int_0^{\pi}\frac{\cos(\phi)[\cos(\phi)-1/\lambda]+\sin^2(\phi)}{[(1/\lambda-\cos(\phi))^2+\sin^2(\phi)]^{1/2}}d\phi~~~0<\lambda\nonumber\\
\end{align}

\begin{figure}
 \centering
\includegraphics[width=.9\columnwidth]{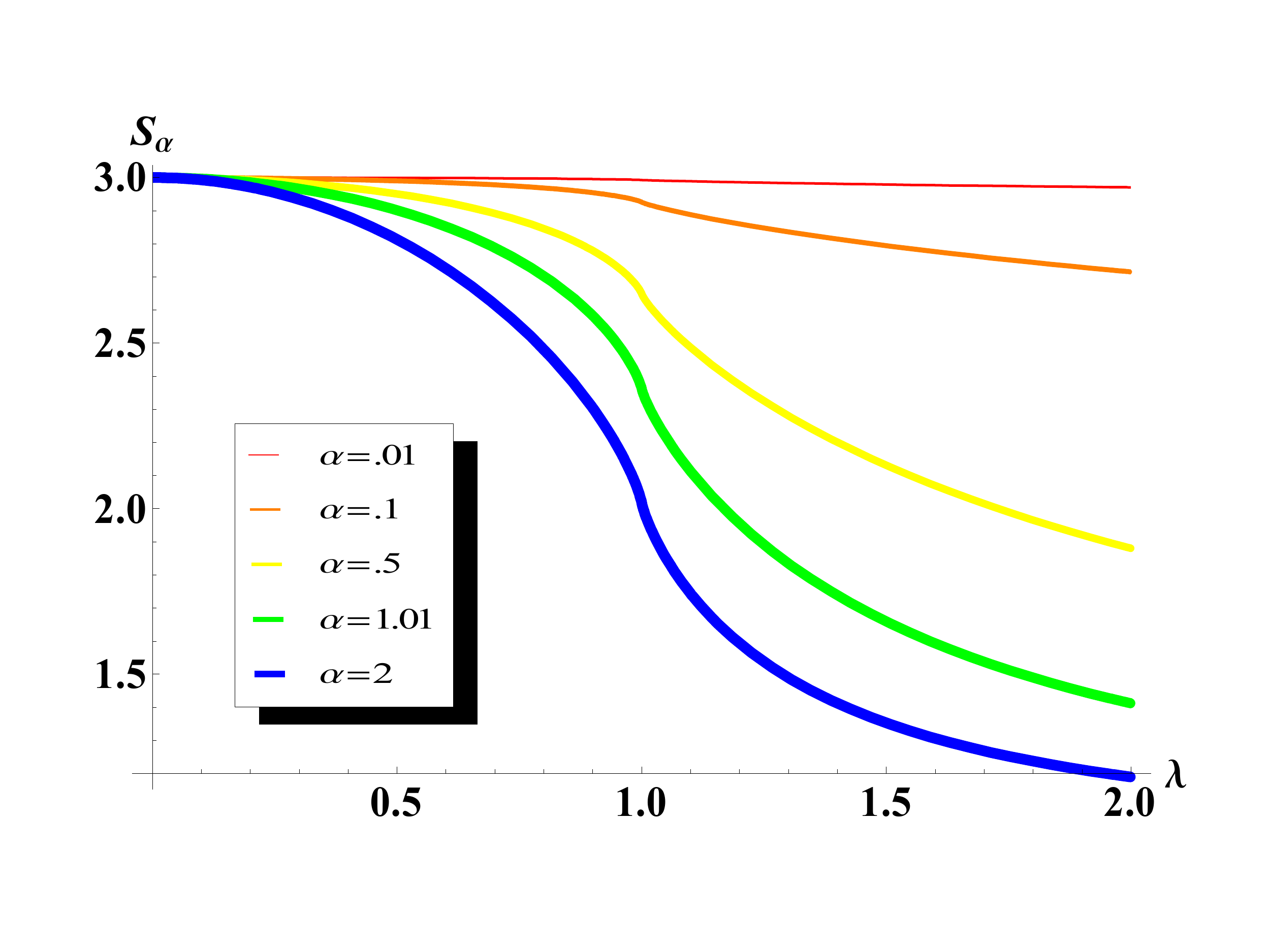}
 \caption{(color online) R\'{e}nyi entropies for a subsystem $A$ of one plaquette (shown in Fig.~\ref{plaquette}), at different values of $\alpha$. All entropies show monotonic behavior in both the phases: they decrease monotonically with increasing correlation length $\xi(\lambda)$ for $\lambda<\lambda_c=1$ while they increase with $\xi(\lambda)$ for $\lambda>1$.}
  \label{singplaqentplot}
\end{figure}

\begin{figure}[t]
 \centering
\includegraphics[width=0.73\columnwidth]{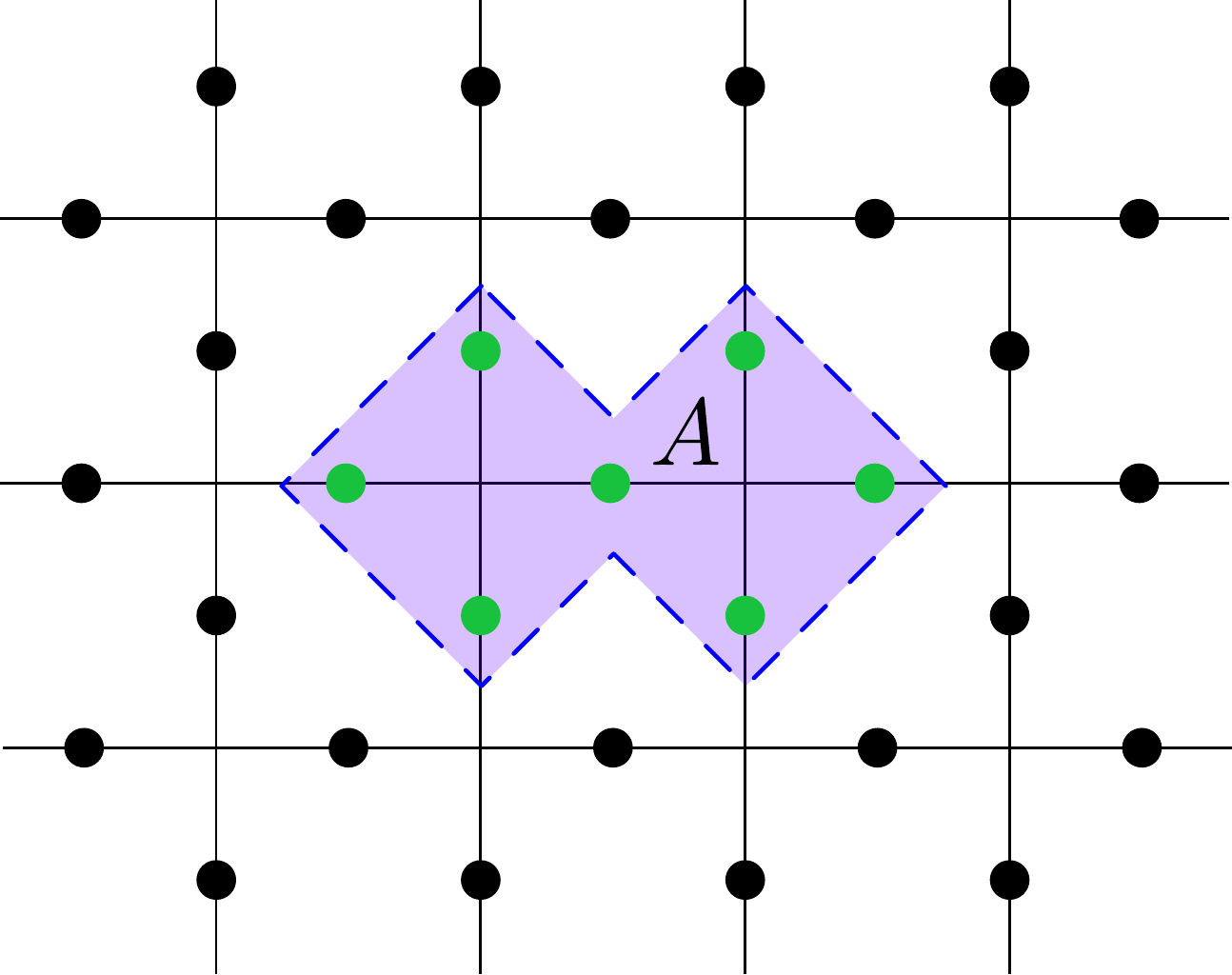}
 \caption{(color online) Subsystem $A$, shown as the shaded region, comprised of a total of seven spins which form two overlapping stars.}
  \label{2starsubsys}
\end{figure}

Thus we can calculate the trace of arbitary powers of the reduced density matrix, $\Tr(\rho_A^\alpha)=\sum_{s_1,s_2,s_3=-1,1}(\rho_A)^{\alpha}_{\bar{s}\bar{s}}$, using which the R\'{e}nyi entropies are given by (with $T(\lambda)=\braket{\hat{\tau}^x_{i}\hat{\tau}^x_{i+1}}$):
\begin{align}
S_{\alpha}(\lambda)&=\frac{1}{1-\alpha}\log[\frac{1}{2^{3\alpha}}\{2(1+T(\lambda))^{2\alpha}+2(1-T(\lambda))^{2\alpha}\nonumber\\
&~~~~~~~~~~~~~~~~~~~~~~~~~~~~~~~~~~~~~~~~~~~~+4(1-T(\lambda)^2)^\alpha\}]
\label{singplaqent}
\end{align}
From the plot of Eq.~(\ref{singplaqent}) in Fig.~(\ref{singplaqentplot}) we observe that  for all values of $\alpha=.01,.1,.5,1.01,2$, the entropies show monotonic behaviour with $\lambda$ in both the phases. While in the Topologically ordered phase, $\lambda<1$, the entropies decrease as we \emph{approach} the quantum critical point, for the disordered region it decreases as we \emph{move away} from it.


\subsubsection{\textbf{General treatment}}
\label{generaltreatment}

On the lattice, we call systems with a `bulk' those that have at least one or more star operators that act on spins exclusively belonging to $A$. This means that the group $G_A$ is non-trivial and the reduced density matrix for the subsystem is not diagonal anymore \cite{hamma:2005b}. Consequently, the analysis of this case is considerably more involved. We refer to \cite{hiz3} for an introduction to the technique used to treat a gauge theory. Since the perturbation we consider is gauge invariant, indeed, we can represent the state as the sum over element of a group, and this makes the calculation possible in the formalism. We can compute exactly the reduced density matrix (See Appendix (\ref{app:bulksys}) for details). Moreover, we can find an exact expression for the purity:
\begin{align}
P(\lambda)&=\frac{|G_B|}{|G|}\sum_{g\in G_A,z \in \mathcal{Z}_A}|\bra{\psi(\lambda)}g z\ket{\psi(\lambda)}|^2,
\label{purity}
\end{align}
where, $\ket{\psi(\lambda)}$ is the ground state of the Hamiltonian (\ref{horizhamil}) and $|G_B|$ is the cardinality of the group of star operators acting \emph{exclusively} in the complement of $A$ i.e. $G_B=\{g\in G|g=\openone_A\otimes g_B\}$. As before,  $G_A$ is the group of spin flips generated by star operators exclusively in $A$ while $\mathcal{Z}_A$ is the group generated by products of $\hat{\sigma}^z$'s acting on spins in A. 
This expression can be generalized to general gauge theories and quantum double models, and to a general R\'enyi entropy of index $\alpha$, and constitutes one of the main results of this paper. 

Although in principle we can calculate the entropies $S_{\alpha}(\lambda)$ for each integer $\alpha$,  we focus on  the $2-$R\'{e}nyi entropy only. In particular, we demonstrate that it has a monotonic behaviour in both the phases. The monotonicity of $S_{2}(\lambda)$ is sufficient to show that all higher entropies obey the same monotonicity because of the continuity of the entropies in $\alpha$ and because of their ordering relation: $S_{\alpha'}\leq S_{\alpha}~\forall \alpha'\geq \alpha$. On the other hand in the Toric code limit at $\lambda=0$, the eigenspectrum is flat with there being $2^5$ equal eigenvalues summing to 1 with the remaining $2^7-2^5=96$ eigenvalues, all zero. Turning on the perturbation has the effect of making some of these zero eigenvalues non-zero which shows up as an increase of $\lim_{\alpha \to 0}S_{\alpha}$ and other R\'{e}nyi entropies with $\alpha$ close to zero. Alternatively put: the Schmidt rank of the state $\ket{\psi}(\lambda)$ increases with $\lambda$ w.r.t. bipartitions with a bulk.

To analyze this case while keeping the presentation simple, we choose a subsystem $A$ which includes the 7 spins of two neighboring stars, Fig.~(\ref{2starsubsys}). For the calculations,  here we  use the symmetric ground state in the $TS_1$ sector. 

The evaluation of the R.H.S of Eq.~(\ref{purity}) again relies on the $\sigma-\tau$ correspondence (\ref{opmap}) and we get for the purity:
\begin{widetext}
\begin{align}
P&=\frac{1}{2^7}\{(1+<\hat{\tau}^x_1\hat{\tau}^x_2>^2)^2(1+3<\hat{\tau}^x_1\hat{\tau}^x_2>^2+2<\hat{\tau}^x_1\hat{\tau}^x_3>^2+<\hat{\tau}^x_1\hat{\tau}^x_4>^2+<\hat{\tau}^x_1\hat{\tau}^x_2\hat{\tau}^x_3\hat{\tau}^x_4>^2+
<\hat{\tau}^z_i>^2+<\hat{\tau}^x_1\hat{\tau}^z_2\hat{\tau}^x_3>^2+<\hat{\tau}^z_2\hat{\tau}^x_3\hat{\tau}^x_4>^2+\nonumber\\
&~~~~~~~~~~~~~~~~~~~~~~~~~~~<\hat{\tau}^x_1\hat{\tau}^z_2\hat{\tau}^x_4>^2+<\hat{\tau}^z_2\hat{\tau}^z_3>^2+<\hat{\tau}^z_2\hat{\tau}^z_3\hat{\tau}^x_2\hat{\tau}^x_3>^2+<\hat{\tau}^z_2\hat{\tau}^z_3\hat{\tau}^x_1\hat{\tau}^x_4>^2+<\hat{\tau}^z_2\hat{\tau}^z_3\hat{\tau}^x_1\hat{\tau}^x_2\hat{\tau}^x_3\hat{\tau}^x_4>)\}
\label{taupurity}
\end{align}
\end{widetext}

\begin{figure}
 \centering
\includegraphics[width=.9\columnwidth]{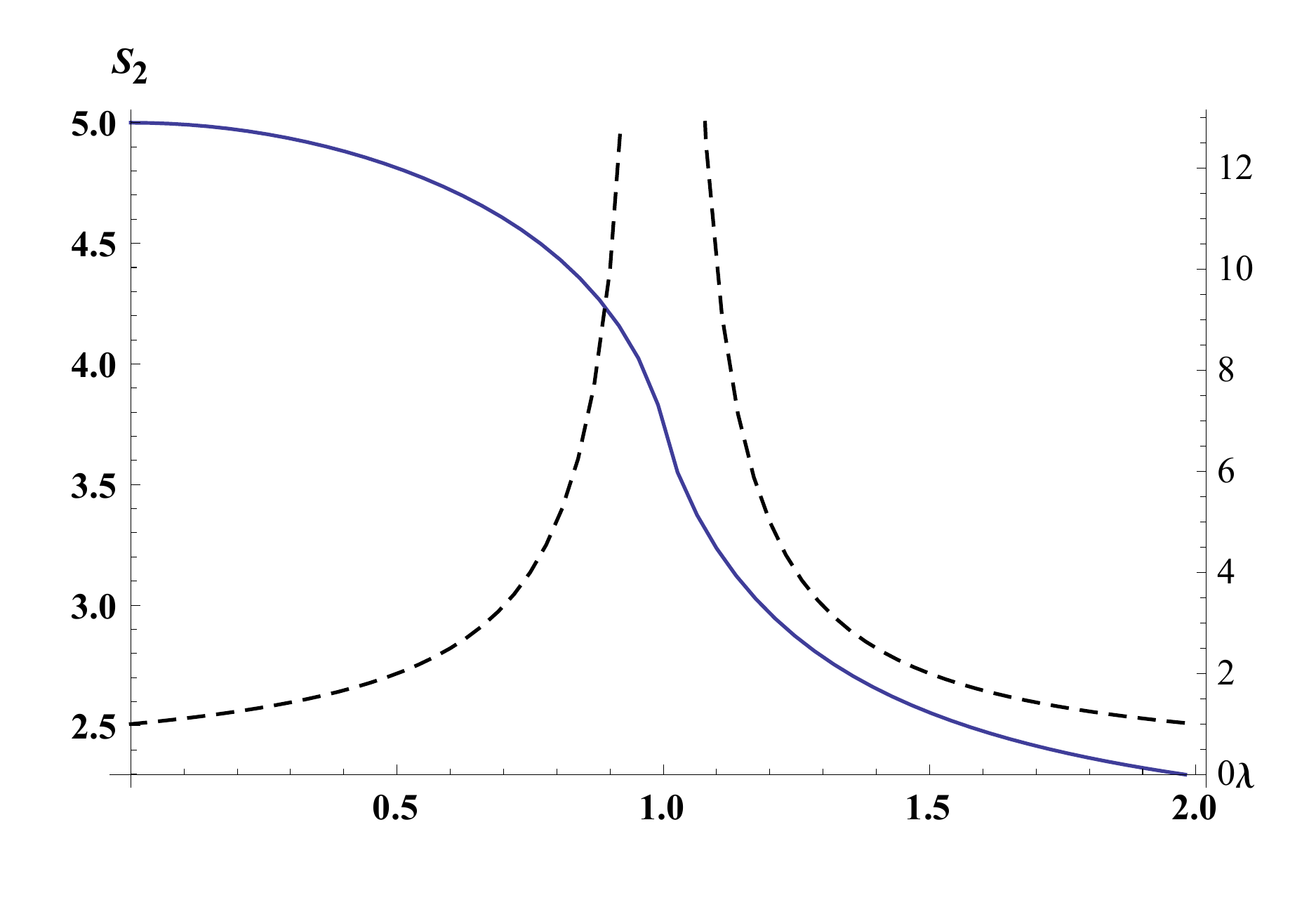}
 \caption{(color online) The 2-R\'{e}nyi entropy of a subsystem comprised of two stars $A$ (shown in Fig.~\ref{2starsubsys}) across the phase transition at $\lambda=1$ for $H=H_{TC}+V_2(\lambda)$. The monotonic behavior in both the phases for $S_2$ implies similar behavior for $S_\alpha~\forall\alpha\geq2$; whereas the general arguments presented in the text imply that  for $\alpha\to0$ they should increase till the quantum critical point. The dotted line is the inverse of the energy gap between the ground and first excited states for the transverse field Ising model to which the perturbed gauge-fixed Toric code Hamiltonian is mapped.}
  \label{2star-ent}
\end{figure}

The 2-R\'{e}nyi entropy  $S_2(\lambda)=-\log(P(\lambda))$ is shown in Fig.~(\ref{2star-ent}). Just as for the \emph{thin} subsystem case, we find similar monotonicity in the approach and departure from the quantum critical point.

\subsection{The Toric-Ising model, $V=V_3$}
\label{sec:toricising}

\begin{figure}[t]
 \centering
\includegraphics[width=0.7\columnwidth]{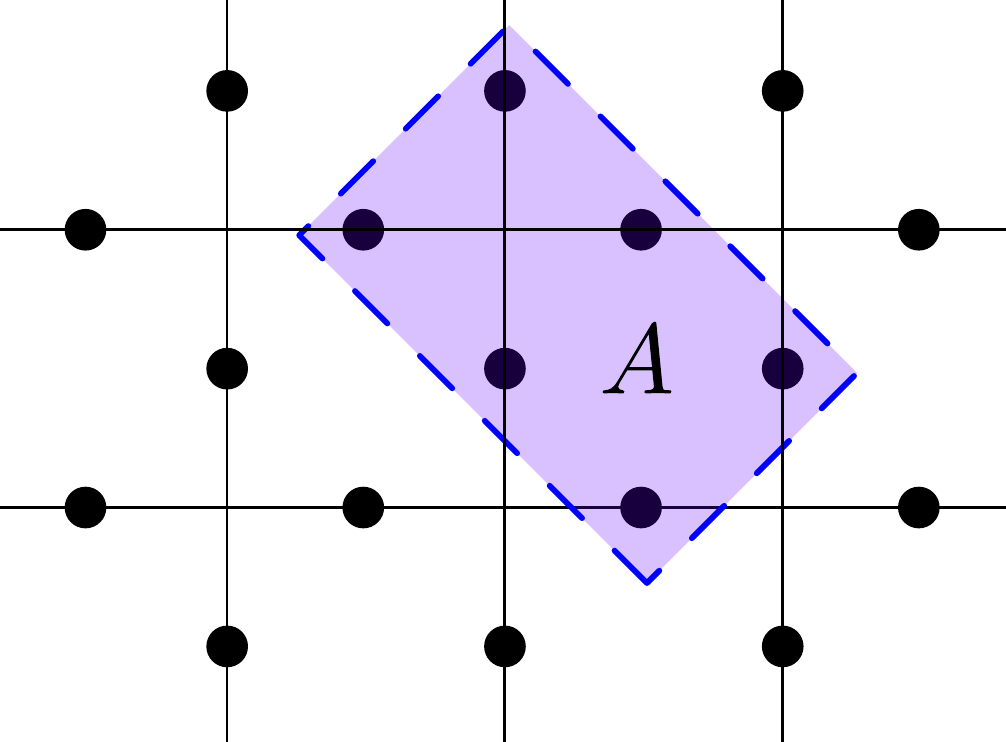}
 \caption{(color online) Subsystem $A$, shown as the shaded region, comprised of a total of six spins which make up the spins on a plaquette and two neighboring spins to its northeast corner.}
  \label{plaqplus2}
\end{figure}

\begin{figure}[t]
 \centering
\includegraphics[width=\columnwidth]{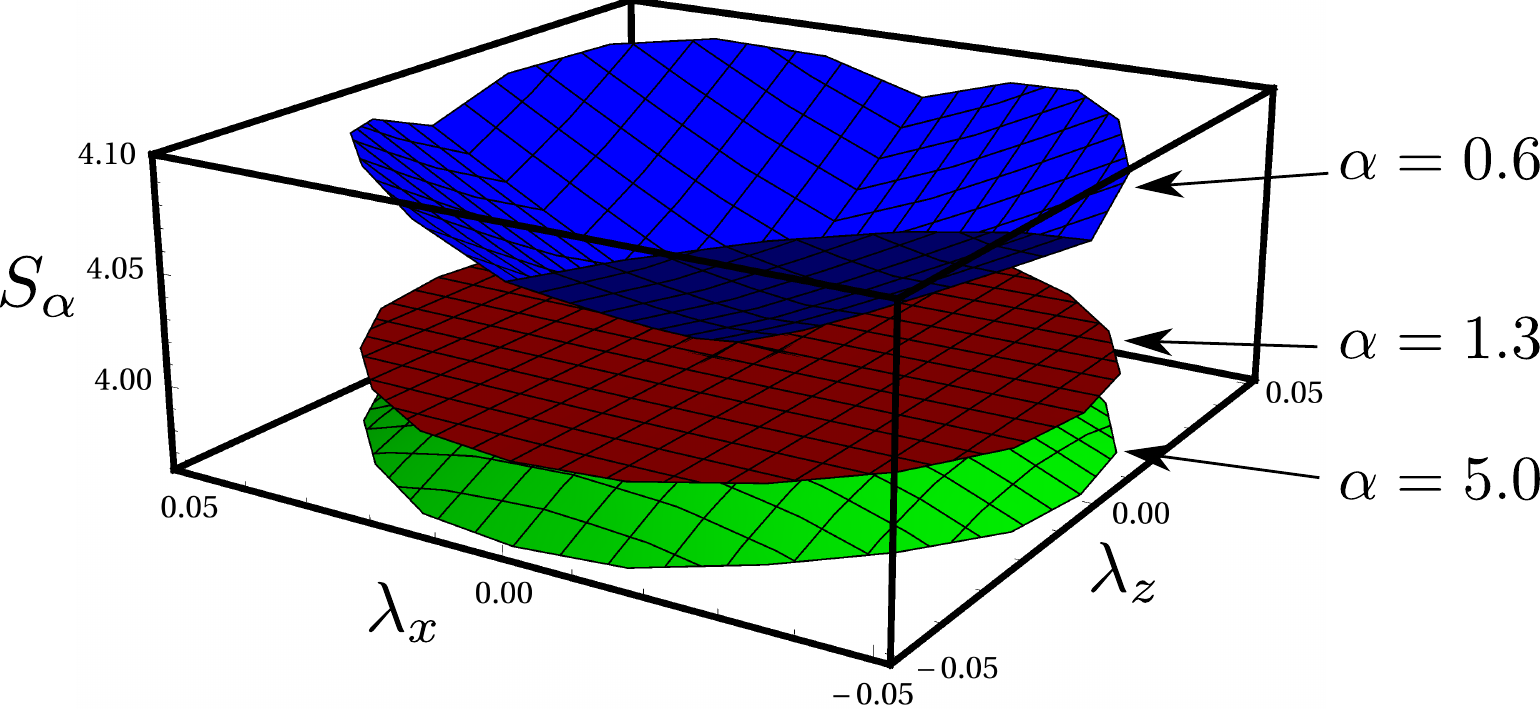}
 \caption{(color online) Behavior of three representative R\'enyi entropies for the Toric-Ising model ($V=V_3(\lambda_x,\lambda_z)$) in the perturbation parameter $(\lambda_x,\lambda_z)$-plane within the topologically ordered phase for subsystem $A$ as shown in Fig.~(\ref{plaqplus2}). For $\alpha=.6$ the entropy increases while for $\alpha=5$ it decreases monotonically with increasing correlation length. The change between these two types of behavior occurs at $\alpha\simeq1.3$, the value of which was identified numerically.}
  \label{fig:toricising}
\end{figure}

Here we consider the subsystem $A$ consisting of a plaquette with two adjoining spins pictured in Fig.\ref{plaqplus2} and numerically show that for the perturbation $V= V_3 (\lambda_x, \lambda_z) = -\sum_{i,\mu=\hat{x},\hat{y}} (\lambda_x \hat{\sigma}^x_i \hat{\sigma}^x_{i+\mu} + \lambda_z \hat{\sigma}^z_i \hat{\sigma}^z_{i+\mu})$, which takes the Toric code Hamiltonian from a TO phase to a ferromagnetic phase, the set of R\'{e}nyi entropies in the TO phase show the splitting behavior. Note that neither the perturbation here nor the choice of the subsystem is fine-tuned. In other words, the lack of differential local convertibility is a robust property of the topologically ordered phase and is universal. Here by universal we mean that this property should hold for all quantum systems that show similar behavior in their entanglement spectrum landscape and correlation length behavior. However, the value of $\alpha$ for the R\'enyi index such that the sign of the derivative $\del_\lambda S_{\alpha}(\lambda)$ changes, is non universal and is numerically found here to be $\alpha\simeq 1.3$, see Fig.\ref{fig:toricising}. The space of the parameters spanned is deep in the topological phase, with $|\lambda_{x,z}|\le 0.05$. For high $\lambda$ values i.e. in the ferromagnetic phase, the sign is found to be the same (not shown in the plot) for every value of the R\'enyi index $\alpha$.
 
Thus even in this model where a phase transition occurs from a TO phase to a ferromagnetic one the latter exhibits differential local convertibility whereas the former does not.

\subsection{Summary of results}
\label{summary}
Here we collect the main results of this section that will help formulate, in the conclusions, the conjecture about the splitting phenomenon of the Renyi's entropies.
\begin{itemize}
\item For perturbations (\ref{sec:chamon}) with \emph{constant} correlation length and \emph{any} bipartition the behaviour of the R\'{e}nyi entropies is monotonic and there is no splitting phenomenon.
\item For perturbations (\ref{sec:horizfield}) with \emph{non-constant} correlation length and \emph{thin} bipartitioning the behaviour of the R\'{e}nyi entropies is monotonic and there is no splitting phenomenon.
\item For perturbations (\ref{sec:horizfield}) with \emph{non-constant} correlation length and \emph{bulk} bipartitioning the Reny's entropies split.
\item For \emph{general} perturbations (\ref{sec:toricising}) the splitting behaviour of the entropies is robust and happens without reference to the size of the subsystem as long as the subsystem has some \emph{bulk}.
\end{itemize}

\section{Discussion and conclusion}
\label{discussion}
In this paper, we have considered a paradigmatic class of topological phases, as those ones arising from the Toric code  with perturbations driven by a set of control parameters $\lambda=(\lambda^1,...,\lambda^n)$. We focused on the case where the  energy gap can vanish, giving rise to a quantum phase transition to a topologically trivial phase (paramagnet). The perturbations studied affect the correlation length $\xi$ of the system that is vanishing for the exact Toric code at $\lambda=0$. 

We have shown that the two phases can be distinguished through their differing local-convertible behavior:  Bipartitioning the system into subsystems $A$ and $B$, the result of the local-convertibility analysis is that two nearby states in the topological quantum phase, generically cannot be connected by Local Operations in $A$ and $B$ augmented with Classical Communications (even in the presence of a catalyst); in the paramagnetic phases, in contrast, the states are locally convertible. This is consistent with the fact that in the topogically trivial phases it is always possible to transform the ground state to a totally factorized state in the physical degrees of freedom by using a local unitary quantum circuit of fixed depth. The locally convertible character of a phase implies it's limited adiabatic computational power since the physical transformation may be simulated using LOCC operations which do not generate quantum coherences between the two parts of the bipartition \cite{cui:2012, franchini:2013}.

Local-convertibility is shown to depend on the manner in which the R\'{e}nyi entropies of the reduced state on a subsystem behave: The non-local convertible phase features a splitting behaviour of the entropies, with their partial derivative along  the control  parameter $\lambda^i$ changing  sign for a particular value of the R\'enyi index $\alpha$. The value of $\alpha$ at which the splitting occurs is instead dependent on the details of the model. The splitting phenomenon  is observed  within the whole topological phase irrespective of the particular form of the perturbation or of the subsystem $A$, unless it is very fine tuned - such as the ones without any \emph{bulk}. For the class of systems we considered, perturbed Toric code models, subsystems with bulk are those that have at least one star operator acting exclusively within it. This implies that the correlation length for local observables (in the subsystem) is non-constant and yet more degrees of freedom contribute to the entanglement spectrum with an increase of correlation length as the perturbation is increased.

 
There is no constraint on how small a subsystem needs to be, aside from the caveats that would qualify a bipartition as fine tuned, for its entropies to show the splitting behavior. This makes the experimental analysis of local-convertibility quite feasible. A schematic of such a protocol is as follows: One first identifies a subsystem small enough to permit complete state tomography with the resources at hand. Second, the system is perturbed by some easy to implement perturbation. State tomography is then done for the subsystem at different values of the perturbation strength. The knowledge of the state at these values yields the R\'{e}nyi entropy plots which establish the state's local-convertibility behavior. By repeating these steps for a few different bipartitions and perturbations one should be able to identify the non locally-convertible behavior, if any, of the global pure state of the whole system. Alternatively, as argued in Sec.~(\ref{generaltreatment}) if the perturbation increases the correlation length then the R\'{e}nyi entropies for small $\alpha\to 0$ values are bound to increase, with increasing perturbation strength, as yet more degrees of freedom contribute to the entanglement spectrum. Thus monotonically decreasing behavior of only the $\alpha=2$ R\'{enyi} entropy in the topologically ordered phase, which guarantees the same behavior for entropies with $\alpha>2$, along with a measurement of the correlation length is sufficient to identify the splitting behavior. The 2-R\'{e}nyi entropy, $S_2(\lambda)=-\log\tr[\rho(\lambda)^2]$, in turn can be determined by purity measurements which directly accessible to experiments \cite{puritynote}.

We hasten to point out that it is our view that the non-LOCC convertibility is typical of states in a phase with no local order parameter, including for e.g. topologically ordered states. Further, that it is a necessary but not sufficient condition exhibited by such states. A case in point is the analysis for cluster states \cite{kalis:2012,son:2011,raussendorf} shown in Fig.~(\ref{fig:cluster}). While these states are topologically trivial they do not have a local order parameter and a perturbative analysis of their R\'{e}nyi entropies shows the characteristic splitting behavior and imply their non local-convertibility.

This phenomenon relies on the structure of the entanglement spectrum around a special point in the phase. Indeed, in the topologically ordered phase of this model there exists an extremal point with a flat entanglement spectrum and zero correlation length, $\xi=0$. As we perturb away from this point, if the correlation length $\xi$ also increases then newer degrees of freedom get involved in the entanglement spectrum as a result of which the lower ($\alpha\to0$) entropies increase, on the other hand the higher $\alpha$ entropies decrease because of the algebraic suppression of the contributions from the new small but non-zero values in the spectrum and loss of contributions from the previously non-zero larger eigenvalues. We comment that since similar phenomenology in the entanglement spectrum is known to be displayed in cluster states \cite{kalis:2012,son:2011,raussendorf}, or more generally in all graph states \cite{nielsen:2006}, similar findings in the R\'{e}nyi entropies response should apply to those as well. Our work here should be seen as supporting a growing body of evidence \cite{cui:apr2013, franchini:2013} that this characteristic perturbative response would hold for a wider class of states such as quantum double models, cluster states and other quantum spin liquids. In the Toric code case knowledge about the ground state degeneracy can additionally distinguish its TO ground states from the latter. Compared to this, ground states of all symmetry broken phases exhibit monotonic behaviour of their R\'{e}nyi entropies with an increase in correlation length, and are thus always locally convertible \cite{franchini:2013}.

In order to compute the R\'enyi entropies for the perturbed toric code, we have resorted to two methods. For general perturbations that break gauge invariance, and also make the system non integrable, we resort to a 2D DMRG method, which can treat infinite cylinders \cite{cincio:2013}. On the other hand, for the gauge invariant perturbation, we find a general expression for the R\'enyi entropies, that can be generalized to every gauge theory \cite{preparation}. Moreover, for a particular form of the perturbation, the system is integrable, and we can find an exact analytical formula for the R\'enyi entropy. This result is technically relevant, and would allow to treat several problems, including stability issues at zero \cite{klich, bravyi, bravyi2} and finite temperature \cite{finiteT, finiteT2,finiteT3,nussinov}, the confinement problem \cite{confinement}, and the identification of relevant correlations \cite{correlations, correlations2}. A very important arena in which this technique can be useful is the dynamical problem \cite{dyn2, spectroscopy, armin}, e.g. the resilience of the splitting property or of topological entropies after a quantum quench \cite{tso,halasz:2012b}. Similarly, this technique can prove useful to probe the resilience to perturbations of measures of topological order based on negativity \cite{negativity1, negativity2}, or symmetry principles \cite{nussinov}.

In perspective, it would also be interesting to see if the local convertibility properties -or failure of thereof- hold for more general TO states without flat entanglement spectra such as fractional quantum Hall states \cite{isakov:2011,sdsarma,tchak} and chiral spin liquids  \cite{papa:2007}.

\acknowledgments
This work was supported in part by the National Basic Research Program of China Grant 2011CBA00300, 2011CBA00301 the National Natural Science Foundation of China Grant 61073174, 61033001, 61061130540.

Research at Perimeter Institute for Theoretical Physics
is supported in part by the Government of Canada through NSERC and
by the Province of Ontario through MRI.

LC acknowledges support from the John Templeton Foundation. PZ is supported by the ARO MURI grant W911NF-11-1-0268 and by NSF grant PHY- 969969. SS and YS would like to thank the hospitality of the Perimeter Institute where most of this work was done while they were visiting graduate students.

\appendix

\section{Calculations for the Castelnovo Chamon model}
\label{app:ccmodel}

\subsection{Derivative of the R\'{e}nyi entropy}
\label{ccrenderiv}
 Differentianting Eq.~(\ref{ccrenyi}) w.r.t $\lambda$ yields:
\begin{widetext}
\begin{align}
\partial_{\lambda}S_{\alpha}(\rho_A)&=\partial_{\lambda}(\frac{1}{(1-\alpha)}\log\mathrm{Tr}[\rho^{\alpha}_A])\nonumber\\
&=\frac{1}{(1-\alpha)\mathrm{Tr}[\rho^{\alpha}_A]}\partial_{\lambda}(\mathrm{Tr}[\rho^{\alpha}_A])\nonumber\\
&= \frac{1}{(1-\alpha)\mathrm{Tr}[\rho^{\alpha}_A]}\partial_{\lambda}[\frac{1}{Z^{\alpha}(\lambda)}\sum_{g\in G}e^{-\lambda E_g}w^{\alpha-1}(\lambda,g)]\nonumber\\
&=\frac{1}{(1-\alpha)\mathrm{Tr}[\rho^{\alpha}_A]}\{ \frac{1}{Z^{\alpha}(\lambda)}\sum_{g\in G}[-E_ge^{-\lambda E_g}w^{\alpha-1}(\lambda,g)+(\alpha-1)w'(\lambda,g)w^{\alpha-2}(\lambda,g)e^{-\lambda E_g}]\nonumber\\
&~~~~~~~~~~~~~~~~~~~~~~~~~~~~~~~~~~~~~~~~~~~~~~~~~~~~~~~~~~~~~~~~~~~~~~~~~~~~~~~~-\frac{\alpha Z'(\lambda)}{Z^{\alpha+1}(\lambda)}\sum_{g\in G}e^{-\lambda E_g}w^{\alpha-1}(\lambda,g)\}\nonumber\\
&=\frac{1}{(1-\alpha)\mathrm{Tr}[\rho^{\alpha}_A]Z^{\alpha}(\lambda)}\{ \sum_{g\in G}[-E_ge^{-\lambda E_g}w^{\alpha-1}(\lambda,g)+(\alpha-1)w'(\lambda,g)w^{\alpha-2}(\lambda,g)e^{-\lambda E_g}]\nonumber\\
&~~~~~~~~~~~~~~~~~~~~~~~~~~~~~~~~~~~~~~~~~~~~~~~~~~~~~~~~~~~~~~~~~~~~~~~~~~~~~~~~-\frac{\alpha Z'(\lambda)}{Z(\lambda)}\sum_{g\in G}e^{-\lambda E_g}w^{\alpha-1}(\lambda,g)\}\nonumber\\
&=\frac{1}{(1-\alpha)\mathrm{Tr}[\rho^{\alpha}_A]Z^{\alpha}(\lambda)}\{ \sum_{g\in G}[-E_ge^{-\lambda E_g}w^{\alpha-1}(\lambda,g)+(\alpha-1)w'(\lambda,g)w^{\alpha-2}(\lambda,g)e^{-\lambda E_g}]\nonumber\\
&~~~~~~~~~~~~~~~~~~~~~~~~~~~~~~~~~~~~~~~~~~~~~~~~~~~~~~~~~~~~~~~~~~~~~~~~~~~~~~~~
+\alpha \avg{E_g}_{Z(\lambda)}\sum_{g\in G}e^{-\lambda E_g}w^{\alpha-1}(\lambda,g)\}\nonumber\\
&=\frac{1}{(1-\alpha)\mathrm{Tr}[\rho^{\alpha}_A]Z^{\alpha}(\lambda)}\sum_{g\in G}[\frac{(\alpha-1)w'(\lambda,g)}{w(\lambda,g)}+\alpha\avg{E_g}_{Z(\lambda)}-E_g]e^{-\lambda E_{g}}w^{\alpha-1}(\lambda,g)
\label{diffren}
\end{align}
\end{widetext}
 In the second last line above we have used the fact that $\frac{Z'(\lambda)}{Z(\lambda)}=-\sum_{g\in G} \frac{E_g e^{-\lambda E_g}}{Z(\lambda)}=-\avg{E_g}_{Z(\lambda)}$. Next we define certain averages that appear in eq.(\ref{diffren}). For any function $f(h,g,k), g\in G, h\in G_A, k\in G_B, f(g)=f(h=\openone_A,g,k=\openone_B)$ we have:
\begin{align}
&\avg{f(g)}_{Z(\lambda)}:= \sum_{g\in G} f(g)\frac{e^{-\lambda E_g}}{Z(\lambda)},~~~~Z(\lambda):=\sum_{g\in G}e^{-\lambda E_g}\nonumber\\
&\avg{f(h,g,k)}_{w(\lambda,g)}:= \sum_{\substack{h\in G_A\\ k\in G_B}} f(h,g,k)\frac{e^{-\lambda E_{hgk}}}{w(\lambda,g)}\nonumber\\
&~~~~~~~~~~~~~~~~~~~~~~~~~~~~~~~~~~~~~w(\lambda,g):=\sum_{\substack{h\in G_A\\ k\in G_B}}e^{-\lambda E_{hgk}}\nonumber\\
&\avg{f(g)}_{\tilde{Z}(\lambda,\alpha)}:=\sum_{g\in G}f(g)\frac{e^{-\lambda E_g}w^{\alpha-1}(\lambda,g)}{\tilde{Z}(\lambda,\alpha)}\nonumber\\
&~~~~~~~~~~~~~~~~~~~~~~~~~~~~~~~~~~~~~\tilde{Z}(\lambda,\alpha):=\sum_{g\in G}e^{-\lambda E_g}w^{\alpha-1}(\lambda,g)
\label{partfun}
\end{align}

Observe now that the term outside the sum in eq.(\ref{diffren}) has in the denominator the product $\mathrm{Tr}[\rho^{\alpha}_A]Z^{\alpha}(\lambda)=\sum_{g\in G}e^{-\lambda E_g}w^{\alpha-1}(\lambda,g)=\tilde{Z}(\lambda,\alpha)$. This implies that the R.H.S. of eq.(\ref{diffren}) is really an average w.r.t. the new partition function $\tilde{Z}(\lambda,\alpha)$ i.e.:
\begin{align}
\partial_{\lambda}S_{\alpha}(\lambda)&=\frac{1}{(1-\alpha)}\sum_{g\in G}T(\alpha,\lambda,g)\frac{e^{-\lambda E_g}w^{\alpha-1}(\lambda,g)}{\tilde{Z}(\lambda,\alpha)}
\label{compact}
\end{align}

where $T(\alpha,\lambda,g):=[\frac{(\alpha-1)w'(\lambda,g)}{w(\lambda,g)}+\alpha\avg{E_g}_{Z(\lambda)}-E_g]$.\\
 Further note that $\frac{w'(\lambda,g)}{w(\lambda,g)}=-\frac{\sum_{\substack{h\in G_A\\k\in G_B}}E_{h,g,k}e^{-\lambda E_{hgk}}}{w(\lambda,g)}=-\avg{E_g}_{w(\lambda,g)}$ is a function of $g\in G$ whereas $\frac{Z'(\lambda)}{Z(\lambda)}=-\sum_{g\in G} \frac{E_g e^{-\lambda E_g}}{Z(\lambda)}=-\avg{E_g}_{Z(\lambda)}$ is independent of $g\in G$. Equation(\ref{compact}) thus takes the form of a sum of averages:
\begin{align}
\partial_{\lambda}S_{\alpha}(\lambda)&=\avg{\avg{E_g}_{w(\lambda,g)}}_{\tilde{Z}(\lambda,\alpha)}+\frac{\alpha}{(1-\alpha)}\avg{E_g}_{Z(\lambda)}\nonumber\\&~~~~~~~~~~~~~~~~~~~~~~~~~~~~~~~~~~~~~~~~~-\frac{1}{(1-\alpha)}\avg{E_g}_{\tilde{Z}(\lambda,\alpha)}
\label{avgsum}
\end{align}


\subsection{Perturbations around the toric code limit}
One can perform a small $\lambda$ expansion of eq.(\ref{avgsum}) to see that the model permits DLOCC for any bipartition for small perturbations to the Toric Code limit of $\lambda=0$. To see this let us note the following:
\begin{align}
Z(\lambda)&\approx\sum_{g\in G}(1-\lambda E_g)=|G|-\lambda\sum_gE_g\nonumber\\
w(\lambda,g)&\approx\sum_{\substack{h\in G_A\\k\in G_B}}(1-\lambda E_{hgk})=|G_A||G_B|-\lambda\sum_{\substack{h\in G_A\\k\in G_B}}E_{hgk}\nonumber\\
\tilde{Z}(\lambda,\alpha)&\approx\sum_{g\in G}(1-\lambda E_g)(|G_A||G_B|-\lambda\sum_{\substack{h\in G_A\\k\in G_B}}E_{hgk})^{\alpha-1}\nonumber\\
&\approx\sum_{g\in G}(|G_A||G_B|)^{\alpha-1}(1-\lambda E_g)(1-\frac{\lambda(\alpha-1)}{|G_A||G_B|}\sum_{\substack{h\in G_A\\k\in G_B}}E_{hgk})\nonumber\\
&\approx(|G_A||G_B|)^{\alpha-1}\sum_{g\in G}(1-\lambda E_g-\frac{\lambda(\alpha-1)}{|G_A||G_B|}\sum_{\substack{h\in G_A\\k\in G_B}}E_{hgk})\nonumber\\
&=(|G_A||G_B|)^{\alpha-1}(|G|-\sum_gE_g-\frac{\lambda(\alpha-1)}{|G_A||G_B|}\sum_g\sum_{\substack{h\in G_A\\k\in G_B}}E_{hgk})\nonumber\\
&=(|G_A||G_B|)^{\alpha-1}(|G|-\alpha\lambda\sum_gE_g)\nonumber\\
\end{align}
Using the weights $(1-\lambda E_g),(1-\lambda E_hgk), (1-\lambda\alpha E_g)$ for the evaluation of the averages w.r.t. $Z(\lambda),w(\lambda,g),\tilde{Z}(\lambda,\alpha)$ respectively we find that :
\begin{align}
\partial_{\lambda}S_{\alpha}(\lambda)&=0+\lambda(C_1\alpha+C_2)+\text{higher order terms in $\lambda$}\nonumber\\
\text{with}~~~~C_1&=\frac{(\sum_g E_g)^2}{|G|^2}-\frac{\sum_g\sum_{\substack{h\in G_A\\k\in G_B}}E_gE_{hgk}}{|G||G_A||G_B|}\nonumber\\
C_2&=-\frac{\sum_g E_g^2}{|G|}+\frac{\sum_g\sum_{\substack{h\in G_A\\k\in G_B}}\sum_{\substack{h'\in G_A\\k'\in G_B}}E_{hgk}E_{h'gk'}}{|G|(|G_A||G_B|)^2}
\end{align}
To prove that $C_1,C_2\leq0$ we note that cosets w.r.t. the subgroup $G_A\times G_B$ of the group $G$ divide the group into disjoint subsets. If $q$ labels these unique subsets then one can write:
\begin{align}
\sum_g\sum_{\substack{h\in G_A\\k\in G_B}}E_gE_{hgk}&\equiv\sum_{q\in Q=G/(G_A\times G_B)}\sum_{\substack{h'\in G_A\\k'\in G_B}}\sum_{\substack{h\in G_A\\k\in G_B}}E_{h'qk'}E_{hh'qk'k}\nonumber\\
&=\sum_{q\in Q=G/(G_A\times G_B)}\sum_{\substack{h'\in G_A\\k'\in G_B}}E_{h'qk'}\sum_{\substack{\tilde{h}\in G_A\\\tilde{k}\in G_B}}E_{\tilde{h}q\tilde{k}}\nonumber\\
&=\sum_{q\in Q=G/(G_A\times G_B)}(\sum_{\substack{h'\in G_A\\k'\in G_B}}E_{h'qk'})^2\nonumber\\
\end{align}
Let us now note that each $E_g\geq 0 \forall g\in G$. Thus to prove that $C_1\leq 0$ one needs to prove that for a collection of $|G|$ positive numbers $E_1,E_2,....,E_{|G|}$ any grouping of $|G_A|\times |G_B|$ numbers such that $mod(|G|,|G_A|\times |G_B|)=0$ yields (with $k=|G|/|G_A|\times |G_B|$)
\begin{widetext}
\begin{align}
\frac{[(E_1+E_2+....+E_{|G_A||G_B|})^2+(E_{|G_A||G_B|+1}+...+E_{2|G_A||G_B|})^2+(E_{(k-1)|G_A||G_B|}+...+E_{k|G_A||G_B|})^2]}{|G||G_A||G_B|}&\geq \frac{(E_1+E_2+....+E_{|G|})^2}{|G|^2}
\label{ineq1}
\end{align}
\end{widetext}

with equality holding iff $E_1=E_2=...=E_{|G|}$. If one represents the sum of the energies in each coset by $S_i,i=1,2,....,k$ then condition (\ref{ineq1}) is equivalent to proving:
\begin{align}
S_1^2+...+S_k^2\geq \frac{(S_1+S_2,+...+S_k)^2}{k}\nonumber\\
\implies (k-1)[S_1^2+...+S_k^2]\geq 2\sum_{i<j}S_iS_j\nonumber
\end{align}
which is the sum of several inequalities all of which are of the form $(S_i^2+S_j^2)\geq 2S_iS_j$. The same inequalities are used to prove $C_2\leq 0$ by noticing that:
\begin{align}
\sum_g\sum_{\substack{h\in G_A\\k\in G_B}}\sum_{\substack{h'\in G_A\\k'\in G_B}}E_{hgk}E_{h'gk'}&=\sum_q\sum_{\substack{h''\in G_A\\k''\in G_B}}\sum_{\substack{h\in G_A\\k\in G_B}}\sum_{\substack{h'\in G_A\\k'\in G_B}}E_{hh''qk''k}E_{h'h''qk''k'}\nonumber\\
&=|G_A||G_B|\sum_q\sum_{\substack{h\in G_A\\k\in G_B}}\sum_{\substack{\tilde{h}\in G_A\\\tilde{k}\in G_B}}E_{hqk}E_{\tilde{h}q\tilde{k}}\nonumber\\
&=|G_A||G_B|\sum_q(\sum_{\substack{h\in G_A\\k\in G_B}}E_{hqk})^2
\end{align}


\subsection{Large-$\lambda$ : Spin Polarized Phase}
For the large-$\lambda$ case note that successive contributions to the partition functions get suppressed by factors of $e^{-2 \lambda }$. This is because the possible lengths of loops increase in steps of two after the shortest non-trivial length of 4 i.e. $E_g=0,4,6,8,10,....$. Although the number of loops of each length increases algebraically in the number of sites in the lattice, the exponential suppression means that we can consider only the maximally contributing term in a proper limit of $\lambda$. Thus,
\begin{align}
Z(\lambda)&=\sum_{g\in G}e^{-\lambda E_g}=1+L^2e^{-4\lambda}+O(e^{-6\lambda})\nonumber
\approx 1+L^2e^{-4\lambda}
\end{align}

The partition function $w(\lambda,g)$ depends on the particular value of the element $g\in G$ and hence admits three possibilities:\\
case(1) When $g=g_A\times g_B \in G_A\times G_B$ we have:
\begin{align}
w(\lambda,g)&=\sum_{\substack{h\in G_A\\k\in G_B}}e^{-\lambda E_{(hg_A)(kg_B)}}\nonumber\\
&=\sum_{\substack{h'\in G_A\\k'\in G_B}}e^{-\lambda E_{(h')(k')}}\nonumber\\
&=\sum_{\substack{h\in G_A\\k\in G_B}}e^{-\lambda (E_{h'}+E_{k'})}\nonumber\\
&=(\sum_{h'\in G_A}e^{-\lambda E_{h'}})(\sum_{k'\in G_B}e^{-\lambda E_{k'}})\nonumber\\
&=(1+n_Ae^{-4\lambda}+O(e^{-6\lambda}))(1+n_Be^{-4\lambda}+O(e^{-6\lambda}))\nonumber\\
&\approx 1+(n_A+n_B)e^{-4\lambda}
\end{align}
where $n_A,n_B$ are respectively the number of independent star operators in $A$ and $B$ - the two parts of the bipartition.\\
case(2) When $g\notin G_A\times G_B$ there are two subcategories of such operators.\\
case(2a) For $g=A_{\partial A} (g_A\times g_B)$ i.e. a product of a single boundary star operator and an element from the subgroup $G_A\times G_B$ the only non-vanishing contribution to $w(\lambda,g)$ comes from a loop of length 4 and thus $w(\lambda,g)=e^{-\lambda 4}$\\
case(2b) For all other loop operators $g\in G$, $w(\lambda,g)=0$ in the limit that we are working in.\\
Thus a complete list of $w(\lambda,g)$ for any $g\in G$ is as follows:
\begin{align}
w(\lambda,g)&=\begin{cases}(1+n_{AB}e^{-4\lambda})~~~\forall g\in G_A\times G_B\\e^{-4\lambda}~~~~~~~~~~~~~~~~~~~~\forall g=A_{\partial A}(g_A\times g_B)\\0~~~~~~~~~~~~~~~~~~~~~~~~~~~\text{otherwise} \end{cases}
\end{align}
with $n_{AB}=n_A+n_B$.\\
At this point let us also evaluate the partition function $\tilde{Z}(\lambda,\alpha)=\sum_{g\in G}e^{-\lambda E_g}w^{\alpha-1}(\lambda,g)$. Note that because of the dependence on $\alpha$ in the different terms of the partition function we get different forms for $\tilde{Z}(\lambda,\alpha)$ for $\alpha>1$ and $\alpha<1$.
\begin{align}
\tilde{Z}(\lambda,\alpha)&=\begin{cases}1+\alpha n_{AB}e^{-4\lambda} ~~~~\text{for}~~~\alpha>1\\ 1+\alpha n_{AB}e^{-4\lambda}+L_{\partial A}e^{-4\lambda\alpha}~~~~\text{for}~~~\alpha<1\end{cases}
\end{align}
where $L_{\partial A}$ is the length of the boundary of the bipartition.\\
Now we evaluate the 3 different expectation values of the loop lengths and find that :
\begin{align}
\braket{E_g}_{Z(\lambda)}&=\frac{4.L^2e^{-4\lambda}}{1+L^2e^{-4\lambda}}\nonumber\\
\braket{E_g}_{\tilde{Z}(\lambda,\alpha)}&=\begin{cases}\frac{4n_{AB}e^{-4\lambda}}{1+\alpha n_{AB}e^{-4\lambda}}~~~\text{for}~~~\alpha>1\\\frac{4(n_{AB}e^{-4\lambda}+L_{\partial A}e^{-4\lambda\alpha})}{1+\alpha n_{AB}e^{-4\lambda}+L_{\partial A}e^{-4\lambda\alpha}}~~~\text{for}~~~\alpha<1\end{cases}\nonumber\\
\braket{\braket{E_g}_{w(\lambda,g)}}_{\tilde{Z}(\lambda,\alpha)}&=\begin{cases}\frac{4n_{AB}e^{-4\lambda}}{1+\alpha n_{AB}e^{-4\lambda}}~~~\text{for}~~~\alpha>1\\\frac{4(n_{AB}e^{-4\lambda}+L_{\partial A}e^{-4\lambda \alpha})}{1+\alpha n_{AB}e^{-4\lambda}+L_{\partial A}e^{-4\lambda \alpha}}~~~\text{for}~~~\alpha<1\end{cases}
\label{expecs}
\end{align}
Using the expressions (\ref{expecs}) in eq.(\ref{avgsum}) we get the derivative of the Renyi entropy in the two domains of $\alpha$ to be:
\begin{align}
\partial_{\lambda}S_{\alpha}(\lambda)&=\frac{4\alpha e^{-4\lambda}}{1-\alpha}[\frac{(L^2-n_{AB})+(\alpha-1)n_{AB}L^2e^{-4\lambda}}{(1+L^2e^{-4\lambda})(1+\alpha n_{AB}e^{-4\lambda})}], ~~\alpha>1\nonumber\\
&=\frac{4\alpha e^{-4\lambda}}{1-\alpha}[\frac{((L^2-n_{AB})-L_{\partial A}e^{-4\lambda(\alpha-1)})}{(1+L^2e^{-4\lambda})(1+\alpha n_{AB}e^{-4\lambda}+L_{\partial A}e^{-4\lambda\alpha})}]\nonumber\\
&~~~~~~~~~~~~~~~~~~~~~~~~~~~~~~~~~~~~~~~~~~~~~~~~~,~~\alpha<1
\end{align}
Note that in the above equation for $\alpha>1$ the numerator is clearly positive for the term in the square bracket whereas the factor $\frac{1}{1-\alpha}$ provides the overall negative sign. For the $\alpha<1$ region that the numerator in the square brackets yields a negative sign can be seen as follows:
\begin{align}
&(L^2-n_{AB})-L_{\partial A}e^{-4\lambda(\alpha-1)}<0\nonumber\\
&\implies \frac{1}{1-\alpha}\log[\frac{L^2-n_{AB}}{L_{\partial A}}]<\lambda\nonumber
&\implies 0<\lambda
\end{align}
which is always true and where we use the fact that $L^2-n_{AB}=L_{\partial A}$.

\section{Calculations for the Toric Code with external field along Horizontal rows}

\subsection{Thin systems}
\label{app:thinsys}
\begin{figure}
 \centering
\includegraphics[width=0.5389\columnwidth]{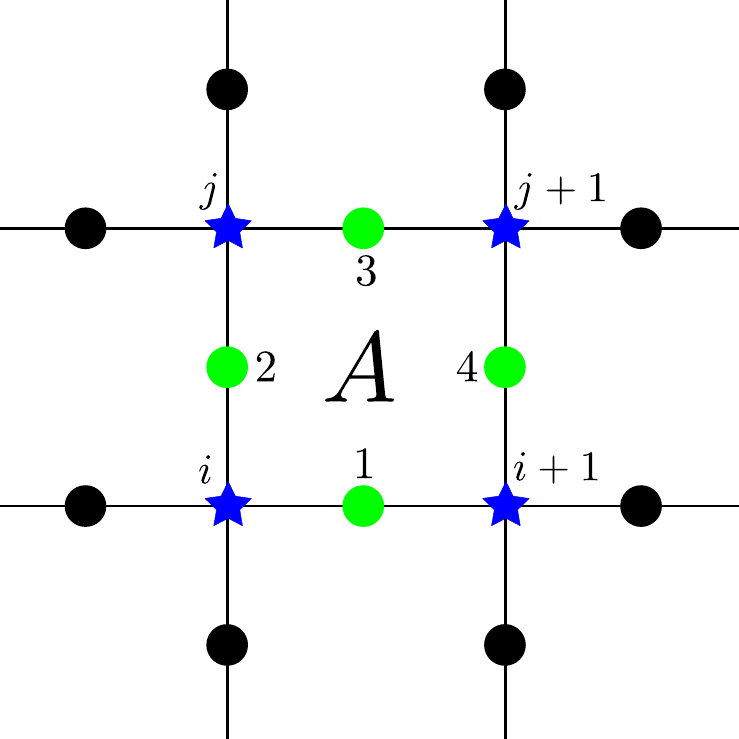}
 \caption{(color online) Subsystem $A$ of one plaquette with the spins (green ovals), $1,2,3,4$, on the edges. The eigenvalues of the reduced density matrix $\rho_A$, involves calculating expectation values of operators on the 4 pseudo-spins (blue stars) $i,i+1,j,j+1$, at the shown vertices.}
  \label{plaquettepseudo}
\end{figure}
To show how the correspondence of the $\sigma$ and $\tau$ operators, (\ref{opmap}), is used we give two examples (the pseudospin operators are the blue stars at the vertices in Fig.~(\ref{plaquettepseudo})):
\begin{align}
\braket{\hat{\sigma}^z_1}&=\braket{\hat{\tau}^x_i\hat{\tau}^x_{i+1}}\nonumber\\
\braket{\hat{\sigma}^z_2\hat{\sigma}^z_3}&=\braket{\hat{\tau}^x_i\hat{\tau}^x_{j}\hat{\tau}^x_j\hat{\tau}^x_{j+1}}=\braket{\hat{\tau}^x_i\hat{\tau}^x_{j+1}}=\braket{\hat{\tau}^x_i}\braket{\hat{\tau}^x_{j+1}}
\end{align}

\subsection{General treatment: systems with a bulk}
\subsubsection{The reduced density matrix}

A state within the gauge theory of TC model with $B_p=1\,\, \forall \, p$ can be written in two different ways:
\begin{align}
  \label{eq:psia}
  \ket{\psi} &= \sum_{g \in G} a(g) g\ket{\Uparrow} \, , \\
  \label{eq:psib}
  \ket{\psi} &= \sum_{z \in \fz} b(z) z \ket{0} \, , \\
  \label{eq:allup}
  \po{i}{z}\ket{\Uparrow} &= \ket{\Uparrow} \quad \forall\,\, i \, , \\
  \label{eq:ground}
  \ket{0}&=\left\vert{G}\right\vert^{-1/2}\sum_{g\in G} g
  \ket{\Uparrow} \, ,
\end{align}
where $\ket{\Uparrow}$ is the state with all spins pointing up in
the $z$-basis and $G$ is the group generated by the $N^2-1$ independent star
operators $A_s$ (or equivalently closed loops of $\po{}{x}$ operators
in the dual lattice).
The ground
state of the TC model is indicated by $\ket{0}$  and $\fz$ is the group generated by all the open
string operators (in the real lattice) of the form $\po{i}{z}\po{j}{z} \cdots \po{k}{z}$.

Combining Eqs.\eqref{eq:psib} and \eqref{eq:ground} we get:
\begin{align}
  \ket{\psi} &= \abs{G}^{-1/2} \sum_{z \in \fz} b(z) \sum_{g \in G} z
  g \up \, .
  \label{eq:psibare} 
\end{align}
Note that $g$ is a product of closed loops of $\sigma^x$ operators and $z$ is a string of $\sigma_z$ operators. If we try to commute these two operators we get a negative
sign for every spin that is common to both of these strings.
Let us introduce the following notation: given $g \in G$ and
$z \in \fz$ we denote by $g \cap z$ the number of spins that gets
acted upon non-trivially by each of these operators.
Thus we arrive at the operator identity:
\begin{align}
   g z = z g (-1)^{g\cap z} \, . 
  \label{eq:opid}
\end{align}
We can also think of $z$ as a set of excitations
of stars rather than a string of $\po{}{z}$ operators as far as $(-1)^{g \cap z}$  is concerned. This is because a star and a string share an odd number
 of spins only if the string has an open end at the position of the
 star. In this picture $g$ and $z$ live in
the same \emph{space}, i.e. the vertices of the real lattice. 
 We refer to $g \cap z$ as the \emph{overlap of g and
 z}, by which we mean the number of stars that are common to
$g=A_{s_1}A_{s_2}\dots$ and $z$ identified 
with the stars at the ends of open strings.

Using Eq.~\eqref{eq:opid} and the fact that $z\up =\up$
Eq.~\eqref{eq:psibare} can be written as:
\begin{align}
   \ket{\psi} = \abs{G}^{-1/2} \sum_{z \in \fz} b(z) \sum_{g \in G}
  (-1)^{g\cap z} g \up \, .
  \label{eq:psi}
\end{align}
The density matrix associated with this pure state is
given by:
\begin{align}
  \rho = \abs{G}^{-1} \sum_{z,z' \in \fz} \bb(z') b(z) \sum_{g,g' \in
  G} (-1)^{g \cap z + g' \cap z'} \ket{x_A x_B} \bra{x'_A x'_B}\, ,
  \label{}
\end{align}
where we have adopted the notation: 
$g\ket{\Uparrow}= \ket{x_A x_B}$.

The reduced density matrix of subsystem $A$ can be obtain by tracing
over the spins in $B$.\begin{align}
  \rho_A = 
  \abs{G}^{-1}\!\! \sum_{z,z' \in \fz} \bb(z') b(z) \! \sum_{g,g' \in
  G} (-1)^{g \cap z + g' \cap z'}  \inner{x'_B}{x_B}
  \ket{x_A}\bra{x'_A}\, .
  \label{}
\end{align}
Note that $\inner{x'_B}{x_B}$ imposes the condition $g'=g \gt$, where
$\gt \in G_A$ and $G_A$ is the subgroup of $G$ generated by star operators
acting non-trivially only on the spins in $A$.  
Thus we can write
\begin{align}
 \rho_A = \abs{G}^{-1} \sum_{z,z' \in \fz} \bb(z') b(z)
  \summ{g \in G}{\gt \in G_A} (-1)^{g \cap z + g \gt \cap z'}
  \ket{x_A}\bra{\gt x_A}\, .
  \label{eq:rhoA}
\end{align}
An expression for the purity of the subsystem $A$ follows directly:
\begin{align}
  P &=\abs{G}^{-2} \summ{z_1, z_2 \in
  \fz}{z'_1, z'_2 \in \fz}\summ{g_1, g_2 \in
  G}{\gt_1, \gt_2 \in G_A} \bb(
  z'_1) b(z_1) \bb( z'_2) b(z_2) \\
  \nonumber
  &\times (-1)^{g_1 \gt_1 \cap z'_1 + g_2 \gt_2
  \cap z'_2 + g_1 \cap z_1  + g_2 \cap z_2 } \inner{\gt_1 x_{1A}}{x_{2A}}\inner{\gt_2
  x_{2A}}{x_{1A}} \, .
  \label{eq:2}
\end{align}
Note that $\inner{\gt_1 x_{1A}}{x_{2A}}$ imposes the condition
$g_{2}=\gt_1 g_1 \gb_1$, where
$\gb \in G_B$ and 
 $G_B$ is the group generated by star operators acting trivially on the spins in subsystem $A$. 
Using this condition we can replace the sum over $g_2$ by a sum over $\gb_1$ and write the last inner product in Eq.\eqref{eq:2} as:
\begin{align}
 \inner{\gt_2 x_{2A}}{x_{1A}} = \inner{\gt_2 g_2}{x_{1A}} =
 \inner{\gt_2 \gt_1 g_1 \gb_1}{x_{1A}} = \inner{\gt_2 \gt_1
 x_{1A}}{x_{1A}}\, ,
\end{align}
where we replaced $x_{2A}$ by $g_{2}$, dropped the $\gb_1$ and
replaced $g_1$ by $x_{1A}$ since none of these changes effect the
spin configuration in subsystem $A$, thus the inner product with
$\ket{x_{1A}}$. This inner
product determines $\gt_2=\gt_1$ and kills the
summation over $\gt_2$. 
After some algebra (also noting that $\gt_1 \gt_1 =\mathbf{1}$) we obtain for the purity:
\begin{align}
  \nonumber
  P &= \abs{G}^{-2} \summ{z_1, z_{2} \in \fz}{z'_1, z'_{2}
  \in \fz}  \bb(
  z'_1) b(z_1) \bb( z'_2) b(z_2) \\
  &\times \summm{g_1 \in G}{\gt_1 \in G_A}{\gb_1 \in
  G_B}(-1)^{g_1 \gt_1 \cap z'_1 + g_1 \gb_1 \cap z'_2 + g_1 \cap z_1 +
   \gt_1 g_1 \gb_1 \cap z_2 }\, .
  \label{eq:temp}
\end{align}
First we focus on the last term. 
If a product of $g$ and $z$ operators are commuted, the result can be expressed
in two different ways. One can apply Eq.~\eqref{eq:opid} to the
products themselves, since any product of $g$'s and $z$'s is another member of the group $G$ or $\fz$ respectively
\begin{align}
  g_1\dots g_k z_1\dots z_l = z_1\dots z_l g_1 \dots g_k
  (-1)^{g_1 \dots g_k \cap z_1 \dots z_l} \, .
  \label{}
\end{align}
However, one can also choose to commute each $g_i$ and $z_j$ one at a
time, picking a sign $\left( -1 \right)^{g_i\cap z_i}$ for each pair.
This procedure results in:
\begin{align}
  g_1\dots g_k z_1\dots z_l = 
  (-1)^{\sum_{i=1}^k\sum_{j=1}^l g_i \cap z_j } \, .
  \label{}
\end{align}
Thus we can manipulate the terms involving powers of $(-1)$ by separating
them and
regrouping back together in different ways. 
We rewrite the last summations in Eq.~\eqref{eq:temp} as:
\begin{align}
  \sum_{\gt_1 \in G_A}&(-1)^{\gt_1 \cap z'_1 + \gt_1\cap z_2}
  \sum_{g_1 \in
  G}(-1)^{g_1 \cap z_1 z'_1 z_2 z'_2} \sum_{\gb_1 \in G_B}(-1)^{\gb_1
  \cap z_2 z'_2} \, .
  \label{eq:saydir}
\end{align}
First, we work on the term appearing in the first sum above. From Eq.\eqref{eq:psib} we have:
\begin{align}
  b(z_2) &= \bra{0} z_2 \ket{\psi}, \quad  \bb(z'_1) = \bra{\psi} z'_1
  \ket{0} \, .
  \label{eq:bees}
\end{align}
Using the above formulae, Eq.\eqref{eq:opid} and the fact that that
$\gt_1 \ket{0} = \ket{0}$
we have:
\begin{align}
  \label{eq:absorbb}
  (-1)^{\gt_1 \cap z_2 +\gt_1 \cap z'_1} \bb(z'_1) b(z_2) &= 
  b(z_2 \gt_1)  \bb(\gt_1 z'_1) \, .
\end{align} 
Next, we work on the last two sums in Eq.\eqref{eq:saydir}.
Let us consider the general expression $\sum_{g\in G_R} (-1)^{g \cap z} $
for an arbitrary subgroup $G_R$ of $G$. 
When phrased in terms of the \emph{overlap}, this summation becomes a problem of combinatorics:
Given $z$, $g\cap z $ does only depend whether $g$ has stars
on the vertices where $z$ has excitations. 
Lets assume that $z$ has $k\ne 0$ excitations in the domain of
$G_R$ (excitations outside $R$ don't effect 
the sum). The sum over $g$ involves
all the combinations of star operators on these $k$ vertices. There
are ${k \choose m}$ elements $g \in G_R$ that have overlap $m$,
because this is the number of ways you can distribute $m$ stars on $k$
vertices. The summation for $z$ with $k\ne 0$ excitations in the
domain of $G_R$ vanishes
since:
\begin{align}
  \sum_{g\in G_R} (-1)^{g \cap z } &=
 \sum_{m=0}^{k} {k \choose m} (-1)^{m} = (1-1)^k=0 \, .
  \label{eq:a}
\end{align}
If, on the other hand, $z$ has no excitations in the domain of
$G_R$ there is no overlap with $g$ and the summation is trivial:
 \begin{align}
  \sum_{g\in G_R} (-1)^{g \cap z} =
  \left\vert{G_R}\right\vert \, . 
   \label{eq:b}
 \end{align}
We can simplify the expression in Eq.~\eqref{eq:saydir} with the help of
Eqs.~(\ref{eq:a}, \ref{eq:b}). The second sum in Eq.~\eqref{eq:saydir}
places the following constraint: $\sum_{g_1 \in G}(-1)^{g_1 \cap z_1 z'_1 z_2 z'_2}
\Rightarrow z'_2 = z_1 z'_1 z_2 $. 
The third sum leads to another constraint $\sum_{\gb_1 \in
G_B}(-1)^{\gb_1 \cap z_2 z'_2} \Rightarrow z_2 z'_2  \in
\fz_A
$. The condition that $z_2z'_2$ not generate any excitations in
$G_B$ is equivalent to saying that only $\sigma_z$ operators in the
subsystem $A$ can be present. 
Using these results to evaluate some of the sums over $z$'s in
Eq.~\eqref{eq:temp} we get:
\begin{align}
  P = \frac{\abs{G} \abs{G_B}}{\left\vert{G}\right\vert^{2}}\!\!
  \sum_{\gt_1 \in G_A} \summ{z_1, z_{2} \in \fz}{\zt_1 \in \fz_A} \!\!\!\!
  \bb(\gt_1 \zt_1  z_1) b(z_1) \bb(\zt_{1} z_{2}) b(z_{2}\gt_{1})\, .
  \label{eq:12}
\end{align}
Finally, from Eq.\eqref{eq:bees} and \eqref{eq:psib} we have
\begin{align}
  \sum_{z \in \fz} & \bb(A z) b(z B) = \sum_{z \in
  \fz}\bra{\psi} A   z \ket{0}\bra{0} z B \ket{\psi} = \bra{\psi} A B
  \ket{\psi}\, ,
  \label{eq:11}
\end{align}
where in the last line we used the fact that $\sum_{z\in \fz}
z\ket{0}\bra{0}z = \id$ within the gauge sector we are working in. 

Using Eq.\eqref{eq:11} in Eq.\eqref{eq:12} we arrive at our final
expression for the subsystem purity:
\begin{align}
  P &=
  \frac{\abs{G_B}}{\left\vert{G}\right\vert} \summ{\gt_1 \in
  G_A}{\zt_1 \in \fz_A} \!\!\!\!
  \bra{\psi} \gt_1 \zt_1 \ket{\psi}
 \bra{\psi} \zt_{1} \gt_{1} \ket{\psi}  = \frac{\abs{G_B}}{\left\vert{G}\right\vert}\!\!\!\! \summ{\gt_1 \in G_A}{\zt_1 \in \fz_A} 
 \left\vert\bra{\psi} \gt_1 \zt_1 \ket{\psi}\right\vert^2 \, .
  \label{eq:result}
\end{align}

Using the technique developed here we also obtained the following, more general result:
\begin{align}
  \nonumber
  \tr_A[&\rho_A^n] =
  \frac{\abs{G_B}^{n-1}}{\left\vert{G}\right\vert^{n-1}}\!\!\!\!\!\!\!\!\! \summ{\gn_1, \dots 
  , \gn_{n-1} \in G_A}{\zn_1, \dots ,
  \zn_{n-1} \in \fz_A} 
  \!\!\!\!\!\!\!\! \bra{\psi} \gn_1 \zn_1\dots \zn_{n-1} \ket{\psi} \bra{\psi} \gn_2
  \zn_1 \gn_1
  \ket{\psi} \\
  &\times \cdots \bra{\psi} \gn_{n-1} \zn_{n-2} \gn_{n-2}
 \ket{\psi}
 \bra{\psi} \zn_{n-1} \gn_{n-1} \ket{\psi} \, .
  \label{eq:generalized}
\end{align}
Note that for a general state (not necessarily within the gauge invariant sector) the trace of the integer
powers of the reduced density matrix would require the measurement of all possible subsystem operators. 
Eq.~\eqref{eq:generalized} shows that for states within the gauge invariant sector the number of
necessary measurements is much smaller, which is a consequence of the gauge condition.

\subsubsection{Evaluation of the purity for a system with 2 adjoining stars}
\label{app:bulksys}

\begin{figure}
 \centering
\includegraphics[width=0.8\columnwidth]{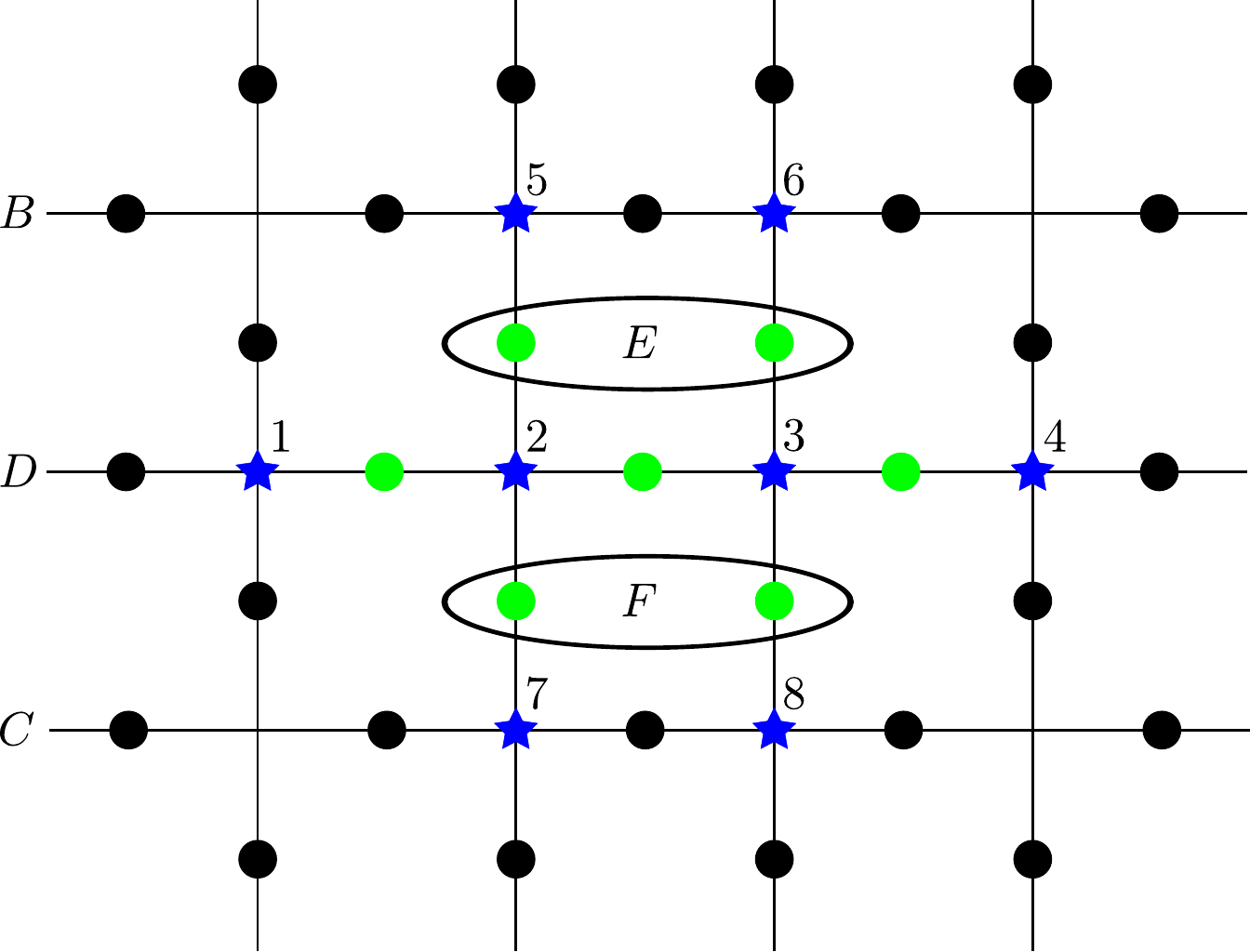}
 \caption{(color online) Subsystem $A$ with a bulk has spins comprised of those that form two adjoining stars (green ovals). The pseudospins (blue stars) that appear in the calculation are labelled by the numbers $1,2,3,4,5,6,7,8$.}
  \label{2starspseudo}
\end{figure}

\begin{widetext}
\begin{table}
{\centering\hfill
\begin{tabular}{ | l | l | l | l |}
  \hline                        
  subtype~$\downarrow$ &$E$&$F$&($E\times F\times A$)\\
\hline
 a & 0 & 0 &  $\openone\times \{\openone,~\sigma^z_{1-2},~\sigma^z_{2-3},~\sigma^z_{1-2}\sigma^z_{2-3},~\sigma^z_{3-4},~\sigma^z_{1-2}\sigma^z_{3-4},~\sigma^z_{2-3}\sigma^z_{3-4},~\sigma^z_{1-2}\sigma^z_{2-3}\sigma^z_{3-4}\}$\\
 & & & $\openone\times\overbrace{\{\openone,~\tau^x_{1}\tau^x_{2},~\tau^x_{2}\tau^x_{3},~\tau^x_{1}\tau^x_{3},~\tau^x_{3}\tau^x_{4},~\tau^x_{1}\tau^x_{2}\tau^x_{3}\tau^x_{4},~\tau^x_{2}\tau^x_{4},~\tau^x_{1}\tau^x_{4}\}}^{Q}$\\
\hline  
b  & 1 & 0 &  $\sigma^z_{5-2}\sigma^z_{6-3}\times \{ \openone,~\sigma^z_{1-2},~\sigma^z_{2-3},~\sigma^z_{1-2}\sigma^z_{2-3},~\sigma^z_{3-4},~\sigma^z_{1-2}\sigma^z_{3-4},~\sigma^z_{2-3}\sigma^z_{3-4},~\sigma^z_{1-2}\sigma^z_{2-3}\sigma^z_{3-4}\}$\\
& & & $\tau^x_{5}\tau^x_{6}\times\overbrace{ \{ \tau^x_{2}\tau^x_{3},~\tau^x_{1}\tau^x_{3},~\openone,~\tau^x_{1}\tau^x_{2},~\tau^x_{2}\tau^x_{4},~\tau^x_{1}\tau^x_{4},~\tau^x_{3}\tau^x_{4},~\tau^x_{1}\tau^x_{2}\tau^x_{3}\tau^x_{4}\}}^{Q}$\\
\hline
c & 0 & 1 &  $\sigma^z_{7-2}\sigma^z_{8-3}\times \{ \openone,~\sigma^z_{1-2},~\sigma^z_{2-3},~\sigma^z_{1-2}\sigma^z_{2-3},~\sigma^z_{3-4},~\sigma^z_{1-2}\sigma^z_{3-4},~\sigma^z_{2-3}\sigma^z_{3-4},~\sigma^z_{1-2}\sigma^z_{2-3}\sigma^z_{3-4}\}$\\
& & & $\tau^x_{7}\tau^x_{8}\times \overbrace{\{ \tau^x_{2}\tau^x_{3},~\tau^x_{1}\tau^x_{3},~\openone,~\tau^x_{1}\tau^x_{2},~\tau^x_{2}\tau^x_{4},~\tau^x_{1}\tau^x_{4},~\tau^x_{3}\tau^x_{4},~\tau^x_{1}\tau^x_{2}\tau^x_{3}\tau^x_{4}\}}^{Q}$\\
\hline
d & 1 & 1 &  $\sigma^z_{5-2}\sigma^z_{6-3}\sigma^z_{7-2}\sigma^z_{8-3}\times \{ \openone,~\sigma^z_{1-2},~\sigma^z_{2-3},~\sigma^z_{1-2}\sigma^z_{2-3},~\sigma^z_{3-4},~\sigma^z_{1-2}\sigma^z_{3-4},~\sigma^z_{2-3}\sigma^z_{3-4},~\sigma^z_{1-2}\sigma^z_{2-3}\sigma^z_{3-4}\}$\\
& & & $\tau^x_{5}\tau^x_{6}\tau^x_{7}\tau^x_{8}\times \overbrace{\{ \openone,~\tau^x_{1}\tau^x_{2},~\tau^x_{2}\tau^x_{3},~\tau^x_{1}\tau^x_{3},~\tau^x_{3}\tau^x_{4},~\tau^x_{1}\tau^x_{2}\tau^x_{3}\tau^x_{4},~\tau^x_{2}\tau^x_{4},~\tau^x_{1}\tau^x_{4}\}}^{Q}$\\
\hline
\end{tabular} \hfill}
\caption{All possible apriori non-zero operators of $type~\romannumeral 1$ arranged into 4 subtypes. For each subtype the first row gives the operators in the $\sigma$-picture and the corresponding operator in the $\tau$-picture appears in the second row. An entry of 1 against $E$ means that both $\hat{\sigma}^z_{5-2}$ and $\hat{\sigma}^z_{6-3}$ appear as factors in the operator product $z\in\mathcal{Z}_A$. }
\label{type1}
\end{table}
\end{widetext}

Working with the symmetric state considerably eases the analytical calculations as all operators that anticommute with the global spin flip (or parity in the fermionic picture), $\prod_i \hat{\tau}^z_i$, have a zero expectation value in the ground state. This implies that many operators in the product: $g z$ that have an odd number of $\hat{\tau}^x_i$ operators in any row, have zero expectation. The expression for purity (\ref{eq:result}) involves expectation values of operators in the $\sigma$-picture.  Our strategy is to calculate the product of operators appearing in Eq.~(\ref{eq:result}) by separating the different contributions based on the number of star operators in the product. Schematically we represent this as:
\begin{align}
\text{All operators of the form~}gz=\overbrace{\text{operator products with no stars}}^{type~\romannumeral 1}\nonumber\\
+\overbrace{\text{operator products with only 1 star}}^{type~\romannumeral 2}\nonumber\\+\overbrace{\text{operator products with both stars}}^{type~\romannumeral 3}
\end{align}

Now we collect all terms of $type~ \romannumeral 1$ as follows. From Fig.~(\ref{2starspseudo}) we find that only those operators which have either both or none of the $\hat{\sigma}^z$ on edges between vertices labelled $(5-2),(6-3)$ in the product contribute. Similarly only those operators that have the product of both or none of $\hat{\sigma}^z$ on edges between vertices labelled $(2-7),(3-8)$ contribute. However all possible products of $\hat{\sigma}^z_i$ on the row of spins labelled $D$ in the same figure are apriori non zero. This means that out of a total of $2^7$  operators of the $type~ \romannumeral 1$ - we need to consider only those that have products of both the $\hat{\sigma}^z_i$'s in the oval marked $E$ or both the  $\hat{\sigma}^z_i$'s in the oval marked $F$ as factors as shown in Fig.~(\ref{2starspseudo}). However all possible products of $\sigma^z_i$'s along the row marked $A$ in the same figure are apriori non-zero. This means that we need to consider a total number of $2^2*2^3$ operators of  $type~ \romannumeral 1$ where the factor $2^2$ comes from the fact that $E,F$ can be turned on(both $\sigma^z_i$'s present in the product) or off(none present) in 4 different ways (subtypes) for each of the $2^3$ operator products of  $\sigma^z_i$'s along row $D$. We can then write down a table corresponding to the possible operators we need to calculate expectation values for, in the $\tau$-picture by using the map (\ref{opmap}). For eg. $\hat{\sigma}^z_{1-2}$ which is an operator on the spin on the edge connecting vertices $1,2$ is mapped to the product $\hat{\tau}^x_{1}\hat{\tau}^x_{2}$. The table(\ref{type1}) tabulates the operators in both the $\sigma$ and $\tau$ pictures. Note that operators of each of the 4 subtypes are products of elements from the group of operators labelled $Q$ which are products of $\hat{\tau}^x$'s only along row $A$ and depending on whether $E$ or $F$ is turned on - product of $\hat{\tau}^x_5\hat{\tau}^x_6$  or/and $\hat{\tau}^x_7\hat{\tau}^x_8$ on rows $B$ and $C$. Because operators of each subtype factorize into operators from the group $Q$, which belong to one particular row, and other operators on adjacent rows, we need to evaluate only 8 correlation functions to determine all expectation values of operators of $type~\romannumeral 1$.

One can similarly tabulate all operators of $type~\romannumeral 2$ and $type~\romannumeral 3$ in the $\sigma$ and $\tau$ representations, which we omit here for the sake of brevity, and evaluate the sum of expectation values in Eq.~(\ref{eq:result}) leading to Eq.~(\ref{taupurity}).

\section{Behavior of R\'{e}nyi entropies for the Cluster state}
\label{app:clusterstate}
\begin{figure}
 \centering
\includegraphics[width=\columnwidth]{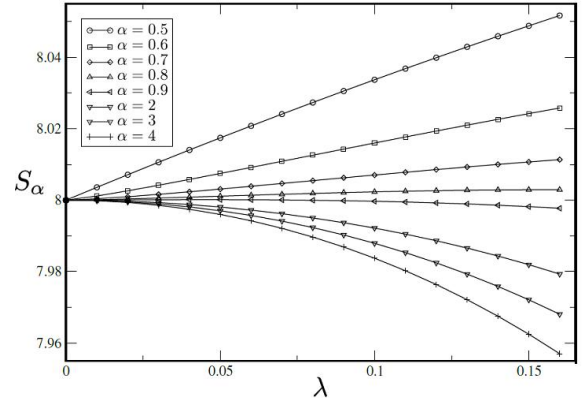}
 \caption{Behavior of representative R\'enyi entropies for a cluster state, that is the ground state of the Hamiltonian $H=-\sum_i\sigma^z_{i-\hat{x}}\sigma^z_{i+\hat{x}}\sigma^z_{i+\hat{y}}\sigma^z_{i-\hat{y}}\sigma^x_{i}-\lambda\sum_{i,\mu=\hat{x},\hat{y}}(\sigma^x_i\sigma^x_{i+\mu}+.5\sigma^z_{i}\sigma^z_{i+\mu})$, w.r.t. the perturbation parameter $\lambda$. The change between monotonically increasing and decreasing behaviors occurs at $\alpha\simeq.8$, the value of which was identified numerically. The  region considered as subsytem $A$ is a $3\times3$ block of spins in a 2-D lattice with infinite extent along the $x$-direction and 5 spins along $y$ with periodic boundary conditions.}
  \label{fig:cluster}
\end{figure}

Cluster states do not possess topological order even though they permit no local order parameter \cite{kalis:2012,son:2011,raussendorf}. A numerical analysis of their local convertibility properties as shown in Fig.~(\ref{fig:cluster}) reveals that the R\'{e}nyi entropies split in their behavior w.r.t. the perturbation and hence cluster states are not locally convertible.

\newpage


\begin{thebibliography}{64}%
\makeatletter
\providecommand \@ifxundefined [1]{%
 \@ifx{#1\undefined}
}%
\providecommand \@ifnum [1]{%
 \ifnum #1\expandafter \@firstoftwo
 \else \expandafter \@secondoftwo
 \fi
}%
\providecommand \@ifx [1]{%
 \ifx #1\expandafter \@firstoftwo
 \else \expandafter \@secondoftwo
 \fi
}%
\providecommand \natexlab [1]{#1}%
\providecommand \enquote  [1]{``#1''}%
\providecommand \bibnamefont  [1]{#1}%
\providecommand \bibfnamefont [1]{#1}%
\providecommand \citenamefont [1]{#1}%
\providecommand \href@noop [0]{\@secondoftwo}%
\providecommand \href [0]{\begingroup \@sanitize@url \@href}%
\providecommand \@href[1]{\@@startlink{#1}\@@href}%
\providecommand \@@href[1]{\endgroup#1\@@endlink}%
\providecommand \@sanitize@url [0]{\catcode `\\12\catcode `\$12\catcode
  `\&12\catcode `\#12\catcode `\^12\catcode `\_12\catcode `\%12\relax}%
\providecommand \@@startlink[1]{}%
\providecommand \@@endlink[0]{}%
\providecommand \url  [0]{\begingroup\@sanitize@url \@url }%
\providecommand \@url [1]{\endgroup\@href {#1}{\urlprefix }}%
\providecommand \urlprefix  [0]{URL }%
\providecommand \Eprint [0]{\href }%
\providecommand \doibase [0]{http://dx.doi.org/}%
\providecommand \selectlanguage [0]{\@gobble}%
\providecommand \bibinfo  [0]{\@secondoftwo}%
\providecommand \bibfield  [0]{\@secondoftwo}%
\providecommand \translation [1]{[#1]}%
\providecommand \BibitemOpen [0]{}%
\providecommand \bibitemStop [0]{}%
\providecommand \bibitemNoStop [0]{.\EOS\space}%
\providecommand \EOS [0]{\spacefactor3000\relax}%
\providecommand \BibitemShut  [1]{\csname bibitem#1\endcsname}%
\let\auto@bib@innerbib\@empty
\bibitem [{\citenamefont {Goldenfeld}()}]{goldenfeld}%
  \BibitemOpen
  \bibfield  {author} {\bibinfo {author} {\bibfnamefont {N.}~\bibnamefont
  {Goldenfeld}},\ }\href {http://lccn.loc.gov/92017055} {\emph {\bibinfo
  {title} {Lectures on phase transitions and the renormalization group}}}\
  (\bibinfo  {publisher} {Addison Wesley, New York, 1992})\BibitemShut
  {NoStop}%
\bibitem [{\citenamefont {Yan}\ \emph {et~al.}()\citenamefont {Yan},
  \citenamefont {Huse},\ and\ \citenamefont {White}}]{yan:2011}%
  \BibitemOpen
  \bibfield  {author} {\bibinfo {author} {\bibfnamefont {S.}~\bibnamefont
  {Yan}}, \bibinfo {author} {\bibfnamefont {D.}~\bibnamefont {Huse}}, \ and\
  \bibinfo {author} {\bibfnamefont {S.}~\bibnamefont {White}},\ }\href
  {http://www.sciencemag.org/content/332/6034/1173} {\bibinfo  {journal}
  {Science, 332, 1173, (2011)}\ }\BibitemShut {NoStop}%
\bibitem [{\citenamefont {Hasan}\ and\ \citenamefont {Kane}()}]{hasan:2010}%
  \BibitemOpen
\bibfield  {journal} {  }\bibfield  {author} {\bibinfo {author} {\bibfnamefont
  {M.}~\bibnamefont {Hasan}}\ and\ \bibinfo {author} {\bibfnamefont
  {C.}~\bibnamefont {Kane}},\ }\href
  {http://rmp.aps.org/abstract/RMP/v82/i4/p3045_1} {\bibinfo  {journal} {Rev.
  Mod. Phys. 82, 3045-3067 (2010)}\ }\BibitemShut {NoStop}%
\bibitem [{\citenamefont {Wen}()}]{wen}%
  \BibitemOpen
\bibfield  {journal} {  }\bibfield  {author} {\bibinfo {author} {\bibfnamefont
  {X.~G.}\ \bibnamefont {Wen}},\ }\href
  {http://ukcatalogue.oup.com/product/9780198530947.do#.UiVHyTakqhQ} {\emph
  {\bibinfo {title} {Quantum field theory of many body systems}}}\ (\bibinfo
  {publisher} {Oxford university press, 2004})\BibitemShut {NoStop}%
\bibitem [{\citenamefont {Flammia}\ \emph {et~al.}()\citenamefont {Flammia},
  \citenamefont {Hamma}, \citenamefont {Hughes},\ and\ \citenamefont
  {Wen}}]{flammia:2009}%
  \BibitemOpen
  \bibfield  {author} {\bibinfo {author} {\bibfnamefont {S.}~\bibnamefont
  {Flammia}}, \bibinfo {author} {\bibfnamefont {A.}~\bibnamefont {Hamma}},
  \bibinfo {author} {\bibfnamefont {T.}~\bibnamefont {Hughes}}, \ and\ \bibinfo
  {author} {\bibfnamefont {X.-G.}\ \bibnamefont {Wen}},\ }\href
  {http://prl.aps.org/abstract/PRL/v103/i26/e261601} {\bibinfo  {journal}
  {Phys. Rev. Lett. 103, 261601 (2009)}\ }\BibitemShut {NoStop}%
\bibitem [{\citenamefont {Briegel}\ and\ \citenamefont
  {Raussendorf}(2001)}]{briegel:2001}%
  \BibitemOpen
\bibfield  {journal} {  }\bibfield  {author} {\bibinfo {author} {\bibfnamefont
  {H.}~\bibnamefont {Briegel}}\ and\ \bibinfo {author} {\bibfnamefont
  {R.}~\bibnamefont {Raussendorf}},\ }\href
  {http://prl.aps.org/abstract/PRL/v86/i5/p910_1} {\bibfield  {journal}
  {\bibinfo  {journal} {Phys. Rev. Lett. 86, 910-913}\ } (\bibinfo {year}
  {2001})}\BibitemShut {NoStop}%
\bibitem [{\citenamefont {Freedman}\ \emph {et~al.}(2002)\citenamefont
  {Freedman}, \citenamefont {Kitaev},\ and\ \citenamefont
  {Wang}}]{freedman:2002}%
  \BibitemOpen
  \bibfield  {author} {\bibinfo {author} {\bibfnamefont {M.~H.}\ \bibnamefont
  {Freedman}}, \bibinfo {author} {\bibfnamefont {A.}~\bibnamefont {Kitaev}}, \
  and\ \bibinfo {author} {\bibfnamefont {Z.}~\bibnamefont {Wang}},\ }\href
  {http://link.springer.com/article/10.1007\%2Fs002200200635} {\bibfield
  {journal} {\bibinfo  {journal} {Commun. Math. Phys. 227, 587 – 603}\ }
  (\bibinfo {year} {2002})}\BibitemShut {NoStop}%
\bibitem [{\citenamefont {Nayak}\ \emph {et~al.}(2008)\citenamefont {Nayak},
  \citenamefont {Simon}, \citenamefont {Stern}, \citenamefont {Freedman},\ and\
  \citenamefont {Sarma}}]{nayak:2008}%
  \BibitemOpen
  \bibfield  {author} {\bibinfo {author} {\bibfnamefont {C.}~\bibnamefont
  {Nayak}}, \bibinfo {author} {\bibfnamefont {S.~H.}\ \bibnamefont {Simon}},
  \bibinfo {author} {\bibfnamefont {A.}~\bibnamefont {Stern}}, \bibinfo
  {author} {\bibfnamefont {M.}~\bibnamefont {Freedman}}, \ and\ \bibinfo
  {author} {\bibfnamefont {S.~D.}\ \bibnamefont {Sarma}},\ }\href
  {http://rmp.aps.org/abstract/RMP/v80/i3/p1083_1} {\bibfield  {journal}
  {\bibinfo  {journal} {Rev. Mod. Phys. 80, 1083–1159}\ } (\bibinfo {year}
  {2008})}\BibitemShut {NoStop}%
\bibitem [{\citenamefont {Stormer}()}]{stormer:1999}%
  \BibitemOpen
  \bibfield  {author} {\bibinfo {author} {\bibfnamefont {H.}~\bibnamefont
  {Stormer}},\ }\href {http://rmp.aps.org/abstract/RMP/v71/i4/p875_1} {\bibinfo
   {journal} {Rev. Mod. Phys. 71, 875-889 (1999)}\ }\BibitemShut {NoStop}%
\bibitem [{\citenamefont {Hamma}\ \emph
  {et~al.}(2005{\natexlab{a}})\citenamefont {Hamma}, \citenamefont
  {Ionicioiu},\ and\ \citenamefont {Zanardi}}]{hamma:2005a}%
  \BibitemOpen
\bibfield  {journal} {  }\bibfield  {author} {\bibinfo {author} {\bibfnamefont
  {A.}~\bibnamefont {Hamma}}, \bibinfo {author} {\bibfnamefont
  {R.}~\bibnamefont {Ionicioiu}}, \ and\ \bibinfo {author} {\bibfnamefont
  {P.}~\bibnamefont {Zanardi}},\ }\href
  {http://www.sciencedirect.com/science/article/pii/S0375960105001544}
  {\bibfield  {journal} {\bibinfo  {journal} {Phys. Lett. A 337, 22-28}\ }
  (\bibinfo {year} {2005}{\natexlab{a}})}\BibitemShut {NoStop}%
\bibitem [{\citenamefont {Hamma}\ \emph
  {et~al.}(2005{\natexlab{b}})\citenamefont {Hamma}, \citenamefont
  {Ionicioiu},\ and\ \citenamefont {Zanardi}}]{hamma:2005b}%
  \BibitemOpen
  \bibfield  {author} {\bibinfo {author} {\bibfnamefont {A.}~\bibnamefont
  {Hamma}}, \bibinfo {author} {\bibfnamefont {R.}~\bibnamefont {Ionicioiu}}, \
  and\ \bibinfo {author} {\bibfnamefont {P.}~\bibnamefont {Zanardi}},\ }\href
  {http://pra.aps.org/abstract/PRA/v71/i2/e022315} {\bibfield  {journal}
  {\bibinfo  {journal} {Phys. Rev. A 71, 022315}\ } (\bibinfo {year}
  {2005}{\natexlab{b}})}\BibitemShut {NoStop}%
\bibitem [{\citenamefont {Kitaev}\ and\ \citenamefont
  {Preskill}()}]{kitaev:2006}%
  \BibitemOpen
  \bibfield  {author} {\bibinfo {author} {\bibfnamefont {A.~Y.}\ \bibnamefont
  {Kitaev}}\ and\ \bibinfo {author} {\bibfnamefont {J.}~\bibnamefont
  {Preskill}},\ }\href {http://prl.aps.org/abstract/PRL/v96/i11/e110404}
  {\bibinfo  {journal} {Phys. Rev. Lett. 96, 110404, 2006}\ }\BibitemShut
  {NoStop}%
\bibitem [{\citenamefont {Levin}\ and\ \citenamefont {Wen}()}]{levin:2006}%
  \BibitemOpen
\bibfield  {journal} {  }\bibfield  {author} {\bibinfo {author} {\bibfnamefont
  {M.}~\bibnamefont {Levin}}\ and\ \bibinfo {author} {\bibfnamefont {X.-G.}\
  \bibnamefont {Wen}},\ }\href
  {http://prl.aps.org/abstract/PRL/v96/i11/e110405} {\bibinfo  {journal} {Phys.
  Rev. Lett. 96, 110405 (2006)}\ }\BibitemShut {NoStop}%
\bibitem [{\citenamefont {Gu}\ and\ \citenamefont {Wen}()}]{gu:2009}%
  \BibitemOpen
\bibfield  {journal} {  }\bibfield  {author} {\bibinfo {author} {\bibfnamefont
  {Z.-G.}\ \bibnamefont {Gu}}\ and\ \bibinfo {author} {\bibfnamefont
  {X.}~\bibnamefont {Wen}},\ }\href
  {http://prb.aps.org/abstract/PRB/v80/i15/e155131} {\bibinfo  {journal} {Phys.
  Rev. B 80, 155131 (2009)}\ }\BibitemShut {NoStop}%
\bibitem [{\citenamefont {Wen}(1995)}]{wen:1995}%
  \BibitemOpen
\bibfield  {journal} {  }\bibfield  {author} {\bibinfo {author} {\bibfnamefont
  {X.-G.}\ \bibnamefont {Wen}},\ }\href
  {http://www.tandfonline.com/doi/abs/10.1080/00018739500101566#.UmBmMVCkp4k}
  {\bibfield  {journal} {\bibinfo  {journal} {Advances in Physics, Vol. 44, 5}\
  } (\bibinfo {year} {1995})}\BibitemShut {NoStop}%
\bibitem [{\citenamefont {Kitaev}()}]{kitaev:2003}%
  \BibitemOpen
  \bibfield  {author} {\bibinfo {author} {\bibfnamefont {A.~Y.}\ \bibnamefont
  {Kitaev}},\ }\href
  {http://www.sciencedirect.com/science/article/pii/S0003491602000180}
  {\bibinfo  {journal} {Ann. Phys. (N.Y.) 303, 1, 2-30 (2003)}\ }\BibitemShut
  {NoStop}%
\bibitem [{\citenamefont {Wen}\ and\ \citenamefont {Niu}(1990)}]{wen:1990}%
  \BibitemOpen
\bibfield  {journal} {  }\bibfield  {author} {\bibinfo {author} {\bibfnamefont
  {X.-G.}\ \bibnamefont {Wen}}\ and\ \bibinfo {author} {\bibfnamefont
  {Q.}~\bibnamefont {Niu}},\ }\href
  {http://prb.aps.org/abstract/PRB/v41/i13/p9377_1} {\bibfield  {journal}
  {\bibinfo  {journal} {Phys. Rev. B 41, 9377–9396}\ } (\bibinfo {year}
  {1990})}\BibitemShut {NoStop}%
\bibitem [{\citenamefont {Chen}\ \emph {et~al.}()\citenamefont {Chen},
  \citenamefont {Gu},\ and\ \citenamefont {Wen}}]{chen:2010}%
  \BibitemOpen
  \bibfield  {author} {\bibinfo {author} {\bibfnamefont {X.}~\bibnamefont
  {Chen}}, \bibinfo {author} {\bibfnamefont {Z.}~\bibnamefont {Gu}}, \ and\
  \bibinfo {author} {\bibfnamefont {X.}~\bibnamefont {Wen}},\ }\href
  {http://prb.aps.org/abstract/PRB/v82/i15/e155138} {\bibinfo  {journal} {Phys.
  Rev. B 82, 155138 (2010)}\ }\BibitemShut {NoStop}%
\bibitem [{\citenamefont {Fradkin}()}]{fradkin}%
  \BibitemOpen
\bibfield  {journal} {  }\bibfield  {author} {\bibinfo {author} {\bibfnamefont
  {E.}~\bibnamefont {Fradkin}},\ }\href
  {http://www.cambridge.org/us/academic/subjects/physics/condensed-matter-physics-nanoscience-and-mesoscopic-physics/field-theories-condensed-matter-physics-2nd-edition}
  {\emph {\bibinfo {title} {Field theories of condensed matter physics}}}\
  (\bibinfo  {publisher} {Cambridge, 2013})\BibitemShut {NoStop}%


\bibitem [{\citenamefont {Haldane}()}]{haldane:1988}%
  \BibitemOpen
  \bibfield  {author} {\bibinfo {author} {\bibfnamefont {F.}~\bibnamefont
  {Haldane}},\ }\href {http://prl.aps.org/abstract/PRL/v61/i18/p2015_1}
  {\bibinfo  {journal} {Phys. Rev. Lett. 61, 2015-2018 (1988)}\ }\BibitemShut
  {NoStop}%

\bibitem [{\citenamefont {Hastings}()}]{hastings:2011}%
  \BibitemOpen
\bibfield  {journal} {  }\bibfield  {author} {\bibinfo {author} {\bibfnamefont
  {M.}~\bibnamefont {Hastings}},\ }\href
  {http://prl.aps.org/abstract/PRL/v107/i21/e210501} {\bibinfo  {journal}
  {Phys. Rev. Lett. 107, 210501 (2011)}\ }\BibitemShut {NoStop}%

\bibitem [{\citenamefont {Castelnovo}\ and\ \citenamefont
  {Chamon}()}]{castelnovo:2006}%
  \BibitemOpen
\bibfield  {journal} {  }\bibfield  {author} {\bibinfo {author} {\bibfnamefont
  {C.}~\bibnamefont {Castelnovo}}\ and\ \bibinfo {author} {\bibfnamefont
  {C.}~\bibnamefont {Chamon}},\ }\href
  {http://prb.aps.org/abstract/PRB/v76/i17/e174416} {\bibinfo  {journal} {Phys.
  Rev. B 76, 174416 (2007)}\ }\BibitemShut {NoStop}%


\bibitem [{\citenamefont {Hamma}\ \emph {et~al.}(2013)\citenamefont {Hamma},
  \citenamefont {Cincio}, \citenamefont {Santra}, \citenamefont {Zanardi},\
  and\ \citenamefont {Amico}}]{hamma:2013}%
  \BibitemOpen
  \bibfield  {author} {\bibinfo {author} {\bibfnamefont {A.}~\bibnamefont
  {Hamma}}, \bibinfo {author} {\bibfnamefont {L.}~\bibnamefont {Cincio}},
  \bibinfo {author} {\bibfnamefont {S.}~\bibnamefont {Santra}}, \bibinfo
  {author} {\bibfnamefont {P.}~\bibnamefont {Zanardi}}, \ and\ \bibinfo
  {author} {\bibfnamefont {L.}~\bibnamefont {Amico}},\ }\href
  {http://prl.aps.org/abstract/PRL/v110/i21/e210602} {\bibfield  {journal}
  {\bibinfo  {journal} {Phys. Rev. Lett. 110, 210602}\ } (\bibinfo {year}
  {2013})}\BibitemShut {NoStop}%
\bibitem [{\citenamefont {Cui}\ \emph {et~al.}()\citenamefont {Cui},
  \citenamefont {Amico}, \citenamefont {Fan}, \citenamefont {Gu}, \citenamefont
  {Hamma},\ and\ \citenamefont {Vedral}}]{cui:apr2013}%
  \BibitemOpen
  \bibfield  {author} {\bibinfo {author} {\bibfnamefont {J.}~\bibnamefont
  {Cui}}, \bibinfo {author} {\bibfnamefont {L.}~\bibnamefont {Amico}}, \bibinfo
  {author} {\bibfnamefont {H.}~\bibnamefont {Fan}}, \bibinfo {author}
  {\bibfnamefont {M.}~\bibnamefont {Gu}}, \bibinfo {author} {\bibfnamefont
  {A.}~\bibnamefont {Hamma}}, \ and\ \bibinfo {author} {\bibfnamefont
  {V.}~\bibnamefont {Vedral}},\ }\href
  {http://prb.aps.org/abstract/PRB/v88/i12/e125117} {\bibinfo  {journal} {Phys.
  Rev. B 88, 125117 (2013)}\ }\BibitemShut {NoStop}%
\bibitem [{\citenamefont {Nielsen}\ and\ \citenamefont
  {Chuang}()}]{nielchuang}%
  \BibitemOpen
\bibfield  {journal} {  }\bibfield  {author} {\bibinfo {author} {\bibfnamefont
  {M.}~\bibnamefont {Nielsen}}\ and\ \bibinfo {author} {\bibfnamefont
  {I.}~\bibnamefont {Chuang}},\ }\href
  {http://www.cambridge.org/us/academic/subjects/physics/quantum-physics-quantum-information-and-quantum-computation/quantum-computation-and-quantum-information}
  {\emph {\bibinfo {title} {Quantum Computation and Quantum Information}}}\
  (\bibinfo  {publisher} {Cambridge, 2000})\BibitemShut {NoStop}%
\bibitem [{\citenamefont {Plenio}\ and\ \citenamefont
  {Jonathan}()}]{jonathanD:1999}%
  \BibitemOpen
  \bibfield  {author} {\bibinfo {author} {\bibfnamefont {M.}~\bibnamefont
  {Plenio}}\ and\ \bibinfo {author} {\bibfnamefont {D.}~\bibnamefont
  {Jonathan}},\ }\href {http://prl.aps.org/abstract/PRL/v83/i17/p3566_1}
  {\bibinfo  {journal} {Phys. Rev. Lett. 83, 3566 (1999)}\ }\BibitemShut
  {NoStop}%
\bibitem [{\citenamefont {Sanders}\ and\ \citenamefont
  {Gour}(2009)}]{sanders:2009}%
  \BibitemOpen
\bibfield  {journal} {  }\bibfield  {author} {\bibinfo {author} {\bibfnamefont
  {Y.~R.}\ \bibnamefont {Sanders}}\ and\ \bibinfo {author} {\bibfnamefont
  {G.}~\bibnamefont {Gour}},\ }\href
  {http://pra.aps.org/abstract/PRA/v79/i5/e054302} {\bibfield  {journal}
  {\bibinfo  {journal} {Phys. Rev. A 79, 054302}\ } (\bibinfo {year}
  {2009})}\BibitemShut {NoStop}%
\bibitem [{\citenamefont {Cui}\ \emph {et~al.}(2011)\citenamefont {Cui},
  \citenamefont {Gu}, \citenamefont {Kwek}, \citenamefont {Santos},
  \citenamefont {Fan},\ and\ \citenamefont {Vedral}}]{cui:2011}%
  \BibitemOpen
  \bibfield  {author} {\bibinfo {author} {\bibfnamefont {J.}~\bibnamefont
  {Cui}}, \bibinfo {author} {\bibfnamefont {M.}~\bibnamefont {Gu}}, \bibinfo
  {author} {\bibfnamefont {L.~C.}\ \bibnamefont {Kwek}}, \bibinfo {author}
  {\bibfnamefont {M.~F.}\ \bibnamefont {Santos}}, \bibinfo {author}
  {\bibfnamefont {H.}~\bibnamefont {Fan}}, \ and\ \bibinfo {author}
  {\bibfnamefont {V.}~\bibnamefont {Vedral}},\ }\href
  {http://arxiv.org/abs/1110.3331} {\bibfield  {journal} {\bibinfo  {journal}
  {arXiv:1110.3331v3}\ } (\bibinfo {year} {2011})}\BibitemShut {NoStop}%
\bibitem [{\citenamefont {Turgut}(2007)}]{turgut:2007}%
  \BibitemOpen
  \bibfield  {author} {\bibinfo {author} {\bibfnamefont {S.}~\bibnamefont
  {Turgut}},\ }\href {http://iopscience.iop.org/1751-8121/40/40/012} {\bibfield
   {journal} {\bibinfo  {journal} {Jour. of Phys. A., 40}\ } (\bibinfo {year}
  {2007})}\BibitemShut {NoStop}%
\bibitem [{\citenamefont {Klimesh}()}]{klimesh:2007}%
  \BibitemOpen
  \bibfield  {author} {\bibinfo {author} {\bibfnamefont {M.}~\bibnamefont
  {Klimesh}},\ }\href {http://arxiv.org/abs/0709.3680} {\bibinfo  {journal}
  {arXiv:0709.3680v1}\ }\BibitemShut {NoStop}%
\bibitem [{\citenamefont {Abanin}\ and\ \citenamefont
  {Demler}(2012)}]{abanin:2012}%
  \BibitemOpen
\bibfield  {journal} {  }\bibfield  {author} {\bibinfo {author} {\bibfnamefont
  {D.~A.}\ \bibnamefont {Abanin}}\ and\ \bibinfo {author} {\bibfnamefont
  {E.}~\bibnamefont {Demler}},\ }\href
  {http://prl.aps.org/abstract/PRL/v109/i2/e020504} {\bibfield  {journal}
  {\bibinfo  {journal} {Phys. Rev. Lett. 109, 020504}\ } (\bibinfo {year}
  {2012})}\BibitemShut {NoStop}%
\bibitem [{\citenamefont {Marshall}\ \emph {et~al.}()\citenamefont {Marshall},
  \citenamefont {Olkin},\ and\ \citenamefont {Arnold}}]{marshall}%
  \BibitemOpen
  \bibfield  {author} {\bibinfo {author} {\bibfnamefont {A.~W.}\ \bibnamefont
  {Marshall}}, \bibinfo {author} {\bibfnamefont {I.}~\bibnamefont {Olkin}}, \
  and\ \bibinfo {author} {\bibfnamefont {B.~C.}\ \bibnamefont {Arnold}},\
  }\href {http://link.springer.com/book/10.1007/978-0-387-68276-1/page/1}
  {\emph {\bibinfo {title} {Inequalities: Theory of majorization and its
  applications}}}\ (\bibinfo  {publisher} {Springer series in statistics,
  2011})\BibitemShut {NoStop}%
\bibitem [{\citenamefont {Aubrun}\ and\ \citenamefont
  {Nechita}(2008)}]{aubrun:2008}%
  \BibitemOpen
  \bibfield  {author} {\bibinfo {author} {\bibfnamefont {G.}~\bibnamefont
  {Aubrun}}\ and\ \bibinfo {author} {\bibfnamefont {I.}~\bibnamefont
  {Nechita}},\ }\href
  {http://link.springer.com/article/10.1007\%2Fs00220-007-0382-4} {\bibfield
  {journal} {\bibinfo  {journal} {Comm. Math. Phys. 278, 1, pp 133}\ }
  (\bibinfo {year} {2008})}\BibitemShut {NoStop}%
\bibitem [{\citenamefont {Bandyopadhyay}\ \emph {et~al.}(2002)\citenamefont
  {Bandyopadhyay}, \citenamefont {Roychowdhury},\ and\ \citenamefont
  {Sen}}]{bandyo:2002}%
  \BibitemOpen
  \bibfield  {author} {\bibinfo {author} {\bibfnamefont {S.}~\bibnamefont
  {Bandyopadhyay}}, \bibinfo {author} {\bibfnamefont {V.}~\bibnamefont
  {Roychowdhury}}, \ and\ \bibinfo {author} {\bibfnamefont {U.}~\bibnamefont
  {Sen}},\ }\href {http://pra.aps.org/abstract/PRA/v65/i5/e052315} {\bibfield
  {journal} {\bibinfo  {journal} {Phys. Rev. A 65, 052315}\ } (\bibinfo {year}
  {2002})}\BibitemShut {NoStop}%
\bibitem [{\citenamefont {Duan}\ \emph {et~al.}(2005)\citenamefont {Duan},
  \citenamefont {Feng}, \citenamefont {Li},\ and\ \citenamefont
  {Ying}}]{duan:2005}%
  \BibitemOpen
  \bibfield  {author} {\bibinfo {author} {\bibfnamefont {R.}~\bibnamefont
  {Duan}}, \bibinfo {author} {\bibfnamefont {Y.}~\bibnamefont {Feng}}, \bibinfo
  {author} {\bibfnamefont {X.}~\bibnamefont {Li}}, \ and\ \bibinfo {author}
  {\bibfnamefont {M.}~\bibnamefont {Ying}},\ }\href {\doibase
  10.1103/PhysRevA.71.042319} {\bibfield  {journal} {\bibinfo  {journal} {Phys.
  Rev. A}\ }\textbf {\bibinfo {volume} {71}},\ \bibinfo {pages} {042319}
  (\bibinfo {year} {2005})}\BibitemShut {NoStop}%
\bibitem [{\citenamefont {Feng}\ \emph {et~al.}(2006)\citenamefont {Feng},
  \citenamefont {Duan},\ and\ \citenamefont {Ying}}]{yuan:2006}%
  \BibitemOpen
  \bibfield  {author} {\bibinfo {author} {\bibfnamefont {Y.}~\bibnamefont
  {Feng}}, \bibinfo {author} {\bibfnamefont {R.}~\bibnamefont {Duan}}, \ and\
  \bibinfo {author} {\bibfnamefont {M.}~\bibnamefont {Ying}},\ }\href {\doibase
  10.1103/PhysRevA.74.042312} {\bibfield  {journal} {\bibinfo  {journal} {Phys.
  Rev. A}\ }\textbf {\bibinfo {volume} {74}},\ \bibinfo {pages} {042312}
  (\bibinfo {year} {2006})}\BibitemShut {NoStop}%
\bibitem [{\citenamefont {Bhatia}()}]{bhatia}%
  \BibitemOpen
  \bibfield  {author} {\bibinfo {author} {\bibfnamefont {R.}~\bibnamefont
  {Bhatia}},\ }\href
  {http://www.springer.com/mathematics/analysis/book/978-0-387-94846-1} {\emph
  {\bibinfo {title} {Matrix Analysis}}}\ (\bibinfo  {publisher} {Springer,
  Graduate text in mathematica, 169})\BibitemShut {NoStop}%


\bibitem [{\citenamefont {Franchini}\ \emph {et~al.}()\citenamefont
  {Franchini}, \citenamefont {Cui}, \citenamefont {Amico}, \citenamefont {Fan},
  \citenamefont {Gu}, \citenamefont {Korepin}, \citenamefont {Kwek},\ and\
  \citenamefont {Vedral}}]{franchini:2013}%
  \BibitemOpen
  \bibfield  {author} {\bibinfo {author} {\bibfnamefont {F.}~\bibnamefont
  {Franchini}}, \bibinfo {author} {\bibfnamefont {J.}~\bibnamefont {Cui}},
  \bibinfo {author} {\bibfnamefont {L.}~\bibnamefont {Amico}}, \bibinfo
  {author} {\bibfnamefont {H.}~\bibnamefont {Fan}}, \bibinfo {author}
  {\bibfnamefont {M.}~\bibnamefont {Gu}}, \bibinfo {author} {\bibfnamefont
  {V.}~\bibnamefont {Korepin}}, \bibinfo {author} {\bibfnamefont {L.~C.}\
  \bibnamefont {Kwek}}, \ and\ \bibinfo {author} {\bibfnamefont
  {V.}~\bibnamefont {Vedral}},\ }\href {http://arxiv.org/abs/1306.6685}
  {\bibinfo  {journal} {arXiv:1306.6685}\ }\BibitemShut {NoStop}%




\bibitem [{\citenamefont {Arovas}\ \emph {et~al.}(1984)\citenamefont {Arovas},
  \citenamefont {Schrieffer},\ and\ \citenamefont {Wilczek}}]{arovas:84}%
  \BibitemOpen
\bibfield  {journal} {  }\bibfield  {author} {\bibinfo {author} {\bibfnamefont
  {D.}~\bibnamefont {Arovas}}, \bibinfo {author} {\bibfnamefont {J.~R.}\
  \bibnamefont {Schrieffer}}, \ and\ \bibinfo {author} {\bibfnamefont
  {F.}~\bibnamefont {Wilczek}},\ }\href {\doibase 10.1103/PhysRevLett.53.722}
  {\bibfield  {journal} {\bibinfo  {journal} {Phys. Rev. Lett.}\ }\textbf
  {\bibinfo {volume} {53}},\ \bibinfo {pages} {722} (\bibinfo {year}
  {1984})}\BibitemShut {NoStop}%
\bibitem [{\citenamefont {Wen}(1995)}]{wen:95}%
  \BibitemOpen
  \bibfield  {author} {\bibinfo {author} {\bibfnamefont {X.-G.}\ \bibnamefont
  {Wen}},\ }\href {\doibase 10.1080/00018739500101566} {\bibfield  {journal}
  {\bibinfo  {journal} {Advances in Physics}\ }\textbf {\bibinfo {volume}
  {44}},\ \bibinfo {pages} {405} (\bibinfo {year} {1995})},\ \Eprint
  {http://arxiv.org/abs/http://www.tandfonline.com/doi/pdf/10.1080/00018739500101566}
  {http://www.tandfonline.com/doi/pdf/10.1080/00018739500101566} \BibitemShut
  {NoStop}%



\bibitem [{\citenamefont {Trebst}\ \emph {et~al.}(2007)\citenamefont {Trebst},
  \citenamefont {Werner}, \citenamefont {Troyer}, \citenamefont {Shtengel},\
  and\ \citenamefont {Nayak}}]{trebst:2007}%
  \BibitemOpen
\bibfield  {journal} {  }\bibfield  {author} {\bibinfo {author} {\bibfnamefont
  {S.}~\bibnamefont {Trebst}}, \bibinfo {author} {\bibfnamefont
  {P.}~\bibnamefont {Werner}}, \bibinfo {author} {\bibfnamefont
  {M.}~\bibnamefont {Troyer}}, \bibinfo {author} {\bibfnamefont
  {K.}~\bibnamefont {Shtengel}}, \ and\ \bibinfo {author} {\bibfnamefont
  {C.}~\bibnamefont {Nayak}},\ }\href
  {http://prl.aps.org/abstract/PRL/v98/i7/e070602} {\bibfield  {journal}
  {\bibinfo  {journal} {Phys. Rev. Lett. 98, 070602}\ } (\bibinfo {year}
  {2007})}\BibitemShut {NoStop}%
\bibitem [{\citenamefont {Hamma}\ and\ \citenamefont
  {Lidar}(2008)}]{hamma:2008}%
  \BibitemOpen
  \bibfield  {author} {\bibinfo {author} {\bibfnamefont {A.}~\bibnamefont
  {Hamma}}\ and\ \bibinfo {author} {\bibfnamefont {D.~A.}\ \bibnamefont
  {Lidar}},\ }\href {http://prl.aps.org/abstract/PRL/v100/i3/e030502}
  {\bibfield  {journal} {\bibinfo  {journal} {Phys. Rev. Lett. 100, 030502}\ }
  (\bibinfo {year} {2008})}\BibitemShut {NoStop}%
\bibitem [{\citenamefont {Jahromi}\ \emph {et~al.}(2013)\citenamefont
  {Jahromi}, \citenamefont {Kargarian}, \citenamefont {Masoudi},\ and\
  \citenamefont {Schmidt}}]{jahromi:2013}%
  \BibitemOpen
  \bibfield  {author} {\bibinfo {author} {\bibfnamefont {S.~S.}\ \bibnamefont
  {Jahromi}}, \bibinfo {author} {\bibfnamefont {M.}~\bibnamefont {Kargarian}},
  \bibinfo {author} {\bibfnamefont {S.~F.}\ \bibnamefont {Masoudi}}, \ and\
  \bibinfo {author} {\bibfnamefont {K.~P.}\ \bibnamefont {Schmidt}},\ }\href
  {http://prb.aps.org/abstract/PRB/v87/i9/e094413} {\bibfield  {journal}
  {\bibinfo  {journal} {Phys. Rev. B 87, 094413}\ } (\bibinfo {year}
  {2013})}\BibitemShut {NoStop}%
\bibitem [{\citenamefont {Dusuel}\ \emph {et~al.}(2011)\citenamefont {Dusuel},
  \citenamefont {Kamfor}, \citenamefont {Or\'{u}s}, \citenamefont {Schmidt},\
  and\ \citenamefont {Vidal}}]{dusuel:2011}%
  \BibitemOpen
  \bibfield  {author} {\bibinfo {author} {\bibfnamefont {S.}~\bibnamefont
  {Dusuel}}, \bibinfo {author} {\bibfnamefont {M.}~\bibnamefont {Kamfor}},
  \bibinfo {author} {\bibfnamefont {R.}~\bibnamefont {Or\'{u}s}}, \bibinfo
  {author} {\bibfnamefont {K.~P.}\ \bibnamefont {Schmidt}}, \ and\ \bibinfo
  {author} {\bibfnamefont {J.}~\bibnamefont {Vidal}},\ }\href
  {http://prl.aps.org/abstract/PRL/v106/i10/e107203} {\bibfield  {journal}
  {\bibinfo  {journal} {Phys. Rev. Lett. 106, 107203}\ } (\bibinfo {year}
  {2011})}\BibitemShut {NoStop}%
\bibitem [{\citenamefont {Nielsen}(1999)}]{nielsen:1999}%
  \BibitemOpen
  \bibfield  {author} {\bibinfo {author} {\bibfnamefont {M.}~\bibnamefont
  {Nielsen}},\ }\href {http://prl.aps.org/abstract/PRL/v83/i2/p436_1}
  {\bibfield  {journal} {\bibinfo  {journal} {Phys. Rev. Lett. 83, 436–439}\
  } (\bibinfo {year} {1999})}\BibitemShut {NoStop}%
\bibitem [{\citenamefont {Castelnovo}\ and\ \citenamefont
  {Chamon}(2008)}]{castelnovo:2008}%
  \BibitemOpen
  \bibfield  {author} {\bibinfo {author} {\bibfnamefont {C.}~\bibnamefont
  {Castelnovo}}\ and\ \bibinfo {author} {\bibfnamefont {C.}~\bibnamefont
  {Chamon}},\ }\href {http://prb.aps.org/abstract/PRB/v77/i5/e054433}
  {\bibfield  {journal} {\bibinfo  {journal} {Phys. Rev. B 77, 054433}\ }
  (\bibinfo {year} {2008})}\BibitemShut {NoStop}%
\bibitem [{\citenamefont {Halasz}\ and\ \citenamefont
  {Hamma}(2012)}]{halasz:2012a}%
  \BibitemOpen
  \bibfield  {author} {\bibinfo {author} {\bibfnamefont {G.}~\bibnamefont
  {Halasz}}\ and\ \bibinfo {author} {\bibfnamefont {A.}~\bibnamefont {Hamma}},\
  }\href {http://pra.aps.org/abstract/PRA/v86/i6/e062330} {\bibfield  {journal}
  {\bibinfo  {journal} {Phys. Rev. A 86, 062330}\ } (\bibinfo {year}
  {2012})}\BibitemShut {NoStop}%
\bibitem [{\citenamefont {Yu}\ \emph {et~al.}(2008)\citenamefont {Yu},
  \citenamefont {Kou},\ and\ \citenamefont {Wen}}]{yu:2008}%
  \BibitemOpen
  \bibfield  {author} {\bibinfo {author} {\bibfnamefont {J.}~\bibnamefont
  {Yu}}, \bibinfo {author} {\bibfnamefont {S.-P.}\ \bibnamefont {Kou}}, \ and\
  \bibinfo {author} {\bibfnamefont {X.-G.}\ \bibnamefont {Wen}},\ }\href
  {http://iopscience.iop.org/0295-5075/84/1/17004/fulltext/} {\bibfield
  {journal} {\bibinfo  {journal} {Europhys. Lett. 84, 17004}\ } (\bibinfo
  {year} {2008})}\BibitemShut {NoStop}%
\bibitem [{\citenamefont {Halasz}\ and\ \citenamefont
  {Hamma}(2013)}]{halasz:2012b}%
  \BibitemOpen
  \bibfield  {author} {\bibinfo {author} {\bibfnamefont {G.}~\bibnamefont
  {Halasz}}\ and\ \bibinfo {author} {\bibfnamefont {A.}~\bibnamefont {Hamma}},\
  }\href {http://prl.aps.org/abstract/PRL/v110/i17/e170605} {\bibfield
  {journal} {\bibinfo  {journal} {Phys. Rev. Lett. 110, 170605}\ } (\bibinfo
  {year} {2013})}\BibitemShut {NoStop}%
\bibitem [{\citenamefont {White}()}]{white:1992}%
  \BibitemOpen
  \bibfield  {author} {\bibinfo {author} {\bibfnamefont {S.}~\bibnamefont
  {White}},\ }\href {http://prl.aps.org/abstract/PRL/v69/i19/p2863_1} {\bibinfo
   {journal} {Phys. Rev. Lett. 69, 2863–2866 (1992)}\ }\BibitemShut {NoStop}%
\bibitem [{\citenamefont {McCulloch}()}]{culloch:2008}%
  \BibitemOpen
\bibfield  {journal} {  }\bibfield  {author} {\bibinfo {author} {\bibfnamefont
  {I.}~\bibnamefont {McCulloch}},\ }\href {http://arxiv.org/abs/0804.2509}
  {\bibinfo  {journal} {arXiv:0804.2509}\ }\BibitemShut {NoStop}%
\bibitem [{\citenamefont {Crosswhite}\ \emph {et~al.}()\citenamefont
  {Crosswhite}, \citenamefont {Doherty},\ and\ \citenamefont
  {Vidal}}]{crosswhite:2008}%
  \BibitemOpen
\bibfield  {journal} {  }\bibfield  {author} {\bibinfo {author} {\bibfnamefont
  {G.~M.}\ \bibnamefont {Crosswhite}}, \bibinfo {author} {\bibfnamefont
  {A.}~\bibnamefont {Doherty}}, \ and\ \bibinfo {author} {\bibfnamefont
  {G.}~\bibnamefont {Vidal}},\ }\href
  {http://prb.aps.org/abstract/PRB/v78/i3/e035116} {\bibinfo  {journal} {Phys.
  Rev. B 78, 035116 (2008)}\ }\BibitemShut {NoStop}%





\bibitem [{\citenamefont {Cincio}\ and\ \citenamefont {Vidal}()}]{cincio:2013}%
  \BibitemOpen
\bibfield  {journal} {  }\bibfield  {author} {\bibinfo {author} {\bibfnamefont
  {L.}~\bibnamefont {Cincio}}\ and\ \bibinfo {author} {\bibfnamefont
  {G.}~\bibnamefont {Vidal}},\ }\href
  {http://prl.aps.org/abstract/PRL/v110/i6/e067208} {\bibinfo  {journal} {Phys.
  Rev. Lett. 110, 067208 (2013)}\ }\BibitemShut {NoStop}%
\bibitem [{\citenamefont {Vidal}(2003)}]{vidal:2003}%
  \BibitemOpen
\bibfield  {journal} {  }\bibfield  {author} {\bibinfo {author} {\bibfnamefont
  {G.}~\bibnamefont {Vidal}},\ }\href
  {http://prl.aps.org/abstract/PRL/v91/i14/e147902} {\bibfield  {journal}
  {\bibinfo  {journal} {Phys. Rev. Lett. 91, 147902}\ } (\bibinfo {year}
  {2003})}\BibitemShut {NoStop}%



\bibitem [{\citenamefont {Chitambar}\ \emph {et~al.}()\citenamefont
  {Chitambar}, \citenamefont {Leung}, \citenamefont {Mancinska}, \citenamefont
  {Ozols},\ and\ \citenamefont {Winter}}]{chitambar:2012}%
  \BibitemOpen
\bibfield  {journal} {  }\bibfield  {author} {\bibinfo {author} {\bibfnamefont
  {E.}~\bibnamefont {Chitambar}}, \bibinfo {author} {\bibfnamefont
  {D.}~\bibnamefont {Leung}}, \bibinfo {author} {\bibfnamefont
  {L.}~\bibnamefont {Mancinska}}, \bibinfo {author} {\bibfnamefont
  {M.}~\bibnamefont {Ozols}}, \ and\ \bibinfo {author} {\bibfnamefont
  {A.}~\bibnamefont {Winter}},\ }\href {http://arxiv.org/abs/1210.4583}
  {\bibinfo  {journal} {arXiv:1210.4583}\ }\BibitemShut {NoStop}%


\bibitem [{\citenamefont {Vidal}\ \emph
  {et~al.}(2009{\natexlab{a}})\citenamefont {Vidal}, \citenamefont {Dusuel},\
  and\ \citenamefont {Schmidt}}]{jvidal:2009}%
  \BibitemOpen
  \bibfield  {author} {\bibinfo {author} {\bibfnamefont {J.}~\bibnamefont
  {Vidal}}, \bibinfo {author} {\bibfnamefont {S.}~\bibnamefont {Dusuel}}, \
  and\ \bibinfo {author} {\bibfnamefont {K.~P.}\ \bibnamefont {Schmidt}},\
  }\href {http://prb.aps.org/abstract/PRB/v79/i3/e033109} {\bibfield  {journal}
  {\bibinfo  {journal} {Phys. Rev. B 79, 033109}\ } (\bibinfo {year}
  {2009}{\natexlab{a}})}\BibitemShut {NoStop}%
\bibitem [{\citenamefont {Vidal}\ \emph
  {et~al.}(2009{\natexlab{b}})\citenamefont {Vidal}, \citenamefont {Thomale},
  \citenamefont {Schmidt},\ and\ \citenamefont {Dusuel}}]{jvidal:2009b}%
  \BibitemOpen
  \bibfield  {author} {\bibinfo {author} {\bibfnamefont {J.}~\bibnamefont
  {Vidal}}, \bibinfo {author} {\bibfnamefont {R.}~\bibnamefont {Thomale}},
  \bibinfo {author} {\bibfnamefont {K.~P.}\ \bibnamefont {Schmidt}}, \ and\
  \bibinfo {author} {\bibfnamefont {S.}~\bibnamefont {Dusuel}},\ }\href
  {http://prb.aps.org/abstract/PRB/v80/i8/e081104} {\bibfield  {journal}
  {\bibinfo  {journal} {Phys. Rev. B 80, 081104(R)}\ } (\bibinfo {year}
  {2009}{\natexlab{b}})}\BibitemShut {NoStop}%
\bibitem [{\citenamefont {Wu}\ \emph {et~al.}(2012)\citenamefont {Wu},
  \citenamefont {Deng},\ and\ \citenamefont {Prokof'ev}}]{wu:2012}%
  \BibitemOpen
  \bibfield  {author} {\bibinfo {author} {\bibfnamefont {F.}~\bibnamefont
  {Wu}}, \bibinfo {author} {\bibfnamefont {Y.}~\bibnamefont {Deng}}, \ and\
  \bibinfo {author} {\bibfnamefont {N.}~\bibnamefont {Prokof'ev}},\ }\href
  {http://prb.aps.org/abstract/PRB/v85/i19/e195104} {\bibfield  {journal}
  {\bibinfo  {journal} {Phys. Rev. B 85, 195104}\ } (\bibinfo {year}
  {2012})}\BibitemShut {NoStop}%
\bibitem [{\citenamefont {Cincio}\ and\ \citenamefont
  {Vidal}(tion)}]{longDMRGtopo}%
  \BibitemOpen
  \bibfield  {author} {\bibinfo {author} {\bibfnamefont {L.}~\bibnamefont
  {Cincio}}\ and\ \bibinfo {author} {\bibfnamefont {G.}~\bibnamefont {Vidal}},\
  }\href@noop {} {} (\bibinfo {year} {\emph{in preparation}})\BibitemShut
  {NoStop}%
\bibitem [{\citenamefont {Karimipour}\ \emph {et~al.}(2013)\citenamefont
  {Karimipour}, \citenamefont {Memarzadeh},\ and\ \citenamefont
  {Zarkeshian}}]{karimipour:2013}%
  \BibitemOpen
  \bibfield  {author} {\bibinfo {author} {\bibfnamefont {V.}~\bibnamefont
  {Karimipour}}, \bibinfo {author} {\bibfnamefont {L.}~\bibnamefont
  {Memarzadeh}}, \ and\ \bibinfo {author} {\bibfnamefont {P.}~\bibnamefont
  {Zarkeshian}},\ }\href {http://pra.aps.org/abstract/PRA/v87/i3/e032322}
  {\bibfield  {journal} {\bibinfo  {journal} {Phys. Rev. A 87, 032322}\ }
  (\bibinfo {year} {2013})}\BibitemShut {NoStop}%
\bibitem [{\citenamefont {Pfeuty}(1970)}]{pfeuty:1970}%
  \BibitemOpen
  \bibfield  {author} {\bibinfo {author} {\bibfnamefont {P.}~\bibnamefont
  {Pfeuty}},\ }\href
  {http://www.sciencedirect.com/science/article/pii/0003491670902708}
  {\bibfield  {journal} {\bibinfo  {journal} {Annals of Physics 57, 79-90}\ }
  (\bibinfo {year} {1970})}\BibitemShut {NoStop}%
\bibitem [{\citenamefont {Barouch}\ and\ \citenamefont
  {McCoy}(1971)}]{barouch:1971}%
  \BibitemOpen
  \bibfield  {author} {\bibinfo {author} {\bibfnamefont {E.}~\bibnamefont
  {Barouch}}\ and\ \bibinfo {author} {\bibfnamefont {B.~M.}\ \bibnamefont
  {McCoy}},\ }\href {http://pra.aps.org/abstract/PRA/v3/i2/p786_1} {\bibfield
  {journal} {\bibinfo  {journal} {Phys. Rev. A 3, 786–804}\ } (\bibinfo
  {year} {1971})}\BibitemShut {NoStop}%
\bibitem [{\citenamefont {Cui}\ \emph {et~al.}(2012)\citenamefont {Cui},
  \citenamefont {Gu}, \citenamefont {Kwek}, \citenamefont {Santos},
  \citenamefont {Fan},\ and\ \citenamefont {Vedral}}]{cui:2012}%
  \BibitemOpen
  \bibfield  {author} {\bibinfo {author} {\bibfnamefont {J.}~\bibnamefont
  {Cui}}, \bibinfo {author} {\bibfnamefont {M.}~\bibnamefont {Gu}}, \bibinfo
  {author} {\bibfnamefont {L.}~\bibnamefont {Kwek}}, \bibinfo {author}
  {\bibfnamefont {M.}~\bibnamefont {Santos}}, \bibinfo {author} {\bibfnamefont
  {H.}~\bibnamefont {Fan}}, \ and\ \bibinfo {author} {\bibfnamefont
  {V.}~\bibnamefont {Vedral}},\ }\href
  {http://www.nature.com/ncomms/journal/v3/n5/full/ncomms1809.html} {\bibfield
  {journal} {\bibinfo  {journal} {Nature commun. 3, 812}\ } (\bibinfo {year}
  {2012})}\BibitemShut {NoStop}%
\bibitem [{\citenamefont {Kalis}\ \emph {et~al.}(2012)\citenamefont {Kalis},
  \citenamefont {Klagges}, \citenamefont {Orus},\ and\ \citenamefont
  {Schmidt}}]{kalis:2012}%
  \BibitemOpen
  \bibfield  {author} {\bibinfo {author} {\bibfnamefont {H.}~\bibnamefont
  {Kalis}}, \bibinfo {author} {\bibfnamefont {D.}~\bibnamefont {Klagges}},
  \bibinfo {author} {\bibfnamefont {R.}~\bibnamefont {Orus}}, \ and\ \bibinfo
  {author} {\bibfnamefont {K.}~\bibnamefont {Schmidt}},\ }\href
  {http://pra.aps.org/abstract/PRA/v86/i2/e022317} {\bibfield  {journal}
  {\bibinfo  {journal} {Phys. Rev. A 86, 022317}\ } (\bibinfo {year}
  {2012})}\BibitemShut {NoStop}%
\bibitem [{\citenamefont {Son}\ \emph {et~al.}(2011)\citenamefont {Son},
  \citenamefont {Amico}, \citenamefont {Fazio}, \citenamefont {Hamma},
  \citenamefont {Pascazio},\ and\ \citenamefont {Vedral}}]{son:2011}%
  \BibitemOpen
  \bibfield  {author} {\bibinfo {author} {\bibfnamefont {W.}~\bibnamefont
  {Son}}, \bibinfo {author} {\bibfnamefont {L.}~\bibnamefont {Amico}}, \bibinfo
  {author} {\bibfnamefont {R.}~\bibnamefont {Fazio}}, \bibinfo {author}
  {\bibfnamefont {A.}~\bibnamefont {Hamma}}, \bibinfo {author} {\bibfnamefont
  {S.}~\bibnamefont {Pascazio}}, \ and\ \bibinfo {author} {\bibfnamefont
  {V.}~\bibnamefont {Vedral}},\ }\href
  {http://iopscience.iop.org/0295-5075/95/5/50001} {\bibfield  {journal}
  {\bibinfo  {journal} {Europhys. Lett. 95, 50001}\ } (\bibinfo {year}
  {2011})}\BibitemShut {NoStop}%
\bibitem [{\citenamefont {Nielsen}(2006)}]{nielsen:2006}%
  \BibitemOpen
  \bibfield  {author} {\bibinfo {author} {\bibfnamefont {M.}~\bibnamefont
  {Nielsen}},\ }\href
  {http://www.sciencedirect.com/science/article/pii/S0034487706800145}
  {\bibfield  {journal} {\bibinfo  {journal} {Rep. on Math. Phys., 57, 1}\ }
  (\bibinfo {year} {2006})}\BibitemShut {NoStop}%
\bibitem [{\citenamefont {Isakov}\ \emph {et~al.}(2011)\citenamefont {Isakov},
  \citenamefont {Hastings},\ and\ \citenamefont {Melko}}]{isakov:2011}%
  \BibitemOpen
  \bibfield  {author} {\bibinfo {author} {\bibfnamefont {S.}~\bibnamefont
  {Isakov}}, \bibinfo {author} {\bibfnamefont {M.}~\bibnamefont {Hastings}}, \
  and\ \bibinfo {author} {\bibfnamefont {R.}~\bibnamefont {Melko}},\ }\href
  {http://www.nature.com/nphys/journal/v7/n10/abs/nphys2036.html} {\bibfield
  {journal} {\bibinfo  {journal} {Nature Phys. 7, 772}\ } (\bibinfo {year}
  {2011})}\BibitemShut {NoStop}%
\bibitem [{\citenamefont {Sarma}\ and\ \citenamefont {Pinczuk}()}]{sdsarma}%
  \BibitemOpen
  \bibfield  {author} {\bibinfo {author} {\bibfnamefont {S.~D.}\ \bibnamefont
  {Sarma}}\ and\ \bibinfo {author} {\bibfnamefont {A.}~\bibnamefont
  {Pinczuk}},\ }\href
  {http://onlinelibrary.wiley.com/doi/10.1002/9783527617258.fmatter/pdf} {\emph
  {\bibinfo {title} {Quantum Hall effects}}}\ (\bibinfo  {publisher} {Wiley,
  New York, 1997 Eds.})\BibitemShut {NoStop}%
\bibitem [{\citenamefont {Chakraborty}\ and\ \citenamefont
  {Pietil\"{a}inen}()}]{tchak}%
  \BibitemOpen
  \bibfield  {author} {\bibinfo {author} {\bibfnamefont {T.}~\bibnamefont
  {Chakraborty}}\ and\ \bibinfo {author} {\bibfnamefont {P.}~\bibnamefont
  {Pietil\"{a}inen}},\ }\href
  {http://link.springer.com/chapter/10.1007\%2F978-3-642-79319-6_4} {\emph
  {\bibinfo {title} {The Quantum Hall effects}}}\ (\bibinfo  {publisher}
  {Springer Series in Solid-State Sciences Volume 85, 1995, pp
  32-38})\BibitemShut {NoStop}%
\bibitem [{\citenamefont {Papanikolaou}\ \emph {et~al.}(2007)\citenamefont
  {Papanikolaou}, \citenamefont {Raman},\ and\ \citenamefont
  {Fradkin}}]{papa:2007}%
  \BibitemOpen
  \bibfield  {author} {\bibinfo {author} {\bibfnamefont {S.}~\bibnamefont
  {Papanikolaou}}, \bibinfo {author} {\bibfnamefont {K.}~\bibnamefont {Raman}},
  \ and\ \bibinfo {author} {\bibfnamefont {E.}~\bibnamefont {Fradkin}},\ }\href
  {http://prb.aps.org/abstract/PRB/v76/i22/e224421} {\bibfield  {journal}
  {\bibinfo  {journal} {Phys. Rev. B 76, 224421}\ } (\bibinfo {year}
  {2007})}\BibitemShut {NoStop}%
  \bibitem{spectroscopy} M. Kamfor, S. Dusuel, J. Vidal, K. P. Schmidt,\href{http://arxiv.org/abs/1308.6150}{arXiv:1308.6150}
  \bibitem{hiz3} Alioscia Hamma, Radu Ionicioiu, Paolo Zanardi, \href{http://pra.aps.org/abstract/PRA/v72/i1/e012324}{Phys.Rev. A 72, 012324 (2005)}
  \bibitem{negativity1} Yirun Arthur Lee, Guifre Vidal, \href{http://arxiv.org/abs/1306.5711}{arXiv:1306.5711}
  \bibitem{negativity2}  Claudio Castelnovo, \href{http://arxiv.org/abs/1306.4990}{arXiv:1306.4990}
  \bibitem{preparation} A. Hamma et al. \emph{in preparation}
  \bibitem{finiteT} C.Castelnovo, and C. Chamon, \href{http://prb.aps.org/abstract/PRB/v76/i18/e184442}{Phys. Rev. B 76, 184442 (2007)}
  \bibitem{tso}D.I. Tsomokos, A. Hamma, W. Zhang, S. Haas, R. Fazio, \href{http://pra.aps.org/abstract/PRA/v80/i6/e060302}{Phys. Rev. A 80, 060302(R) (2009)}
  \bibitem{finiteT2} M.B. Hastings, \href{http://prl.aps.org/abstract/PRL/v107/i21/e210501}{Phys. Rev. Lett. 107, 210501 (2011)}
  \bibitem{finiteT3} D. Mazac, and A. Hamma, \href{http://www.sciencedirect.com/science/article/pii/S0003491612000723}{Ann. Phys. 327, 2096 (2012)}
  \bibitem{dusuel2} M. D. Schulz, S. Dusuel, R. Orus, J. Vidal, K. P. Schmidt, \href{http://iopscience.iop.org/1367-2630/14/2/025005}{New J. Phys. 14, 025005 (2012)}
  \bibitem{grover}Tarun Grover, Ari M. Turner, Ashvin Vishwanath, \href{http://prb.aps.org/abstract/PRB/v84/i19/e195120}{Phys. Rev. B 84, 195120 (2011)}
  \bibitem{cirac}J. Ignacio Cirac, Didier Poilblanc, Norbert Schuch, Frank Verstraete, \href{http://prb.aps.org/abstract/PRB/v83/i24/e245134}{Phys. Rev. B 83, 245134 (2011)}
  \bibitem{dyn2}S.V. Isakov, P. Fendley, A.W.W. Ludwig, S. Trebst, M. Troyer, \href{http://prb.aps.org/abstract/PRB/v83/i12/e125114}{Phys. Rev. B 83, 125114 (2011)}
\bibitem{confinement}K. Gregor, David A. Huse, R. Moessner, S. L. Sondhi, 	\href{http://iopscience.iop.org/1367-2630/13/2/025009}{New J.Phys.13:025009 (2011)}
\bibitem{correlations} Hao Wang, B. Bauer, M. Troyer, V. W. Scarola, \href{http://prb.aps.org/abstract/PRB/v83/i11/e115119}{Phys. Rev. B 83,115119 (2011)}
\bibitem{armin}Armin Rahmani, Claudio Chamon, \href{http://prb.aps.org/abstract/PRB/v82/i13/e134303}{Phys. Rev. B 82, 134303 (2010)}
\bibitem{klich} I. Klich, \href{http://www.sciencedirect.com/science/article/pii/S000349161000093X}{Annals of Physics, Volume 325, Issue 10, p. 2120-2131 (2010)}
\bibitem{bravyi}S. Bravyi, M. B. Hastings, \href{http://link.springer.com/article/10.1007\%2Fs00220-011-1346-2#}{Commun. Math. Phys. 307, 609 (2011)}
\bibitem{bravyi2}Sergey Bravyi, Matthew Hastings, Spyridon Michalakis, \href{http://scitation.aip.org/content/aip/journal/jmp/51/9/10.1063/1.3490195}{J. Math. Phys. 51 093512 (2010)}
\bibitem{nussinov}Zohar Nussinov and Gerardo Ortiz, \href{http://www.sciencedirect.com/science/article/pii/S0003491608001711}{Annals of Physics, Volume 324, Issue 5, May 2009, Pages 977-1057}
\bibitem{correlations2} Zohar Nussinov and Gerardo Ortiz, \href{http://prb.aps.org/abstract/PRB/v77/i6/e064302}{Phys. Rev. B 77, 064302 (2008)}
\bibitem [{\citenamefont {Raussendorf}\ \emph {et~al.}(2005)\citenamefont
  {Raussendorf}, \citenamefont {Bravyi},\ and\ \citenamefont
  {Harrington}}]{raussendorf}%
  \BibitemOpen
  \bibfield  {author} {\bibinfo {author} {\bibfnamefont {R.}~\bibnamefont
  {Raussendorf}}, \bibinfo {author} {\bibfnamefont {S.}~\bibnamefont {Bravyi}},
  \ and\ \bibinfo {author} {\bibfnamefont {J.}~\bibnamefont {Harrington}},\
  }\href {\doibase 10.1103/PhysRevA.71.062313} {\bibfield  {journal} {\bibinfo
  {journal} {Phys. Rev. A}\ }\textbf {\bibinfo {volume} {71}},\ \bibinfo
  {pages} {062313} (\bibinfo {year} {2005})}\BibitemShut {NoStop}%
\bibitem{puritynote} The purity $P=\tr(\rho^2)$ of a density matrix $\rho\in \mc{H}_V$ can be expressed as the measurement of a Hermitian operator by taking two copies of the state and measuring the swap operator acting on a doubled Hilbert space $P=\tr_{V,V'}(\rho\otimes\rho~S_{V,V'}),~\mc{H}_V\cong\mc{H}_{V'},~S_{V,V'}^\dagger S_{V,V'})=\openone_{V,V'}$
\bibitem{globalnature} `global nature' refers to, in the case of topological order, long-range entanglement in the sense of Ref.~\cite{chen:2010} i.e. one that cannot be removed via local quantum circuits of finite depth. In the case of cluster states it refers to the fact that such states, although not possessing a local order parameter, have distant parts entangled \cite{raussendorf}.
 \end{thebibliography}
\end{document}